\documentclass{jfm}
\usepackage{graphicx}
\usepackage{epstopdf}
\usepackage{amsmath}
\usepackage{siunitx}

\usepackage{natbib}
\usepackage{booktabs}
\usepackage{subcaption}
\usepackage{float}
\usepackage{svg}
\usepackage{pgfplots}
\usepackage{pdflscape} 
\usepackage{rotating}
\usepackage{svg}
\usepackage{comment}
\usepackage[normalem]{ulem}
\newcommand{\blue}[1]{{\color{black}{#1}}}
\usepackage{tikz}
\usetikzlibrary{shapes.geometric}
\DeclareRobustCommand{\redstar}{%
  \tikz[baseline=-0.6ex]\node[star, star points=6, inner sep=0pt, minimum size=2mm, fill=red] {};%
}

\shorttitle{Acoustic-flow interaction over an acoustic liner}
\shortauthor{A. Paduano, F. Scarano, J. Cordioli, D. Casalino and F. Avallone}

\title{On the impact of the turbulent grazing flow development on the
acoustic response of an acoustic liner}

\author{A. Paduano\aff{1}
  \corresp{\email{angelo.paduano@polito.it}},
  F. Scarano\aff{1},
 J. Cordioli\aff{2},
 D. Casalino\aff{3}
 \and F. Avallone\aff{1}}

\affiliation{\aff{1}Department of Mechanical and Aerospace Engineering, Polytechnic of Turin, Turin
\aff{2}Department of Mechanical Engineering, Federal University of Santa Catarina, Florianópolis
\aff{3}Flow Physics and Technology Department, Delft University of Technology, Delft}

\begin{document}

\maketitle

\begin{abstract}
The interaction between acoustic waves and turbulent grazing flow over an acoustic liner is investigated using Lattice-Boltzmann Very-Large-Eddy simulations. A single-degree-of-freedom liner with 11 streamwise-aligned cavities is studied in a grazing flow impedance tube. The conditions replicate reference experiments from the Federal University of Santa Catarina. The influence of grazing flow (with a centerline Mach \blue{number} of 0.32), acoustic wave amplitude, frequency, and propagation direction relative to the mean flow is analysed.  Impedance is computed using both  \blue{ direct (i.e. the in-situ method)} and \blue{model-fitting inference (i.e. the mode-matching method)} methods. The \blue{former} reveals strong spatial variations; however, averaged values throughout the sample show minimal differences between upstream and downstream propagating waves, in contrast to \blue{what is obtained with the latter} method. Flow analyses reveal that the orifices displace the flow away from the face sheet, with this effect amplified by acoustic waves and dependent on the wave propagation direction. Consequently, the boundary layer displacement thickness ($\delta^*$) increases along the streamwise direction compared to a smooth wall and exhibits localised humps downstream of each orifice. The growth of $\delta^*$ alters the flow dynamics within the orifices by weakening the shear layer at downstream positions. This influences the acoustic-induced mass flow rate through the orifices \blue{at equal Sound Pressure Level},  suggesting that acoustic energy is dissipated differently along the liner. The \blue{asymmetry of the flow field experienced by the acoustic wave, depending on its propagation direction,} highlights the need to consider a spatially evolving turbulent flow when studying the acoustic–flow interaction and measuring impedance. 

\end{abstract}


\section{Introduction}

Acoustic liners are components of aircraft engines adopted to reduce noise \blue{(Figure \ref{fig:scheme_jfm} (a))}. They are usually installed in the engines' intake and core jet section. The recent development of ultra-high bypass ratio engines, characterised by a larger fan diameter compared to traditional high-bypass ratio engines, has significantly increased the contribution of fan noise to the overall engine noise. This noise source consists of two main components: a tonal component at the Blade-Passing Frequency (BPF) and its harmonics, and a broadband component generated by turbulence impingement of the fan wake on the stator, which arises from fan/stator proximity \citep{Mallat1989ARepresentation, Hughes2011TheIntegration, Casalino2018TurbofanMethod}.

Various classes of liners with customizable sound absorption properties exist, ranging in complexity. Conventional acoustic liners consist of honeycomb cavities enclosed between a perforated face sheet and a rigid backplate \citep{Motsinger19914Treatment}, a configuration commonly referred to as a Single Degree of Freedom (SDOF) liner \blue{(Figure \ref{fig:scheme_jfm} (b))}. They operate on the principle of a Helmholtz resonator to dissipate incident acoustic energy. The resonant frequency of the liner is typically tuned to coincide with the fan BPF or its harmonics, making SDOF liners particularly suitable for fan noise attenuation. 
The geometry of SDOF liners is defined by five key parameters: the number of orifices, their diameter \( d \), the thickness \( \tau \) of the facesheet, and the cavity area \( A \) and depth \( \blue{\zeta} \). In the absence of grazing flow, the resonant frequency is expressed as \citep{PantonResonantresonators1975}:

\begin{equation}
f_{0} = \frac{a_0}{2 \pi} \sqrt{\frac{S}{V_c (\tau + \tau^*) + P}},
\label{eq:resonance_frequency}
\end{equation}
\\
where \( a_0 \) is the speed of sound, \( S \) is the orifice area, and \( V_c = A \blue{\zeta} \) is the volume of the cavity. The terms \( P = \frac{1}{3} \blue{\zeta}^2 A \) and \( \tau^* \approx 0.8\sqrt{S/\pi} \) are end-corrections, accounting for the oscillatory motion of not only the fluid medium within the neck of the orifice but also a small portion of fluid inside the cavity and outside the orifice.
\blue{\begin{figure}
    \centering
    \includegraphics[width=0.8\linewidth]{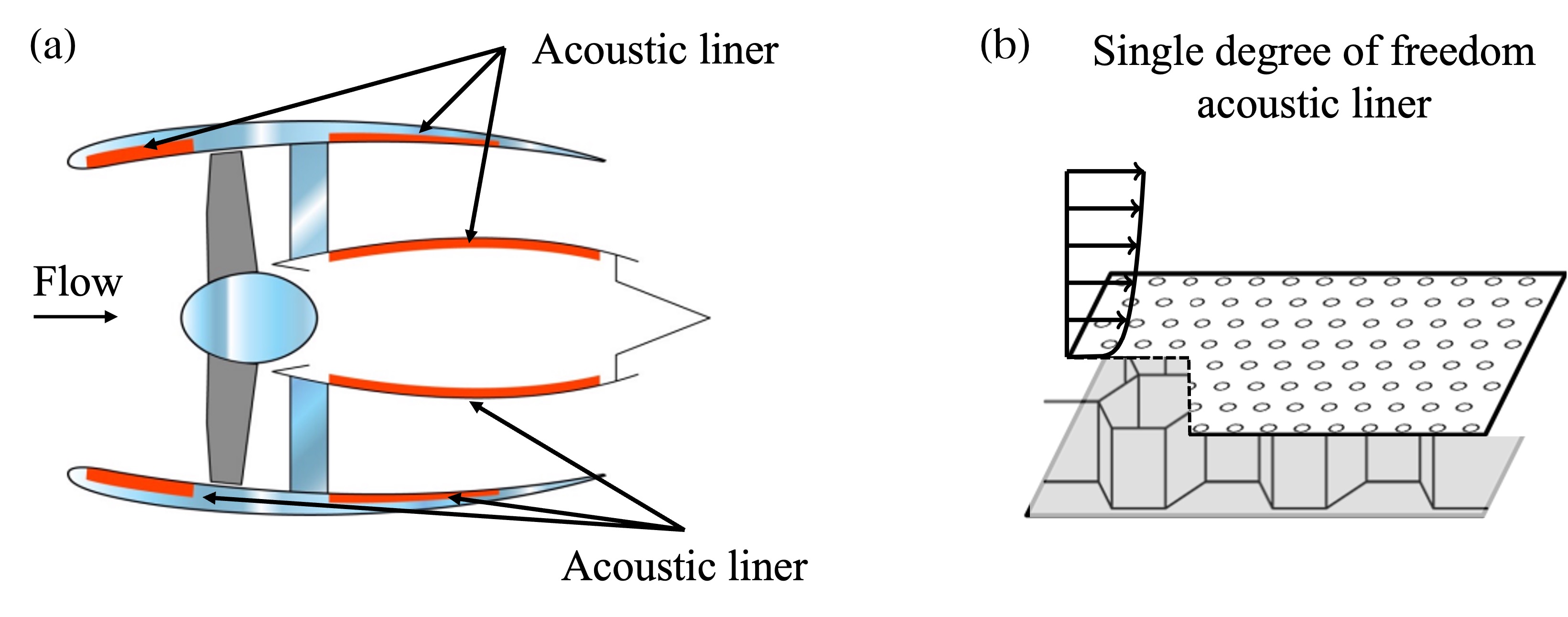}
    \caption{\blue{(a) Turbofan engine scheme with detail on location of acoustic liners. (b) Sketch of a single degree of freedom acoustic liner.}}
    \label{fig:scheme_jfm}
\end{figure}}
In the realm of classical acoustics, SDOF liners are commonly regarded as locally reacting, meaning their response hinges solely on the local Sound Pressure Level (SPL) and acoustic particle velocity rather than the angle of incidence of the acoustic wave \citep{Rienstra2004AnAcoustic, Motsinger19914Treatment}. Essentially, this \blue{implies that the wavelength of the acoustic wave is significantly larger than the backing cavity width, thus no wave motion in the liner is possible in the transverse direction \citep{CUMMINGS19769,NILSSON1980}.}

A SDOF liner dissipates acoustic energy by viscosity at the orifice side walls and vortex shedding. The liner \blue{it is said to operate predominantly} in the linear regime when the incident acoustic wave has a low SPL, conventionally below \SI{130}{dB} \blue{\citep{SCARANO2025119568}}. In this case, the acoustic energy is mainly dissipated through viscous effects along the internal walls of the liner's orifices, where a laminar boundary layer develops \citep{tam_microfluid_2000}. 
As the SPL increases, the dissipation mechanism is dominated by the formation of jets and vortex shedding at the mouths of the orifices \citep{howe_absorption_1984, Leon2019Near-wallFlow, tam_computational_2010, Zhang_Bodony_2012}. This regime is referred to as non-linear.
The acoustic energy is converted into turbulent kinetic energy, associated with the rotational motion of the vortices, which is subsequently dissipated as heat through viscous effects \citep{tam_microfluid_2000}. According to \citet{tam_microfluid_2000}, vortex shedding is amplified at frequencies close to the liner's resonance, but it is not influenced by the angle of incidence of the acoustic waves. In a more recent experimental study, in the absence of flow, \cite{tang_piv_2024} showed the formation of multi-scale vortex structures excited by acoustic waves. Similar findings were also found numerically by \cite{Zhang2016NumericalLayers} in the presence of grazing flow. They described how the flow field is linked to the acoustic response of an acoustic liner and how this changes if the boundary layer is laminar or turbulent. It was found that the impact of the boundary layer state (i.e., laminar or turbulent) is more pronounced at low SPL. 

A widely used approach to characterise a liner is to utilise a spatially homogeneous quantity named acoustic impedance. It is defined in the frequency domain as:
\begin{equation}
\hat{Z}(\omega)= \frac{\hat{p}}{{\hat{\textbf{v}}} \cdot \textbf{n}} = \theta + i\chi,
\end{equation}
\\
where $\hat{p}$ is the acoustic complex pressure \blue{such that $p(\mathbf{x},t)=\Re(\hat{p}\, e^{i\omega t})$}, ${\hat{\textbf{v}}} \cdot \textbf{n}$ is the acoustic particle velocity normal to the surface (with $\textbf{n}$ denoting the unit normal vector pointing outward from the surface). \blue{ The quantity $\theta$ is the resistance and $\chi$ the reactance. Under this convention, $\chi>0$ corresponds to mass-like behaviour, while negative reactance indicates spring-like behaviour \citep{Rienstra2004AnAcoustic}}.
Although impedance is an intrinsic property of the liner’s surface and shall remain independent of the duct geometry in which the liner is tested, several studies have highlighted its sensitivity to the eduction technique, i.e., the method used to calculate impedance \citep{AvalloneF2024OnDatabase}, and the flow profile within the duct \citep{Quintino2025}. The presence of a grazing flow introduces additional complexities to the impedance eduction process \citep{Schultz2021} since impedance depends not only by geometrical and acoustic parameters like orifice diameter, face sheet thickness, cavity depth, SPL, but also by the flow Mach number and the boundary layer displacement thickness, \(\delta^{*}\) \citep{NAYFEH1974413, Jones2002EffectsImpedance, Temiz2015ThePlates, Bonomo2023AProfiles, Quintino2025}.

Even though the physics of acoustic liners is well-known when they are exposed solely to an acoustic wave \citep{Melling1973THELEVELS, Tam2000MicrofluidLiners}, a gap persists in our knowledge when the liners operate in the presence of both an acoustic wave and grazing turbulent flow \citep{Murray2012DevelopmentSPL, Zhang2016NumericalLayers, Kooijman2008AcousticalGeometry, Avallone2021Acoustic-inducedLayer, Shahzad2023TurbulenceLiners}. \citet{Hersh1979EFFECTORIFICES} studied experimentally a multiple-orifice Helmholtz resonator with grazing ﬂow. They found that the reactance depends on the orifice spacing, in particular when the orifices are aligned parallel to the grazing ﬂow direction. The interaction between the acoustic-induced flow field and the grazing flow was visualised for the first time by \citet{KennethBaumeister1975NASAOrifice}. They identified the presence of a vortex at the upstream edge of the orifice neck, leading to a reduction in the effective inflow area. A more recent experimental study by \citet{Leon2019Near-wallFlow}, using particle image velocimetry, showed how the near-orifice flow is altered in the presence of a grazing acoustic wave and turbulent flow. Their study highlighted the ratio between the acoustic velocity and the shear velocity as a key parameter governing the transition between linear and non-linear operating regimes. These experiments provided valuable insights into the flow near the wall; however, the flow dynamics inside the orifice remain largely inaccessible to experimental observation and can only be comprehensively investigated through high-fidelity numerical simulations.

Several computational studies have investigated the physics of the flow within the orifice. Early efforts focused on simplified geometries in the absence of grazing flow \citet{Tam2000MicrofluidLiners}, later progressing to cases incorporating grazing flow \citep{Avallone2019Lattice-boltzmannFlow, Avallone2021Acoustic-inducedLayer, Zhang2011NumericalFlow}.
Fully three-dimensional numerical simulations of sound interacting with laminar and turbulent boundary layers were performed by \citet{Zhang2016NumericalLayers}. While these computational studies provide valuable descriptions of the flow and acoustic fields within the orifice, they remain confined to single-resonator configurations and do not account for boundary layer modifications induced by the presence of multiple cavities. In the recent study by \cite{Shahzad2023TurbulenceLiners}, three-dimensional direct numerical simulations were conducted on channel flow with a full acoustic liner mounted on the walls. The work provided an in-depth characterisation of how the turbulent grazing flow is altered in the presence of the liner, with particular emphasis on the flow development within the orifices and the added aerodynamic drag. However, the absence of an acoustic source raises important questions regarding how incident acoustic waves modify the grazing flow and the flow topology within the orifices, and how these modifications influence the acoustic impedance. \cite{Tam2014ExperimentalDrag}  performed 2D simulations of an array of eight-orifice slit resonators in the presence of acoustic waves. Particular attention was given to replicating the spatially developing flow in the Grazing Flow Impedance Tube (GFIT) at NASA Langley through a tuned eddy viscosity model along the duct. Despite the computed impedance showing similar trends as in the experiments, there were differences in the resistance and reactance values, especially at the lowest frequencies. In a more recent effort, \citet{Pereira2021ValidationFlow} conducted Lattice-Boltzmann Very Large Eddies Simulations (LB-VLES) of an array of eleven resonators, each composed of eight orifices. The computational setup was designed to replicate the Federal University of Santa Catarina (UFSC) test rig. Different techniques were employed to educe impedance. Comparisons of impedance with experimental results revealed discrepancies up to a factor of two, highlighting that geometrical variations between the experimental sample and the one investigated numerically can affect the results \citep{Bonomo2022ASPL,Paduano2024}.

The utilisation of acoustic impedance to assess liner performance has spurred extensive investigation, leading to several semi-empirical models aimed at predicting this quantity.  \cite{Hersh1979EFFECTORIFICES} were among the first to devise a semi-empirical model to predict orifice resistance and reactance as a function of incident SPL, frequency, and orifice geometry.  \cite{Howe1996EmendationHalf-plane} formulated an expression for the Rayleigh conductivity of an aperture subjected to a high-Reynolds-number flow, and \cite{Cummings1987TheField} established a connection between acoustic resistance and discharge coefficient via a quasi-steady model. A semi-empirical model to account for linear and non-linear effects was developed by \cite{Yu2008ValidationData}.

Studies have also shown that impedance can vary depending on the direction of acoustic wave propagation when a grazing flow is present \citep{Auregan2011}. This observation challenges the conventional assumption of locally reacting liners. Some of these impedance variations have been attributed to simplified approximations of the boundary layer within the Ingard-Myers boundary condition applied in impedance eduction techniques \citep{Ingard1959InfluenceTransmission,Myers1980OnFlow}. This boundary condition assumes an infinitely thin vortex sheet along the liner surface. Extensive research efforts aimed to incorporate boundary layer profiles with small but finite thicknesses into the impedance boundary conditions \citep{Brambley2011, RIENSTRA_DARAU_2011, Auregan2001InfluenceWall}. However, all these studies are based on the assumption of homogeneous liner impedance. In practice, liners consist of numerous small perforations and the significant variation that the boundary layer undergoes when interacting with the liner should be taken into account.

Based on the existing literature, it is a fact that the impact of the grazing flow development and its integral quantity must be accounted for. The literature still lacks a quantitative analysis of the acoustic and fluid dynamic fields over a multi-cavity acoustic liner when both the acoustic waves and the flow are grazing. This could provide the basis for developing robust semi-empirical models for impedance estimation and clarifying the role of acoustic wave direction in the presence of grazing flow.
Furthermore, a clearer understanding of near-wall flow–acoustic interactions could help in the development of more effective acoustic liner geometries.

This study presents the results from high-fidelity numerical simulations conducted using LB-VLES on a nominal geometry that has been experimentally characterised \citep{Quintino2025}. The objective is to investigate the interaction between the acoustic field and the turbulent boundary layer, with a particular focus on the near-wall flow and in-orifice dynamics that influence the liner’s acoustic response. To explore these effects, a wide range of acoustic waves, varying in amplitude, frequency, and propagation direction, have been simulated. As a result, a comprehensive and open-access database has been developed, which is a valuable benchmark for future studies on acoustic–flow interactions over acoustic liners.

The paper is organized as follows: Section \ref{sec:Methodology} summarizes the methodology and the post-processing techniques, Section \ref{sec:Computationalsetup} describes the computational setup and the grid convergence study, Section \ref{sec:Acoustic_results} discusses the acoustic results, Section \ref{sec:aerodynamic_results} discusses fluid dynamic findings, Section \ref{sec:acousitc_induced_flow} delves into acoustic-induced flow within the orifices and Section \ref{sec:conclusion} draws main conclusions.

\section{Methodology}
\label{sec:Methodology}
\subsection{Flow solver}
\blue{The commercial software 3DS Simulia PowerFLOW$^{\copyright~}$
 version 6 has been used. The solver is based on the Lattice Boltzmann Method (LBM). A comprehensive introduction to the method can be found in \citet{Succi2001}. In the LBM framework, the fluid is described at a mesoscopic level through particle distribution functions, whose statistical moments yield the macroscopic quantities such as density, momentum, and energy.

The method originates from the continuous Boltzmann equation, which reads:
\begin{equation}
    \frac{\partial \mathbf{g}}{\partial t} + \boldsymbol{\xi} \cdot \frac{\partial \mathbf{g}}{\partial \mathbf{x}} + \mathbf{F} \cdot \frac{\partial \mathbf{g}}{\partial \mathbf{\xi}} = \Omega(g)
\end{equation}
where $\mathbf{g} (\boldsymbol{\xi}, \mathbf{x}, t)$ is the particle distribution function, giving the mass density of particles located within the mesoscopic volume $d\mathbf{x}$ around $\mathbf{x}$ and in the infinitesimal time interval $(t, t +dt)$ having a microscopic velocity within $(\boldsymbol{\xi},\boldsymbol{\xi}+ d\boldsymbol{\xi})$. The left-hand side represents free-streaming and the effect of external forces $\mathbf{F}$. $\Omega(g)$ is the collision operator and describes the interaction between particles. The Bhatnagar-Gross-Krook (BGK) model \citep{Bhatnagar1954ASystems} is adopted thanks to its simplicity:
\begin{equation}
    \Omega(g) = - \frac{1}{\tau} (g - g^{eq}),
\end{equation}

\noindent where $\tau$ is the relaxation time and $g^{eq}$ is the equilibrium distribution function derived from the Maxwell-Boltzmann equilibrium distribution.
In the LBM formulation, the continuous distribution function is discretised in velocity space, yielding a finite set of discrete distributions $\mathbf{g_i}$ associated with discrete velocities $\boldsymbol{\xi_i}$. The particle transport and collisions are solved on a Cartesian mesh (lattice); the discrete volume elements are called voxels (vx). The D3Q19 lattice scheme is employed, where “D3” refers to three spatial dimensions and “Q19” to the number of discrete velocity directions \citep{Qian1992LatticeEquation}.

Hence, the macroscopic flow quantities density $\rho$ and velocity $\mathbf{u}$ are obtained by discrete integration:
\begin{equation}
    \rho (\mathbf{x},t) = \sum_i \mathbf{g_i}( \mathbf{x}, t), \hspace{10mm} \rho \mathbf{u}(\mathbf{x},t) = \sum_i \boldsymbol{\xi_i} \mathbf{g_i}(\mathbf{x}, t) 
\end{equation}}

A VLES approach has been employed to resolve only the larger turbulence scales. The sub-grid scales are accounted for by adding a turbulent relaxation time to the viscous relaxation time using a turbulence model, based on the two-equation Renormalisation Group Theory (RNG) $k-\epsilon$ \citep{Yakhot1986Renormalization-groupTurbulence} given by:

\begin{equation}
    \tau_{\text{eff}} = \tau + C_{\mu} \frac{k^2/ \epsilon}{(1+\eta^2)^{1/2}},
\end{equation}
\\
where $C_{\mu}\!=\!0.09$ and $\eta$ are a combination of the local strain, local vorticity, and local helicity parameters. The term $\eta$ allows for mitigation of the sub-grid scale viscosity, in the presence of large resolved vortical structures \citep{Texeira1998IncorporatingMethod}.

It should be pointed out that the usage of the $k- \epsilon$ RNG model under the LBM-VLES framework differs significantly from its application in Reynolds-Averaged Navier-Stokes (RANS) simulations. In RANS, the Reynolds stress tensor is computed directly using the turbulence model to solve a closure problem. In contrast, under the LBM-VLES framework, the turbulence model modifies the relaxation properties of the Boltzmann equation, thereby influencing the local eddy viscosity. The Reynolds stresses are a consequence of the computed turbulent chaotic motion and not a model add-on to the governing flow equations.   This implementation enables the development of large turbulent eddies in the simulation domain and recovers to some extent a non-linear constitutive form of the Reynolds stresses.

The solver uses an extended turbulent wall model that dynamically incorporates the presence of a pressure gradient (PGE-WM) \citep{Texeira1998IncorporatingMethod}. 
This model takes into account the effect of the pressure gradient by rescaling the length-scale $y^+$, in the generalised law-of-the-wall \citep{Launder1974TheFlows}, by a scaling parameter $A$:

\begin{equation}
    u^+ = \frac{1}{k} ln \Biggl( \frac{y^+}{A} \Biggr) + B,
    \label{eq:lawofthewall}
\end{equation}
\\
where $B$ and $k$ are constants, $y^+\!=\!(u_{\tau} y)/\nu$ and $A$ is a function of the pressure gradient.
The parameter $A$ captures the physical consequence of the velocity profile slowing down and expanding due to the pressure gradient. It is defined as proposed by  \cite{Texeira1998IncorporatingMethod}:

\begin{align}
    A = 1 +\frac{\beta | \frac{d p}{d s} |}{\tau_{w}}, \quad  \boldsymbol{u} \cdot \frac{dp}{ds} > 0,\\
    A = 1, \quad \text{otherwise};
\end{align}
where $\tau_w$ is the wall shear stress, $dp/ds$ is the streamwise pressure gradient, $\mathbf{u}$ is the streamwise velocity, and $\beta$ is a length of the same order as the unresolved near-wall region.
\subsection{Impedance Measurement Techniques}
Two techniques have been used to compute impedance: the Mode Matching (MM) \citep{Elnady2004AnMeasurements} and the in-situ \citep{Dean1974AnDucts}. 

\label{sec:ImpedanceTechnique}

\subsubsection{Mode matching method}
The MM is an inverse impedance eduction method. It is based on minimising the difference between a computed acoustic field and measurements by iteratively varying the liner impedance. This method was first proposed by \citet{Elnady2004AnMeasurements} and subsequently validated by \citet{Elnady2009ValidationImpedance}. 

\begin{figure}
    \centering
    \includegraphics[width=\linewidth]{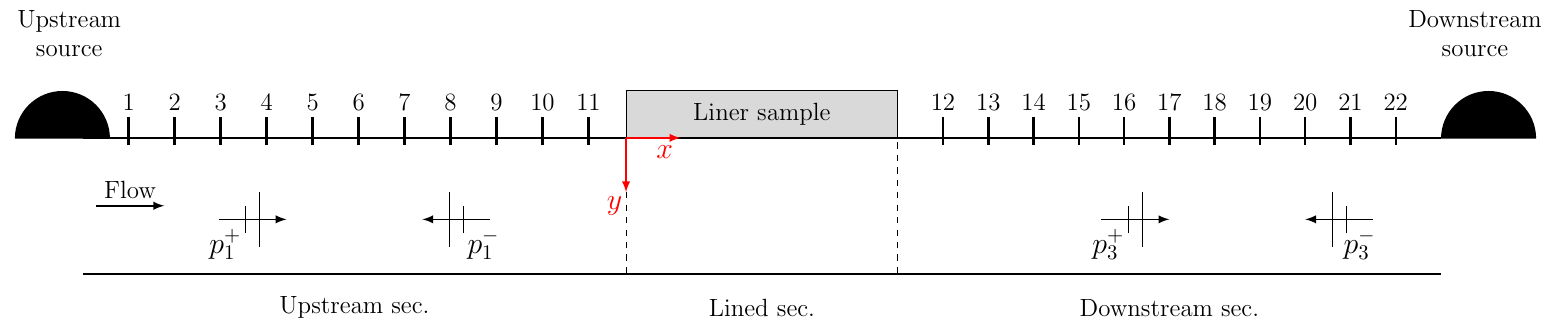}
    \caption{\blue{Representation of the acoustic field in the mode matching method and schematic view of the test rig.}}
    \label{fig:MM_scheme}
\end{figure}

This method requires pressure measurements upstream and downstream of the liner. \blue{Figure \ref{fig:MM_scheme} shows a schematic representation of probes location inside the duct}. The measured acoustic pressure at the rigid wall, opposite the lined one, \blue{can be written as :
\begin{equation}
    p(x,y,z)=\sum_{q=1}^{Q} A_q^{+}\,\Phi_q^{+}(y,z)\,\mathrm{e}^{-ik_{xq}^{+}x}
        +\sum_{q=1}^{Q} A_q^{-}\,\Phi_q^{-}(y,z)\,\mathrm{e}^{-ik_{xq}^{-}x},
\end{equation}
where superscripts \(+\) and \(-\) denote incident and reflected waves propagating in the positive and negative \textit{x}-directions, \(q\) is the modal index, \(A_q^{\pm}\) are the modal amplitudes, and \(k_{xq}\) is the axial wavenumber of mode \(q\), obtained from the dispersion relation and  $\Phi$ is the mode shape. The method assumes that only plane waves propagate in the hard-wall sections, which is justified since the channel has an infinite width and the frequencies considered are below the first cut-on frequency. Under this assumption, \(p_1^+\) represents the plane wave excited by the active source, \(p_3^-\) is the reflected plane wave at the duct termination, and \(p_1^-\) and \(p_3^+\) represent the scattered plane waves at the liner edges. 

Since the setup employs acoustic sponge layers, no acoustic waves are reflected at the duct termination. Therefore, \(p_3^- = 0\) when the acoustic source is located upstream, and \(p_1^+ = 0\) when it is located downstream.
}
Viscothermal losses are accounted for by correcting the plane wave axial wavenumber:
\begin{equation}
    k^{\pm}_{x1} = \frac{\pm \omega K_0}{(1 \pm K_0 M)},
\end{equation}
where $K_0$ is the first-order Kirchhoff solution given by:
\begin{equation}
    K_0 = 1 + \frac{1 - i}{Sh \sqrt{2}} \Biggl( 1 + \frac{ \gamma -1}{\sqrt{Pr}} \Biggr ),
\end{equation}
and $Sh = r\sqrt{\omega/\nu}$ is the shear wavenumber, \textit{r} is the hydraulic radius, $\gamma = 1.4$ is the heat capacity ratio and $Pr = 0.7$ is the Prandtl number. 
By using eleven microphones in each section, an over-determined system is obtained. Solving this system of equations in a least-squares sense for each set of microphones gives the amplitude of the waves propagating forward and backwards in the channel; these are used as inputs for the MM model. An initial guess on impedance for the lined section is obtained using the semi-empirical model by \citet{Yu2008ValidationData}. Then the impedance is obtained by minimising a cost function using the Levenberg-Marquardt algorithm \citep{Levenberg1944ASquares, Marquardt1963AnParameters}. The cost function is defined as:
\begin{equation}
    F(Z) = \sum_{i=1}^{22} \Biggl | \frac{p_{i}^{meas} - p_{i}^{analytic}(Z)}{p_{i}^{meas}}\Biggr|.
\end{equation}

Once the convergence criterion is satisfied, the liner impedance is obtained.

\subsubsection{In-situ technique}
The in-situ technique, also known as the two-microphones method, was first proposed by \citet{Dean1974AnDucts}. It requires unsteady pressure measurements at the face sheet and the bottom of the cavity. This method provides a point-wise measurement of impedance. It is based on the following key assumptions:
the wavelength of the incident acoustic wave is significantly larger than the cavity width.
The walls of the cavity are considered to be sufficiently thick, resulting in the liner being locally reactive;
any wave entering the cavity is assumed to be reflected at the backplate.
Therefore, the acoustic pressure of the standing wave within the cavity is the sum of the incident and reflected ones. Using the linearised momentum equation, it is possible to calculate the acoustic-induced velocity and, subsequently, the impedance as:

\begin{equation}
    Z_{f} = \frac{Z}{Z_0} = -i \tilde{H}_{fb} \frac{1}{\sin(k\zeta)},
    \label{eq:insitu-equation}
\end{equation}

\noindent where $Z_{0}$ is the characteristic impedance of air, $\tilde{H}_{fb}$ is the transfer function defined as the ratio between the pressure measured at the face sheet $\tilde{p}_{f}$ and at the backplate $\tilde{p}_{b}$, $\zeta$ is the depth of the cavity, and $k\!=\!\omega/c_0$ is the free-field wavenumber, with $\omega$ being the acoustic wave angular frequency.
This technique has been widely used for estimating the impedance of acoustic liners in the presence of grazing flow \citep{Schuster2012AMeasurements, Zhang2016NumericalLayers}. Unlike impedance eduction techniques, this approach does not require a flow-based boundary condition for capturing near-wall acoustic-flow interactions. However, studies have underlined the sensitivity of this technique to the sampling position \citep{Avallone2021Acoustic-inducedLayer}. 

\begin{figure}
    \centering
    \includegraphics[width=0.8\linewidth]{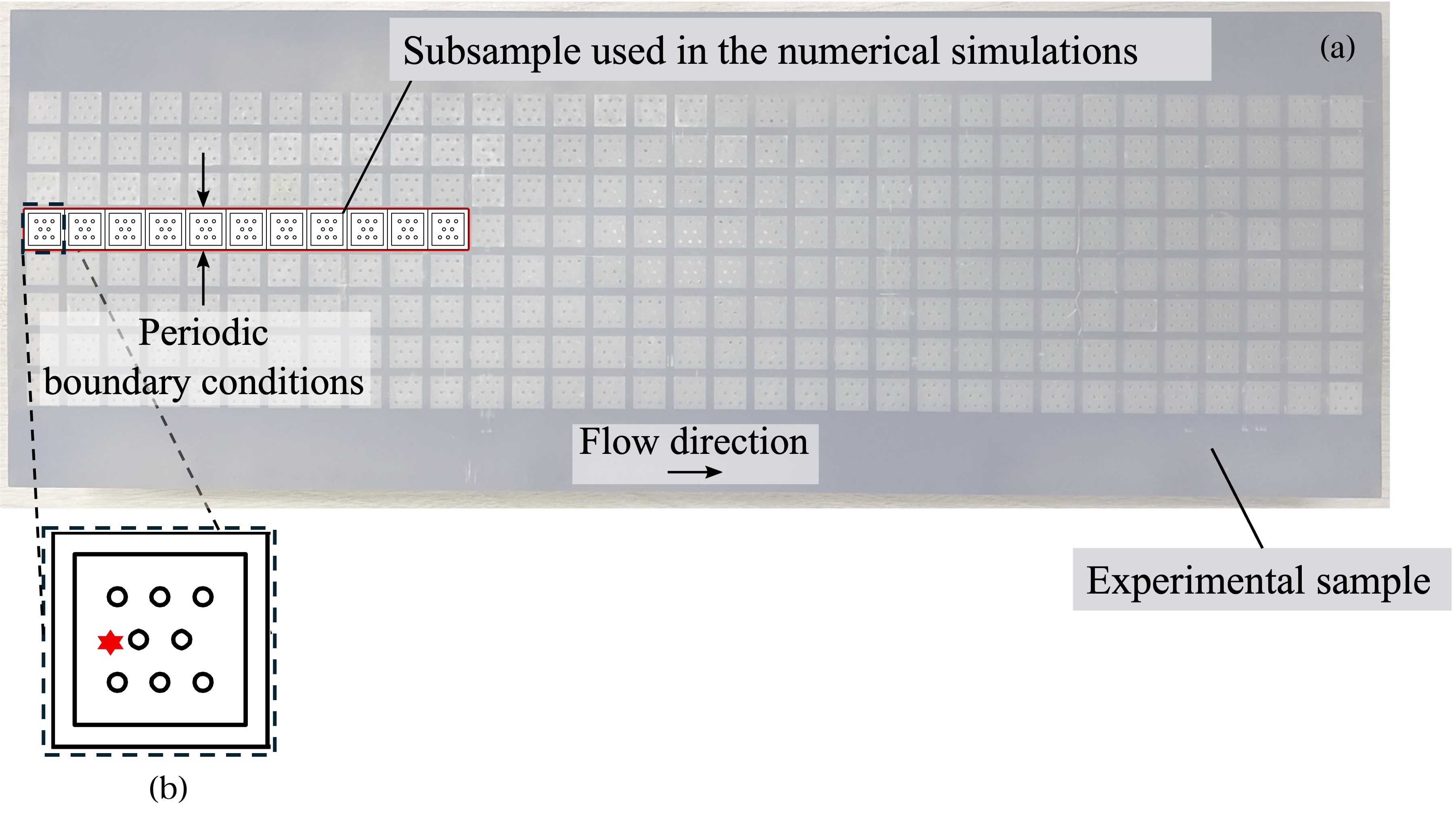}
    \caption{\blue{(a)} Comparison between the real UFSC sample and the modelled geometry for the simulations.\blue{ (b) Detail of the sampling location for the in-situ technique for both experiments and simulations.  \redstar\ Denotes the face sheet probe.}}
    \label{fig:linertested}
\end{figure}

To ensure a robust comparison with experimental data, impedance values from the in-situ method have been sampled at the same locations as in the experiments. Figure \ref{fig:linertested} \blue{(b)} presents a schematic representation of the in-situ sampling position. Additionally, since simulations provide access to the pressure values over the entire liner surface, the minimum, maximum, and mean values across the cavities have also been extracted and analysed.

\subsection{Triple Decomposition}
\label{sec:triplevaluedecomp}
To describe the influence of the grazing flow on the in-orifice flow dynamics, examining the acoustic-induced velocity profiles is crucial. Extracting this information requires isolating the coherent acoustic-induced velocity field from the stochastic turbulent fluctuations. The triple decomposition method is employed to separate these contributions \citep{Avallone2021Acoustic-inducedLayer}.

The method consists of the following steps:
the time series extracted from the LB-VLES simulations are initially phase-locked with the incoming acoustic wave; the resultant phase-locked velocity components are denoted as $\tilde{u}$, $\tilde{v}$, $\tilde{w}$. These phase-locked fields are then averaged, yielding the corresponding mean velocity components, indicated as $U, V, W$. The acoustic-induced velocity components are subsequently determined by subtracting the phase-averaged fields from the phase-locked fields, yielding $\tilde{\tilde{u}}, \tilde{\tilde{v}}, \tilde{\tilde{w}}$. 
This method provides a straightforward approach for separating the acoustic-induced and the turbulent flow fields. Although it is not effective when the acoustic excitation involves broadband or non-tonal signals, because it is not possible to phase-lock the signals, the present study considers only tonal plane waves, making this approach well-suited.

\section{Computational setup}
\label{sec:Computationalsetup}

\subsection{Computational domain}
\begin{figure}
    \centering
    \includegraphics[width=\textwidth]{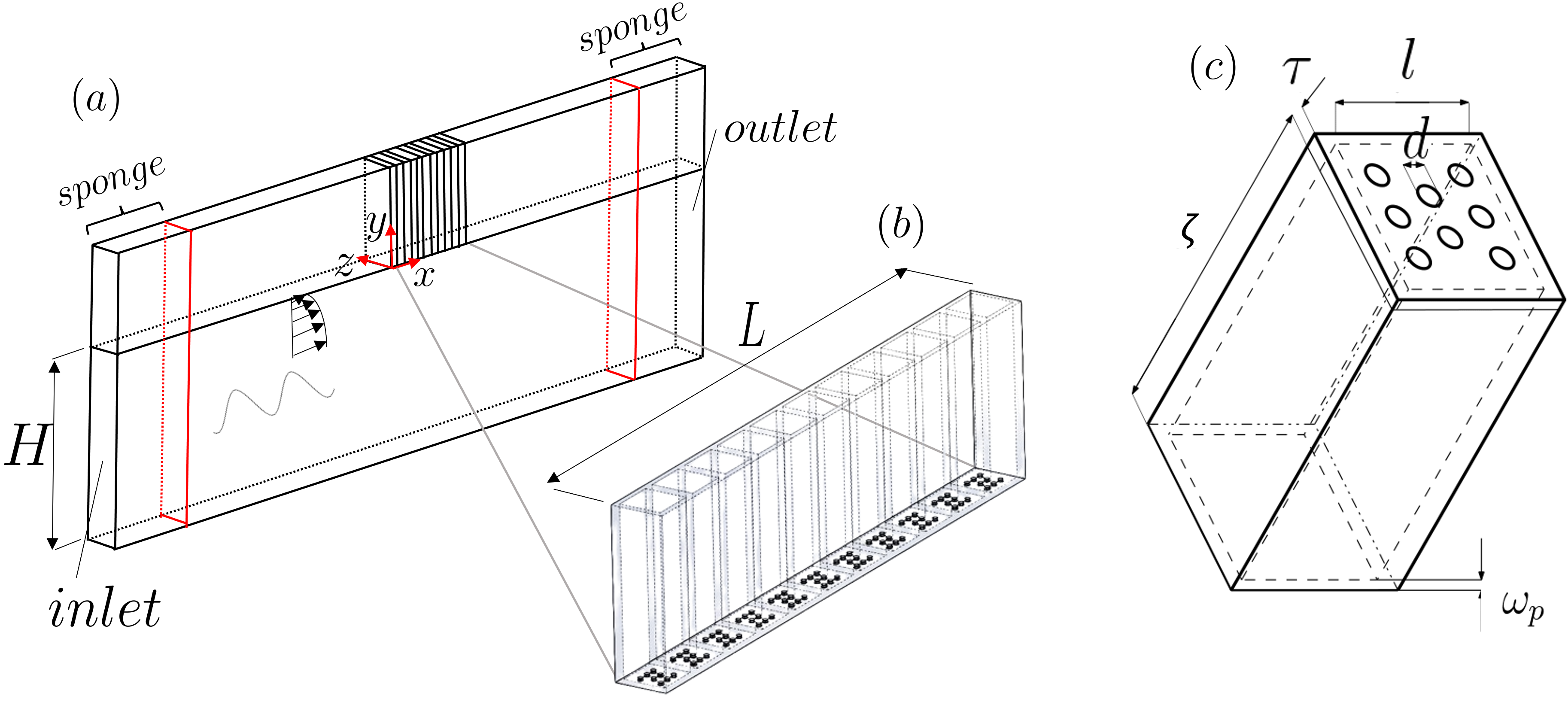}
    \caption{Schematic representation of the full computational domain.}
    \label{fig:entirecomputationaldomain}
\end{figure}

The computational domain is illustrated in Figure~\ref{fig:entirecomputationaldomain}. The coordinate system is defined as follows: \( x \) denotes the streamwise direction, \( y \) the wall-normal direction, and \( z \) the spanwise direction. In this work, the following velocity notation is used: \( U, V, W \): time-averaged streamwise, wall-normal, and spanwise velocity components, respectively; \( u, v, w \): instantaneous streamwise, wall-normal, and spanwise velocity components; \( u' = u - U, \; v' = v - V, \; w' = w - W \): velocity fluctuations relative to the mean.

The liner is placed in the middle of the channel at the top wall of a duct with a rectangular cross-section. 
Each cavity has a square cross-section of $l\!=\!8.46d$ and a depth of $\zeta=\!\blue{32.56d}$, where $d=1.17$~mm is the orifice diameter. 
Each cavity has eight orifices, partition walls  of thickness $\blue{w_p\!=\!1.08d}$, and the face sheet thickness of $\tau\!=\!0.46d$. 
These dimensions result in a percentage of open area for the entire sample of 5.5\%.
The cross-section of the channel has a height of $H = 2h = 40$ mm and a width equal to $l+w_p$.
In the upstream region of the channel, a zig-zag trip was added on both the top and bottom walls. Its size and position were manually adjusted to match the experimental velocity profile upstream of the liner. The zig-zag trip was placed at $x = -1367d$ upstream of the liner, where $x=0$ marks the start of the liner. The zig-zag trip had a height of $0.21d$ and a length of $1.71d$.

To achieve a quasi-anechoic condition and prevent acoustic reflections at the channel's termination, the fluid viscosity was significantly increased using sponge regions as shown in Figure \ref{fig:entirecomputationaldomain}. In these regions, the viscosity was increased by a factor of one hundred following an exponential law.
All walls in the computational domain were treated as adiabatic. A uniform velocity boundary condition was applied at the inlet, corresponding to a Mach number of M = 0.3, which results in a centerline velocity of $U_0 = 110$ m/s (M=0.32) in the lined section. A pressure boundary condition was set at the outlet.

The computational domain was designed to replicate the experimental setup of the UFSC Liner Test Rig~\citep{Bonomo2022ASPL}. The design of the sample, though resembling the one presented in a previous numerical study \citep{Pereira2021ValidationFlow}, exhibits variations in terms of the face sheet thickness, orifice diameter, and shape of the edges of the orifice, which were slightly rounded as discovered from the 3D scanning of the tested liner sample \citep{Quintino2025}.  Due to manufacturing limitations, the scan also identified variability in these parameters along the sample. Therefore, averaged values of orifice diameter and face sheet thickness were used to construct the liner sample for the numerical simulations. However, a few differences between the real and simulated liners should be noted. 
First, the simulated liner is represented by a single row of eleven cavities, while the one tested in the experiments features an 8$\times$33 cavity grid, as shown in Figure \ref{fig:linertested}. However, both configurations maintain the same number of orifices per cavity, ensuring consistency in porosity. A second key difference is that in the UFSC Liner Test Rig, the duct has a rectangular cross-section of $100 \times 40$ ~mm$^2$. In contrast, the simulation assumes periodic boundary conditions on both sides of the duct. Previous studies \citep{Tam2014ExperimentalDrag} have shown that this choice has minimal influence on the acoustic response, supporting the validity of the comparison with the experimental results.

\subsection{Computational grid design}
\begin{figure}
    \centering
    \includegraphics[width=0.8\linewidth]{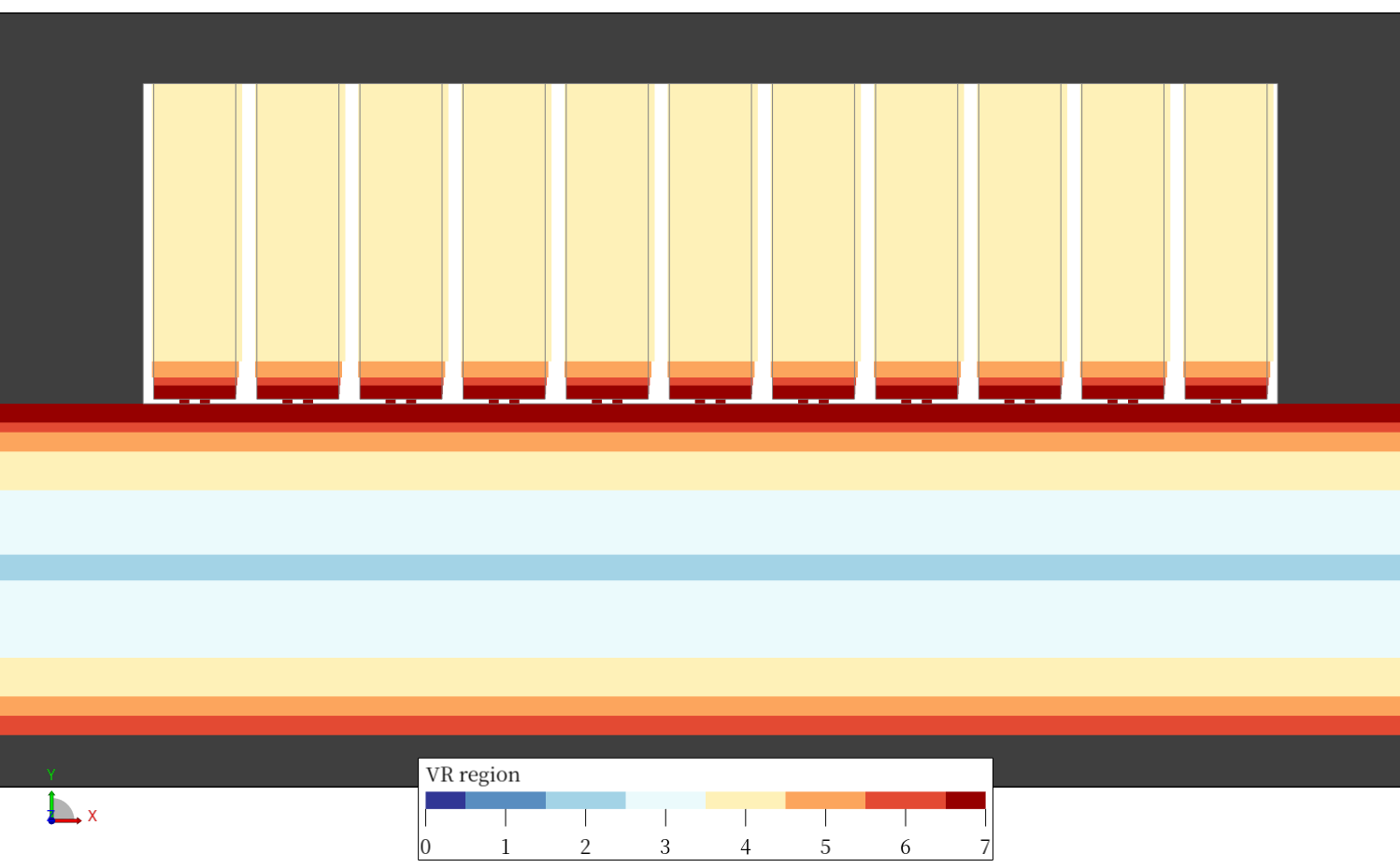}
    \caption{Schematic representation of the VR regions.}
    \label{fig:VR_regions}
\end{figure}

A Variable Resolution (VR) scheme was adopted. Details of the VR regions are provided in Figure \ref{fig:VR_regions}. The VR regions are symmetric with respect to the center of the channel.
The finest resolution, VR = 7, was used to discretise the entire face sheet, the orifices, and portions of the backing cavities. Each subsequent resolution level was defined by doubling the cell size of the previous level. Within the orifice, the minimum grid spacing was \blue{$\Delta z_{min} = \Delta y_{min} = \Delta x_{min} = 0.0207d$}, yielding a resolution of \blue{$\approx 48$ vx/$d$} for the fine mesh. \blue{This results in a spacing expressed in wall units equal to $\Delta x^+= \Delta y^+ = \Delta z^+=6.9$. Where $\Delta x^+ = \Delta x u_{\tau} /\nu$ and $u_{\tau} = \sqrt{\tau_w / \rho} = 4.2 $ m/s refers to the friction velocity of the smooth reference surface.}

\subsection{Simulation strategy and test cases}
In the presence of the grazing flow, the flow developed spatially in the channel flow until statistical convergence was achieved. After achieving statistical convergence, the acoustic simulations were performed. Starting from the flow-only converged solution, an instantaneous flow field was saved and modified by superimposing a plane acoustic wave with a specified frequency and amplitude using the \textit{OptydB} toolkit. This modified flow field was used as the initial condition for the subsequent simulations, which include the acoustic wave \citep{Avallone2019Lattice-boltzmannFlow}.

While this approach effectively reduced computational costs, especially when analysing multiple configurations, it introduced a change in the initial condition that required a few acoustic cycles for the solution to reach a statistically steady state. For each configuration, the plane acoustic wave was at least 10 wavelengths long. The downside of this approach is that the length of the channel must be long enough to accommodate 10 acoustic wave wavelengths for the lowest frequency of interest. However, this approach is also beneficial since it allows the presence of acoustic sponges, thus minimising the impact of acoustic reflections at the boundary of the computational domain. \blue{Details on the computational cost are reported in Appendix \ref{app:computational cost}.}

Twenty different simulations were performed and grouped into seven sets, as shown in Table \ref{tab:testmatrix}. Sets (A) and (B) consist of simulations with the acoustic liner, without grazing flow, and with plane acoustic waves with amplitudes equal to 130 dB and 145 dB, and three frequencies (800 Hz, 1400 Hz, and 2000 Hz). The acoustic wave was located upstream of the liner. These frequencies were chosen to investigate the acoustic liner's behaviour near the resonance frequency and at frequencies above and below it. The SPL was calculated using the standard reference pressure of $20 \times 10^{-6} $\ Pa. 

For the cases with grazing flow, the centerline Mach number was set to 0.32, as in the reference experiments. This corresponds to a bulk Reynolds number of $Re_b = 2.7 \times 10^{5}$ and a friction Reynolds number of $Re_{\tau} \approx 3500$.
Set (C) includes two simulations: one of a turbulent channel flow with smooth walls and the other of a turbulent channel flow with the liner mounted in the upper wall. All other sets account for the presence of grazing turbulent flow and acoustic waves.
Set (D) contains simulations with different frequencies, at an SPL of \SI{130}{dB}, and an upstream-located acoustic source, i.e., propagating in the same direction as the mean flow. Set (E) includes simulations with different frequencies at a fixed SPL of \SI{145}{dB} and an upstream acoustic source. Sets (F) and (G) include simulations with different frequencies, and downstream acoustic sources at an SPL of \SI{130}{dB} and \SI{145}{dB}, respectively.

\begin{table}
    \centering
    \caption{List of the simulations carried out}
    \begin{tabular}{ccccc}
        \toprule
        \textbf{Mach} & \textbf{Configuration} & \textbf{Acoustic Source Pos.} & \textbf{Amplitude (dB)} & \textbf{Frequency (Hz)} \\
        \midrule

        \multicolumn{5}{c}{\textbf{Set (A)}} \\
        0.00 & Lined & Upstream & 130 & 800 \\
        0.00 & Lined & Upstream & 130 & 1400 \\
        0.00 & Lined & Upstream & 130 & 2000 \\

        \addlinespace
        \multicolumn{5}{c}{\textbf{Set (B)}} \\
        0.00 & Lined & Upstream & 145 & 800 \\
        0.00 & Lined & Upstream & 145 & 1400 \\
        0.00 & Lined & Upstream & 145 & 2000 \\

        \multicolumn{5}{c}{\textbf{Set (C)}} \\
        \addlinespace
        0.32 & Smooth & - & - & - \\
        0.32 & Lined  & - & - & - \\

        \addlinespace
        \multicolumn{5}{c}{\textbf{Set (D)}} \\
        0.32 & Lined & Upstream & 130 & 800 \\
        0.32 & Lined & Upstream & 130 & 1400 \\
        0.32 & Lined & Upstream & 130 & 2000 \\

        \addlinespace
        \multicolumn{5}{c}{\textbf{Set (E)}} \\
        0.32 & Lined & Upstream & 145 & 800 \\
        0.32 & Lined & Upstream & 145 & 1400 \\
        0.32 & Lined & Upstream & 145 & 2000 \\

        \addlinespace
        \multicolumn{5}{c}{\textbf{Set (F)}} \\
        0.32 & Lined & Downstream & 130 & 800 \\
        0.32 & Lined & Downstream & 130 & 1400 \\
        0.32 & Lined & Downstream & 130 & 2000 \\

        \addlinespace
        \multicolumn{5}{c}{\textbf{Set (G)}} \\
        0.32 & Lined & Downstream & 145 & 800 \\
        0.32 & Lined & Downstream & 145 & 1400 \\
        0.32 & Lined & Downstream & 145 & 2000 \\

        \bottomrule
    \end{tabular}
    \label{tab:testmatrix}
\end{table}

\subsection{Validation of the numerical approach}

To validate the computational results, first the mean flow profile for the smooth wall configuration was compared with the experimental data from \citet{Bonomo2022ASPL} and \citet{Vallikivi2015TurbulentNumber} using three different mesh resolutions: coarse (\blue{12} vx/\(d\)), medium (\blue{24} vx/\(d\)) and fine (\blue{48} vx/\(d\)). Literature data were also used because velocity fluctuation measurements were not available from the experiments conducted by \citet{Bonomo2022ASPL}.

\begin{figure}
  \centering

  \begin{subfigure}{0.4\textwidth}
    \centering
    \includegraphics[width=\linewidth]{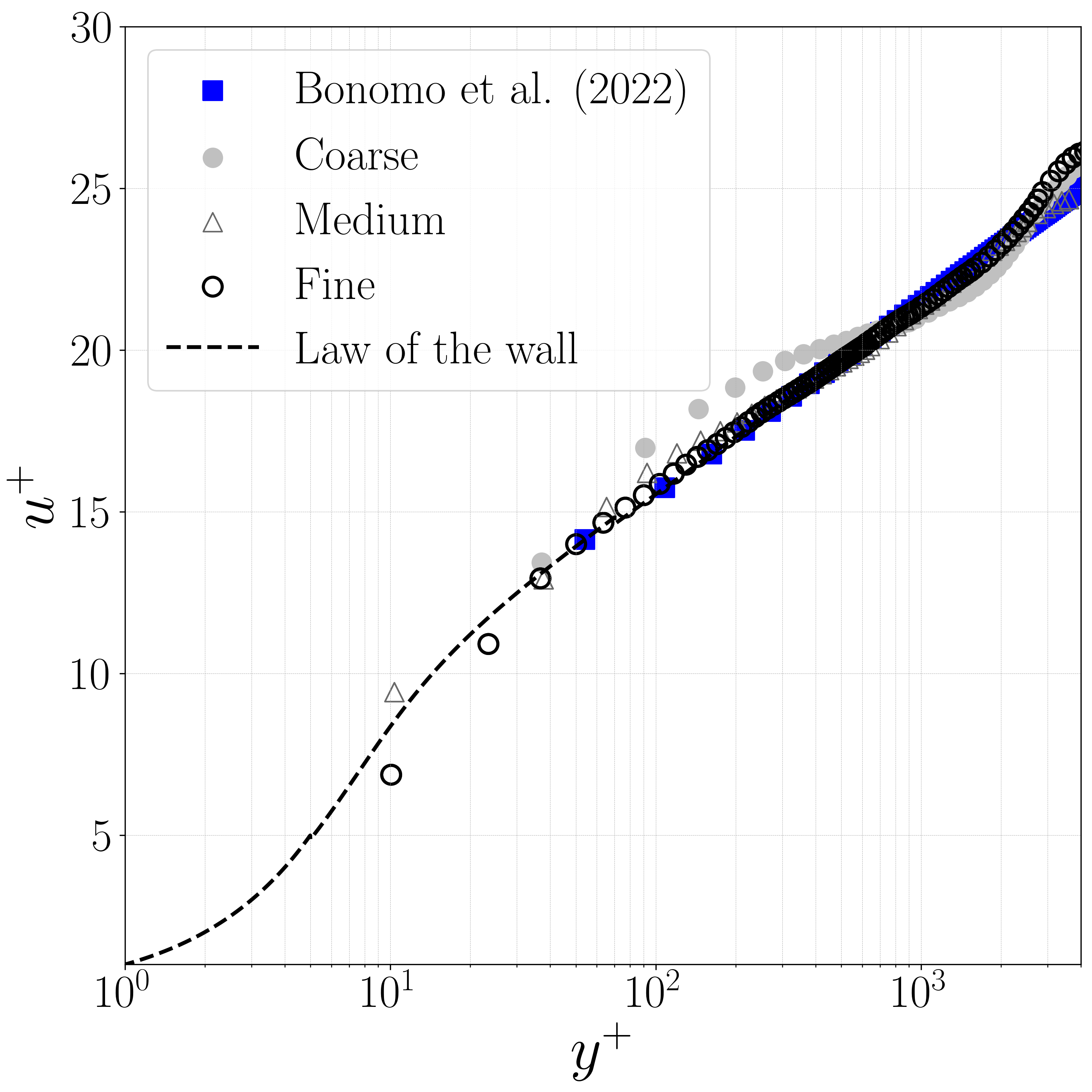}
    \caption{}
    \label{fig:flowwallcoordinate}
  \end{subfigure}\hspace{0.05\textwidth}
  \begin{subfigure}{0.4\textwidth}
    \centering
    \includegraphics[width=\linewidth]{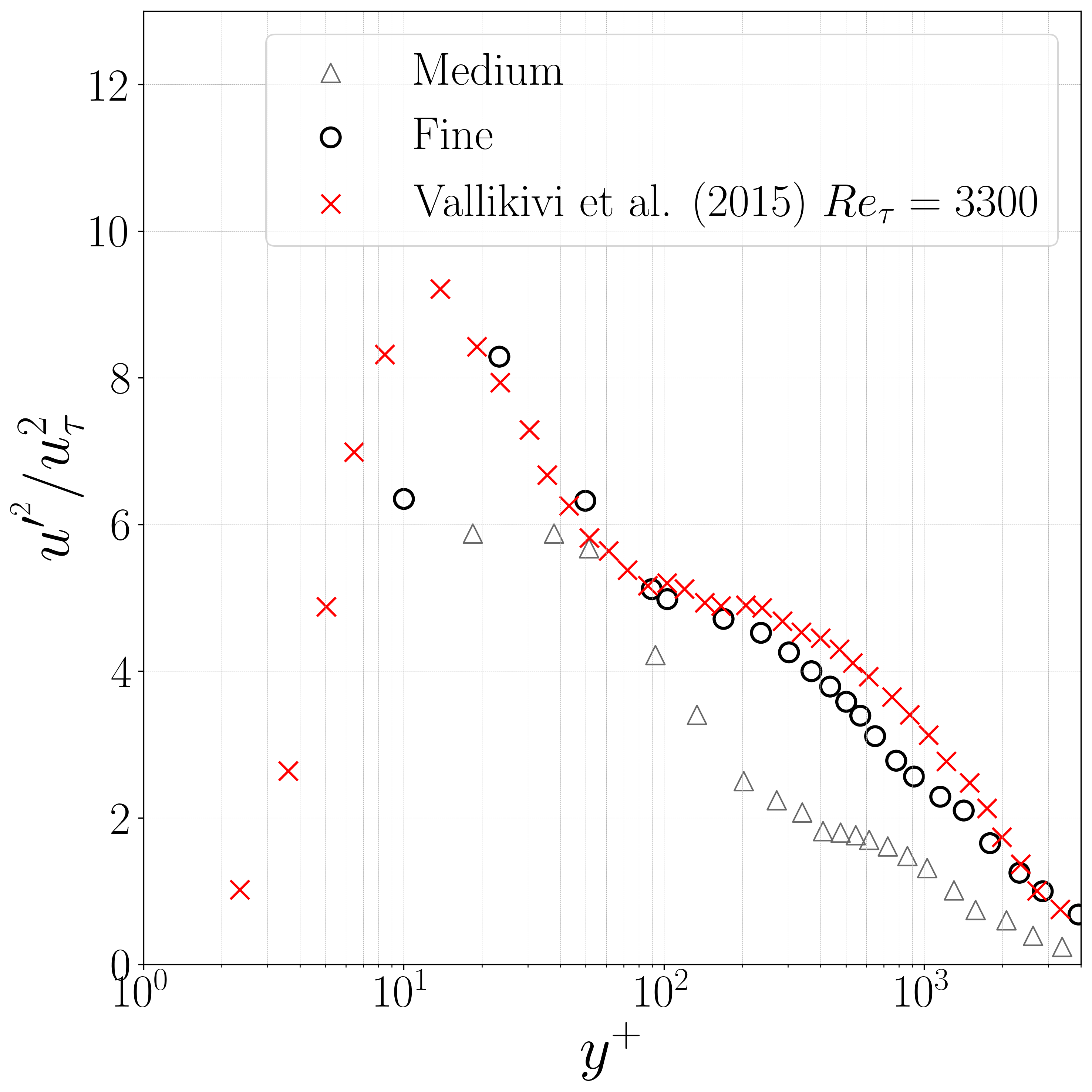}
    \caption{}
    \label{fig:fluctuationwallcoordinate}
  \end{subfigure}
\caption{(a)  Mean velocity profile comparison with experimental data  by \cite{Bonomo2022ASPL} (b) streamwise velocity variance compared with experiments by  \cite{Vallikivi2015TurbulentNumber}.}
    \label{fig:mesh_convergence}

\end{figure}

Data are presented in wall units ($y^+ = yu_{\tau}/\nu$ and $u^+=U/u_{\tau}$). In the experiments, the friction velocity was computed using the Clauser chart technique \citep{Clauser1954TurbulentGradients}. In the simulations, $u_\tau$ was computed from the wall shear stress calculated using the extended wall model described in Section \ref{sec:Methodology}. Figure \ref{fig:mesh_convergence} (a) shows the mean velocity profile. The data from experiments by \cite{Bonomo2022ASPL} and simulations at the three resolutions are compared with the law-of-the-wall using constants $k=0.37$ and $B=3.7$, as recommended for turbulent channel flows \citep{Nagib2008}. A good agreement between experiments and simulations is found.

Figure \ref{fig:mesh_convergence} (b) presents measurements of the streamwise Reynolds stress derived from simulations, compared with experimental data reported by \citet{Vallikivi2015TurbulentNumber} at a friction Reynolds number of  $Re_{\tau} \approx 3300$. Values were made non-dimensional using the local  $u_{\tau}^{2}$. The profile of $u'^2$ for the fine simulation is similar to the experiments at a comparable Reynolds number.  Further validation of the simulation was performed by comparing the friction coefficient $C_f$, computed from the streamwise pressure drop, against the experimental data of \citet{Schultz2013}. The comparison, presented in Appendix \ref{app:Friction coefficient for the smooth wall case}, confirms the reliability of the numerical predictions.

Based on these results, the fine-resolution grid was selected for all subsequent analyses.

\section {Acoustic field and impedance distribution}
\label{sec:Acoustic_results}
Before analysing the flow field and examining how the acoustic liner and acoustic waves influence the streamwise evolution of the flow within the channel, we first describe the SPL distribution along the liner and the impedance values.

\subsection{SPL decay}

\begin{figure}
    \centering
    \includegraphics[width=\linewidth]{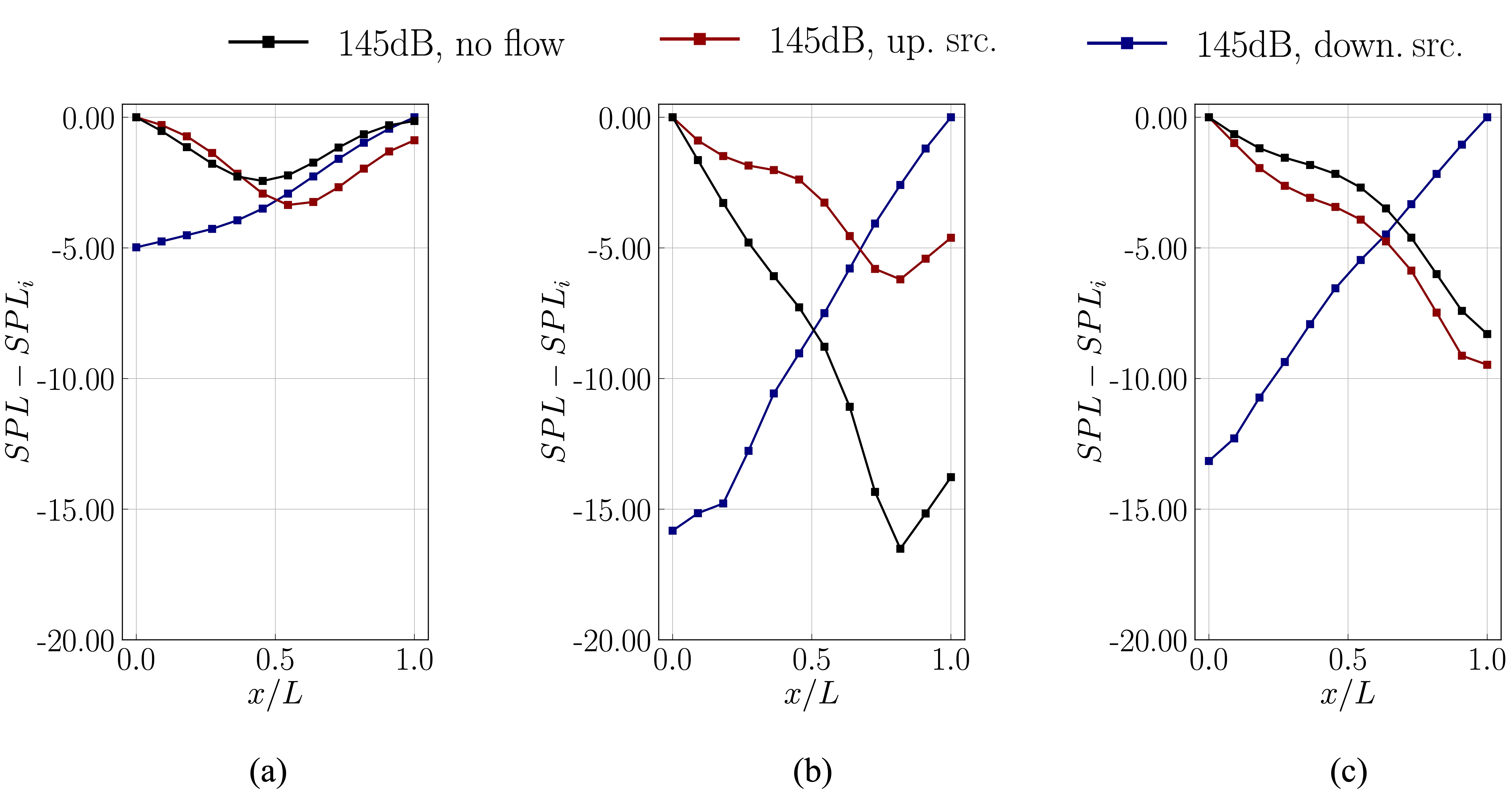}
    \caption{SPL decay along the channel centerline ($y/h = -1$) for an incident acoustic wave with SPL = 145 dB under different flow conditions:  (a) $f$ = 800 Hz, (b) $f$ = 1400 Hz and (c) $f$ = 2000 Hz. The SPL is normalised with respect to the incident SPL ($SPL_{i}$).}
    \label{fig:spl_decay_145}
    \end{figure}

The analysis starts with the SPL distributions at the centerline of the channel (Figure \ref{fig:spl_decay_145}), where the acoustic pressure fluctuations are almost not affected by the flow-induced ones. \blue{The SPL distribution is plotted with respect to the incident $SPL_i$ for each case. Although the SPLs were set equal for both simulations, minor deviations of about 2~dB ($\approx 50$ Pa) at the onset of the liner are present, which might be caused by different reflection of the acoustic waves at the wall impedance discontinuity depending on the propagation direction in the presence of flow \citep{Yves_JASA_2019}}. The shape of the SPL curves depends on the presence of grazing flow, the direction of acoustic wave propagation, and its frequency. A relevant observation is that the SPL decays more when the source is located downstream, i.e., propagates against the mean flow, independently of the frequency of the acoustic wave. Furthermore, in this case, the SPL curves show a more monotonic behaviour compared to cases where the acoustic wave propagates in the same direction as the mean flow. 

\begin{figure}
    \centering
    \includegraphics[width=\linewidth]{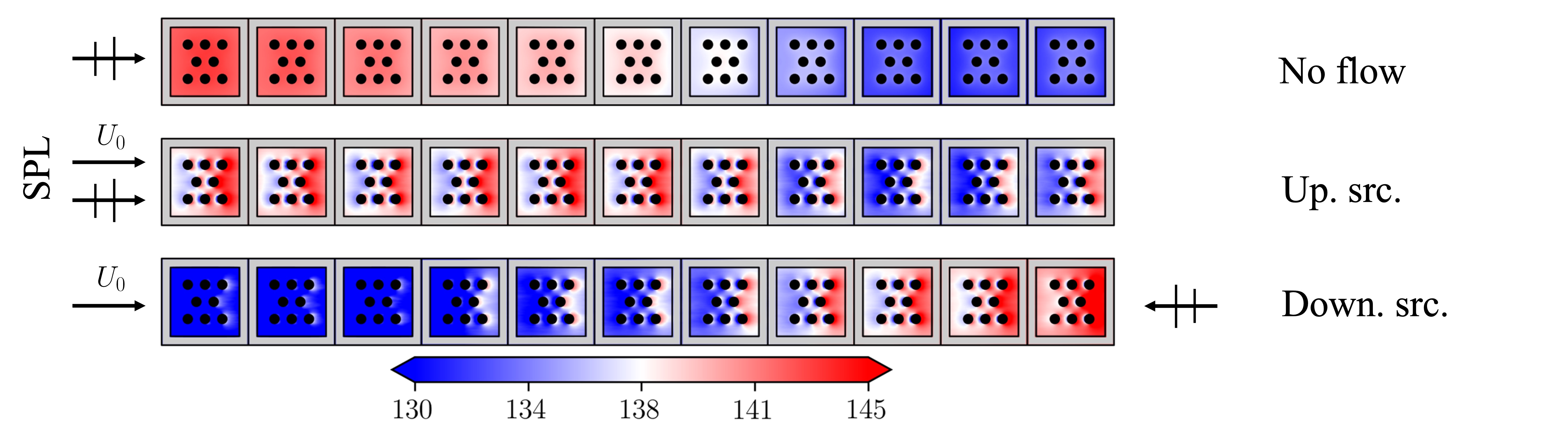}
    \caption{SPL along the liner's surface for the case with SPL = \SI{145}{dB} and $f$ = 1400 Hz under three conditions: without grazing flow, with grazing flow and upstream acoustic source, and with grazing flow and downstream acoustic source.}
    \label{fig:fig31}
\end{figure}

Figure \ref{fig:fig31} shows the distribution of SPL on the liner surface for the cases with an incoming acoustic plane acoustic wave with a frequency of 1400 Hz and SPL of 145 dB, with and without grazing flow. The SPL on the surface results from two main contributions: pressure fluctuations induced by the acoustic wave and those generated by the turbulent grazing flow. \blue{The interaction of the grazing flow with liner's orifices affects both wall-normal and streamwise velocity fluctuations near the orifices, as will be detailed in section \ref{sec:impactofspl}. This results in localized variations of the surface pressure fluctuations, which contribute to the spatial distribution of SPL shown in the figure. } This \blue{might} explain why the local SPL can exceed that of the imposed acoustic wave alone, consistent with the findings of \cite{Roncen2025}, who emphasised that turbulence-induced pressure fluctuations cannot be neglected.

The SPL surface distributions differ from those at the channel centerline. Without grazing flow, there is an almost monotonic SPL reduction on the surface. On the other hand, in the presence of a grazing flow, there is a reduction of the SPL upstream of the orifice and an increase downstream of it, independently of the propagation direction of the acoustic wave. This trend persists across all tested frequencies and SPL levels, including the lower SPL case (\SI{130}{dB}), though results are omitted here for brevity. The cause of this redistribution \blue{might lie} in the near-wall hydrodynamic field around the orifices.

\subsection{Surface distribution of impedance components obtained with the in-situ method}

\begin{figure}
    \centering
    \includegraphics[width=\linewidth]{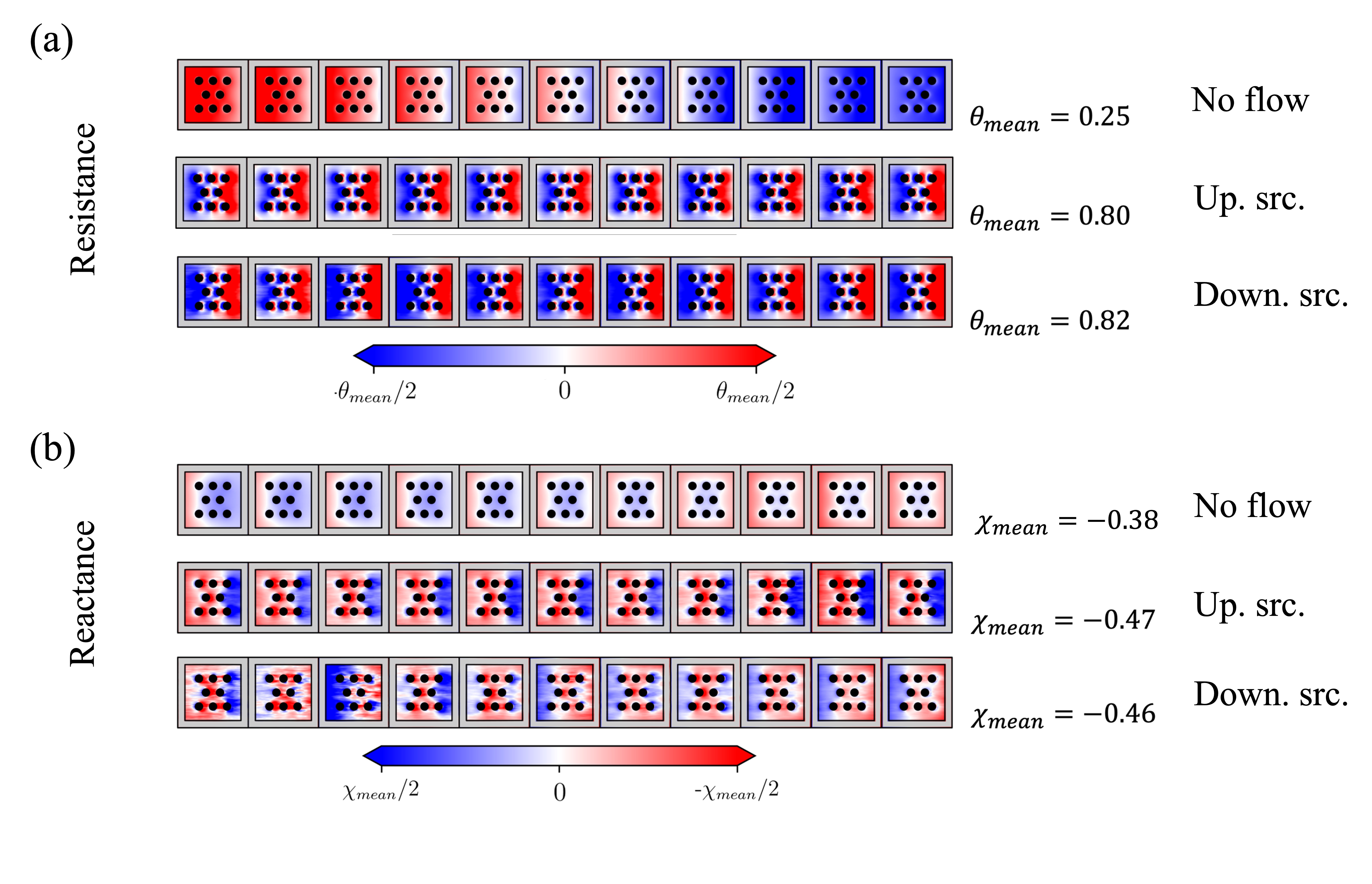}
    \caption{(a) Resistance and (b) reactance along the liner's surface obtained with in-situ for the case with plane acoustic wave with SPL = \SI{145}{dB} and $f$ = 1400 Hz under three conditions: without grazing flow, with grazing flow and upstream acoustic source, and with grazing flow and downstream acoustic source.}
    \label{fig:fig31_impedance}
\end{figure}

Resistance and reactance are computed using the in-situ method as described above. The reference microphone at the backplate is always located at the center of each cavity backplate, while the face sheet microphone is moved to generate contour plots. These contours are shown in Figure \ref{fig:fig31_impedance} for an incident acoustic plane wave with SPL = \SI{145}{dB} and $f$ = 1400 Hz, with and without grazing flow. To highlight the location-dependent impedance, the figures display the changes in resistance and reactance relative to the mean value calculated on the entire surface. 

In the absence of the grazing flow, the resistance decreases in the propagation direction of the acoustic wave because of a reduction of the SPL. The resistance decreases along each cavity. Then, a jump in the resistance value is present when changing the cavity because the reference microphone at the backplate changes. 
Conversely, in the presence of a grazing flow, the spatial distribution of the resistance sees larger variations over each cavity; there is a strong reduction of the resistance when changing cavity; around each orifice, the resistance is always lower upstream of the orifice and higher downstream. The variation of resistance is up to a factor of  3, suggesting that the selection of the sampling location is crucial. \blue{This behaviour might be caused by the local surface SPL, which depends on the local amplitude of the acoustic wave and the surface pressure fluctuations altered by the presence of the orifices, as it will be described later.} The pattern remains unchanged with respect to the direction of acoustic wave propagation, although the mean value over the surface slightly varies.

\begin{figure}
    \centering
    \includegraphics[width=\linewidth]{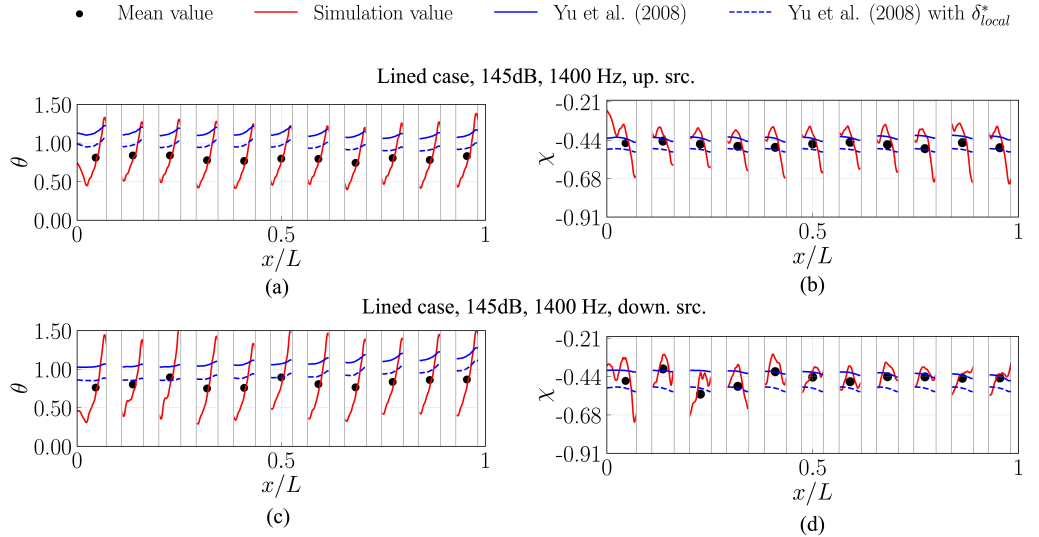}
    \caption{Streamwise distribution of resistance and reactance compared with the semi-empirical model by \cite{Yu2008ValidationData} using different values of $\delta^{*}$ and the local SPL. Data shown for SPL = 145 dB and frequency of 1400 Hz. (a, b) Upstream acoustic source, (c, d) downstream acoustic source.}
    \label{fig:Insitu_line_plot}
    
\end{figure}

Data extracted at the centerline of the liner sample are shown in Figure
\ref{fig:Insitu_line_plot} for both upstream (a, b) and downstream (c, d) acoustic wave locations. In this figure, the resistance predicted using the semi-empirical model by \citet{Yu2008ValidationData} is reported using the local SPL and the value of the boundary layer displacement thickness $\delta^*$ obtained from the smooth wall simulation upstream of the liner and the local one obtained from the lined simulation (which will be detailed in the next section).
 The line plot in Figure
\ref{fig:Insitu_line_plot} (a, c) highlights that the semi-empirical prediction from \citet{Yu2008ValidationData} approaches the mean value of resistance if the local values of $\delta^*$ are used. The reason will be explained in the following.

Similar observations can be made for the reactance, for which \blue{the cavity-averaged values (black dots in the figure)}  \blue{weakly depend on the acoustic wave propagation direction. These results contrast with findings reported in the literature for impedance educed using model-fitting inference methods \citep{Spillere2020ExperimentallyDownstream} and may be explained by the fact that the in-situ approach relies on surface pressure measurements and does not employ boundary conditions to model the near-wall flow–acoustic interaction.}

These findings emphasise that the local amplitude of the acoustic and flow profile must be considered when measuring with the in-situ method. However, the semi-empirical model does not capture the high resistance values observed in the numerical data at the beginning and end of each cavity.

\subsection{Average impedance and comparison with experimental data}

\begin{figure}
  \centering
\includegraphics[width=0.8\textwidth]{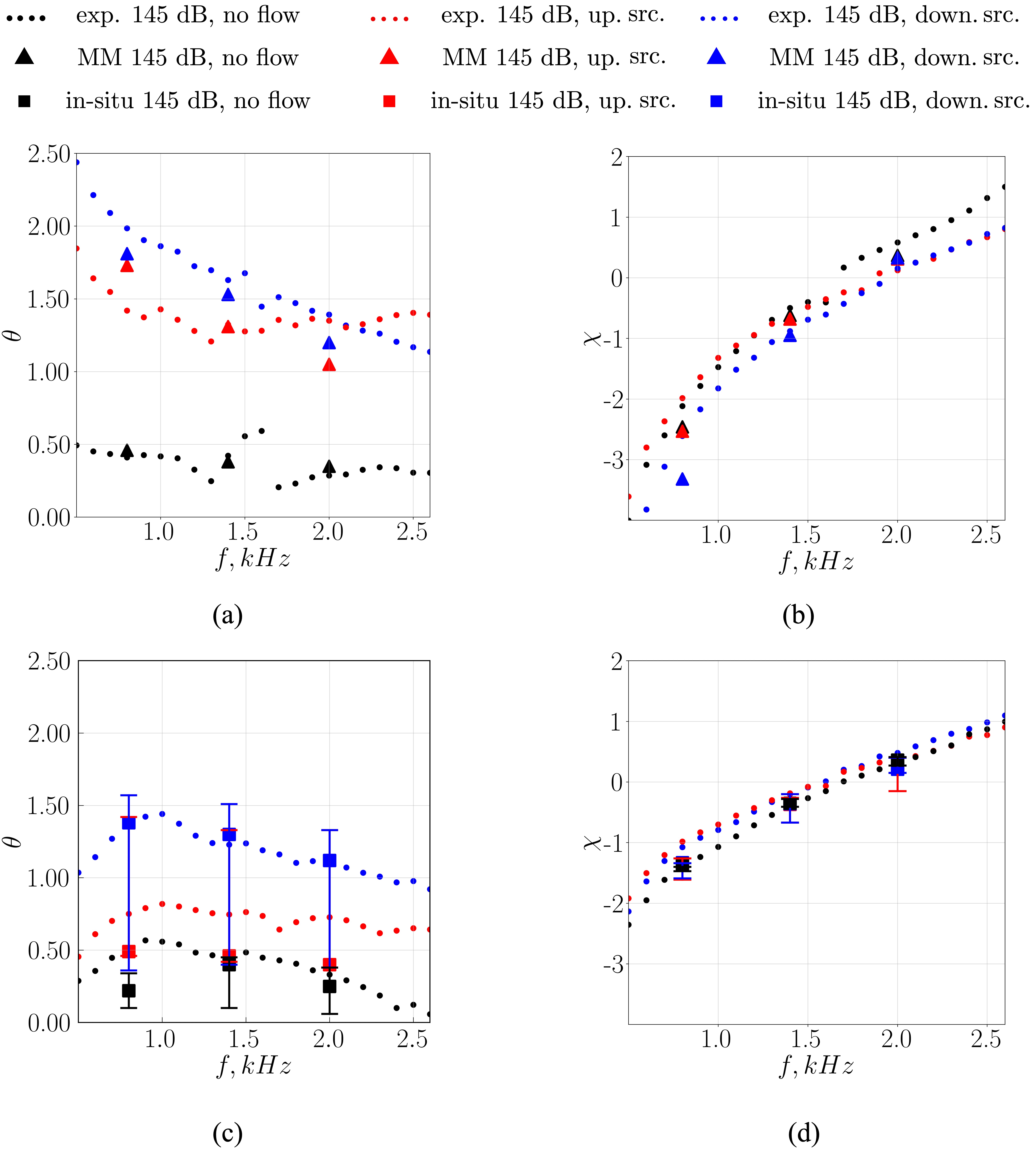}

  \caption{Comparison of (a, c) resistance and (b, d) reactance components of impedance for acoustic wave amplitude equal to \SI{145}{dB}. (a, b) MM and (c, d) in-situ results.}
\label{fig:impedance_145}
\end{figure}

While numerical simulations allow for mapping the impedance distribution on the entire liner sample, the average value is often used to characterise the acoustic liner. Therefore, the impedance obtained using the MM method is reported in Figure \ref{fig:impedance_145} (a, b) for the cases with SPL equal to \SI{145}{dB} and the three frequencies investigated, with and without grazing flow and both acoustic wave propagation directions. Experimental results from  \cite{Bonomo2022ASPL} are also reported. In Figure \ref{fig:impedance_145} (c, d), the in-situ results are compared with the experimental data at a similar sampling location (Figure \ref{fig:linertested} (b)); the bars indicate the minimum and maximum values obtained over the liner's surface. Numerical and experimental results show a reasonable agreement with similar trends. However, as pointed out above, the length of the liner sample differs between experiments and simulations, thus challenging the comparison at the lowest frequency investigated.

While the two methods yield roughly similar resistance values without flow, the resistance values obtained in the presence of flow, both in the experiments and numerical simulations, differ. The largest value obtained from the in-situ method is lower than the average one obtained with the MM method. However, both methods show that the resistance increases with the grazing flow and is lower when the acoustic source is located upstream. 

The reactance shows smaller variation with and without flow than the resistance \blue{for both techniques}. \blue{This result agrees with previous findings in the literature \citep{Spillere2020ExperimentallyDownstream} and can be explained by the fact that, in the mass-spring analogy, the reactance corresponds to the balance between mass and stiffness, which is more directly related to the cavity depth than to the flow presence \citep{PantonResonantresonators1975,Jones2002EffectsImpedance}. According to the Goodrich model \citep{Yu2008ValidationData} the reactance is therefore predominantly influenced by the cavity depth and by the SPL. Grazing flow affects the reactance only via the orifice end correction—i.e., the oscillating fluid volume in and around the orifice—thereby causing the observed minor variations in the reactance curve. However,} the differences between cases are more evident in the results from the MM method. In particular, the MM method shows that the frequency at which the reactance is equal to zero shifts to higher values in the presence of the flow, and this does not depend on the propagation direction of the acoustic wave. This frequency shift is attributed to the flow-induced blockage effect, which effectively reduces the face sheet porosity \citep{Tam2000MicrofluidLiners}.
Furthermore, in the low-frequency range, the reactance values are higher for the upstream acoustic source than for the downstream one. It is important to note that the slope of the reactance curve differs between the MM and the in-situ method. \blue{This highlights that the educed impedance is inherently dependent on the eduction technique employed, as each method relies on distinct physical assumptions and boundary conditions.}

\begin{figure}
    \centering
        \includegraphics[width=0.8\textwidth]{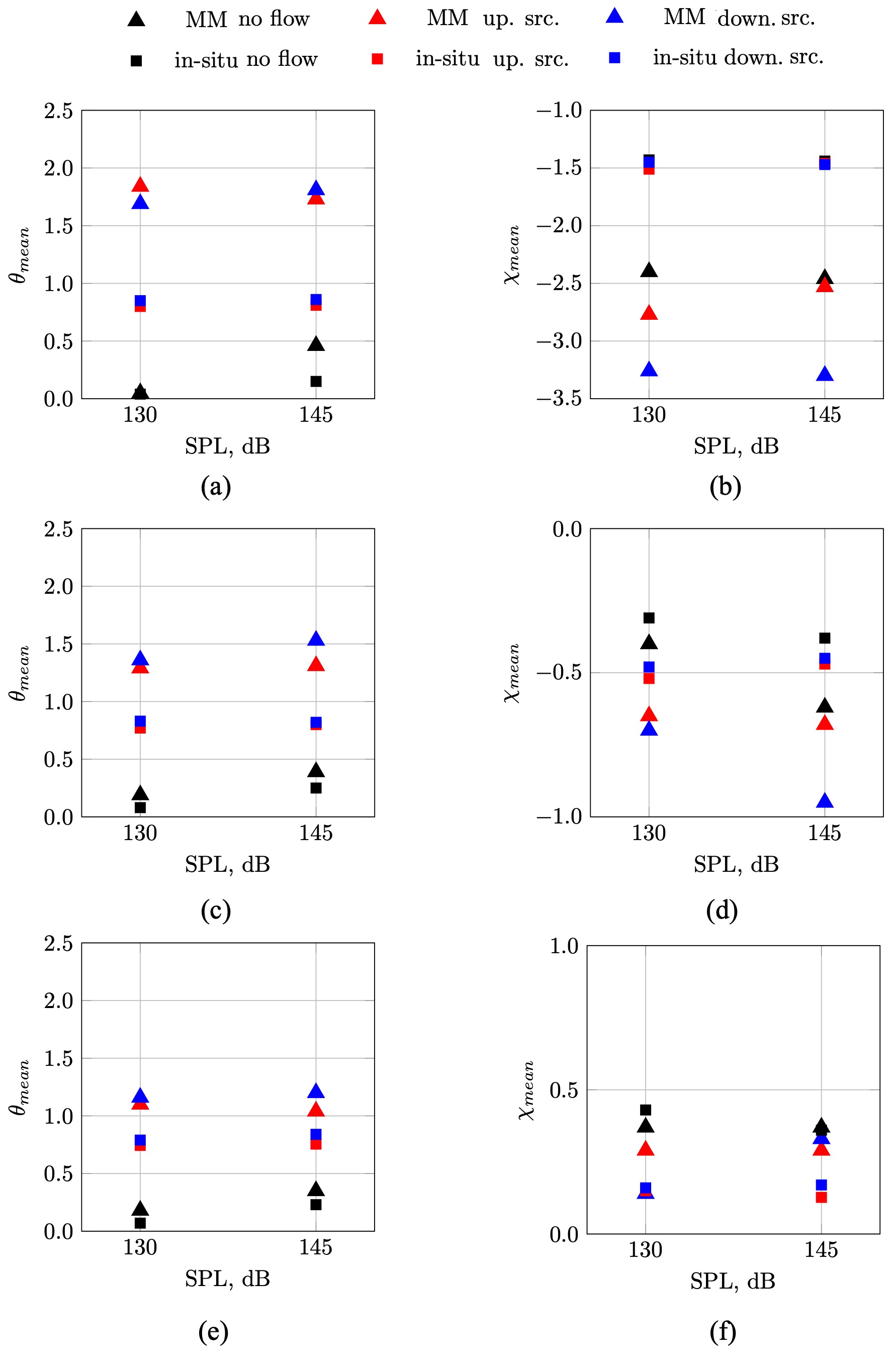}
    \caption{(a, c, e) Resistance and (b, d, f) reactance obtained from the MM and in-situ techniques. The in-situ results have been obtained as an average over the entire liner's surface. Results refer to a plane acoustic wave at varying SPLs, with and without grazing flow. (a, b) Incident acoustic wave with $f = 800$ Hz, (c, d) $f = 1400$ Hz and (e, f ) $f = 2000$ Hz.}
    \label{fig:Impedance_different_SPLs}
\end{figure}

Finally, the effect of the acoustic wave SPL is reported for the frequencies equal to \SI{800}{Hz}, \SI{1400}{Hz} and \SI{2000}{Hz} in Figure \ref{fig:Impedance_different_SPLs}.  In this case, the in-situ results are reported as the average over the entire liner surface for the sake of comparison with the MM method.

As expected, without the grazing flow, increasing the SPL results in a rise in resistance. This reflects a transition in the liner's behaviour from an \blue{almost} linear to a non-linear regime \citep{Melling1973THELEVELS,tam_microfluid_2000}. In the presence of flow, the resistance varies less by varying the SPL for both the MM and in-situ techniques. Similar observations can be made for the reactance. This observation is consistent with previous findings in the literature, particularly with the trend in resistance predicted by the semi-empirical model proposed by \citet{Yu2008ValidationData}. \blue{This effect may be attributed to the increase in pressure fluctuations caused by the turbulent flow, as reported by \citet{RONCEN2025119058}, which is not accounted for in the eduction techniques. Indeed, both flow- and SPL-induced effects are primarily related to the wall-normal particle velocity; increases in turbulent fluctuations or SPL both contribute to an increase in the liner’s resistance \citep{RONCEN2025119058}.} 

Notably, when considering the mean resistance obtained with the in-situ method, at the highest SPL, values in the presence of grazing flow vary less with varying the direction of propagation of the acoustic wave with respect to the ones obtained with the MM technique. This might be caused by the fact that impedance obtained from the in-situ technique is not dependent on any flow-based boundary condition. Finally, this figure shows that for both SPLs, the surface-averaged in-situ results are different from the MM ones.

\section{Aerodynamic results}
\label{sec:aerodynamic_results}
The presence of a grazing flow influences the impedance of acoustic liners. In this section, we describe how the spatial development of the turbulent flow is affected by the transition from the solid to the acoustically-treated surface, by the acoustic liner streamwise length, and even by the acoustic-induced velocity fluctuations.

\subsection{Instantaneous flow}
\begin{figure}
    \centering
    \begin{subfigure}[b]{0.75\textwidth}
        \centering
        \includegraphics[width=\textwidth]{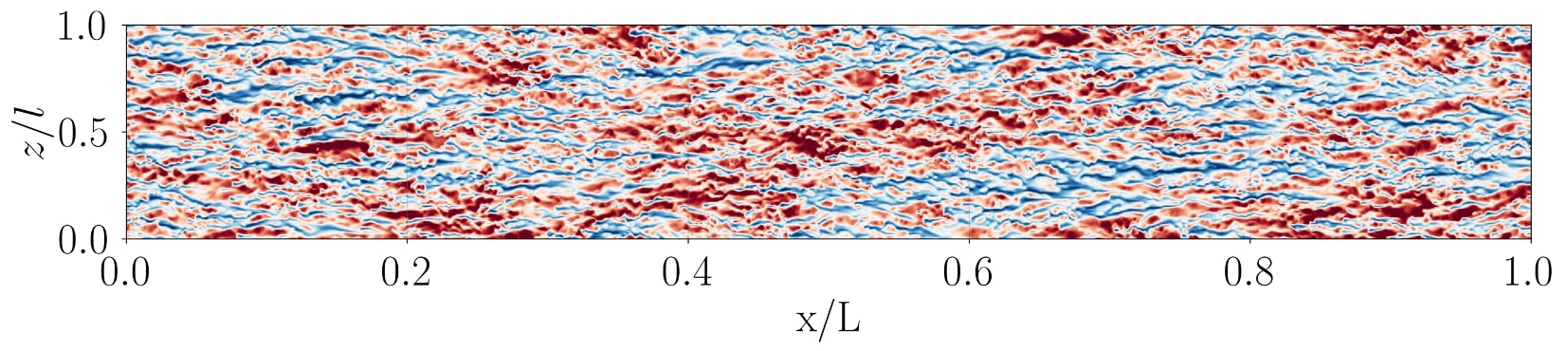}
        \caption{}
        \label{fig:subfig1}
    \end{subfigure}
    \begin{subfigure}[b]{0.2\textwidth}
        \centering
        \raisebox{0.8cm}{ 
            \includegraphics[width=0.6\linewidth]{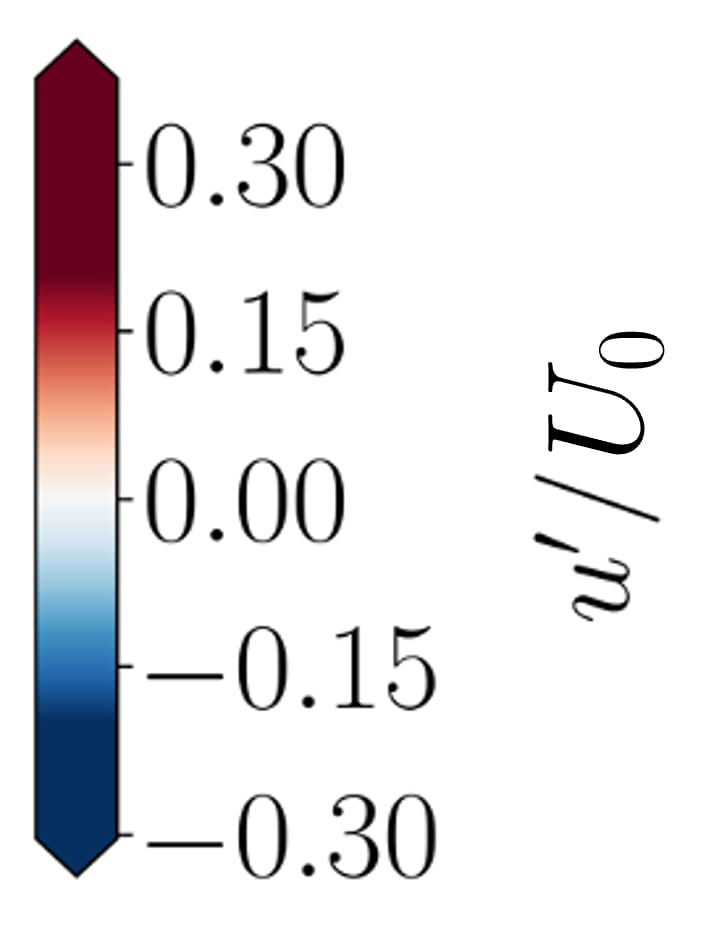}
        }
        \end{subfigure}
    \begin{subfigure}[b]{0.75\textwidth}
        \centering
        \includegraphics[width=\textwidth]{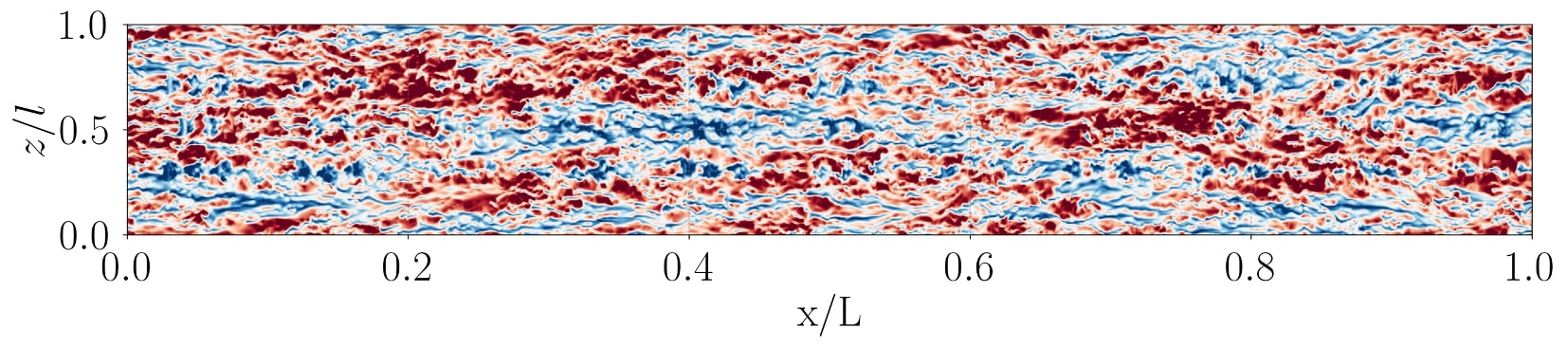}
        \caption{}
        \label{fig:subfig2}
    \end{subfigure}
    \begin{subfigure}[b]{0.2\textwidth}
        \centering
        \raisebox{0.8cm}{ 
            \includegraphics[width=0.6\linewidth]{Figure_new/colorbar_fig13ab.jpg}
        }
    \end{subfigure}
    \begin{subfigure}[b]{0.75\textwidth}
        \centering
        \includegraphics[width=\textwidth]{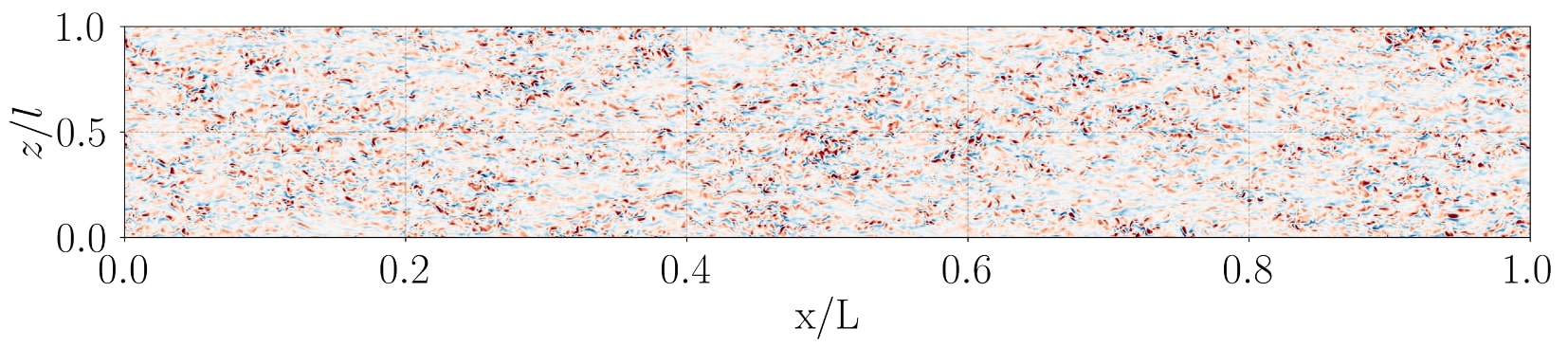}
        \caption{}
        \label{fig:subfig3}
    \end{subfigure}
     \begin{subfigure}[b]{0.2\textwidth}
        \centering
        \raisebox{0.8cm}{ 
            \includegraphics[width=0.6\linewidth]{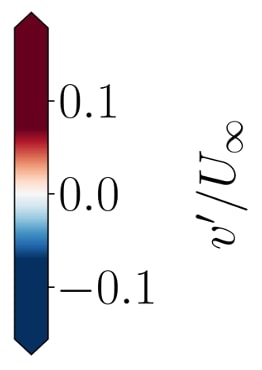}
        }
    \end{subfigure}
    \begin{subfigure}[b]{0.75\textwidth}
        \centering
        \includegraphics[width=\textwidth]{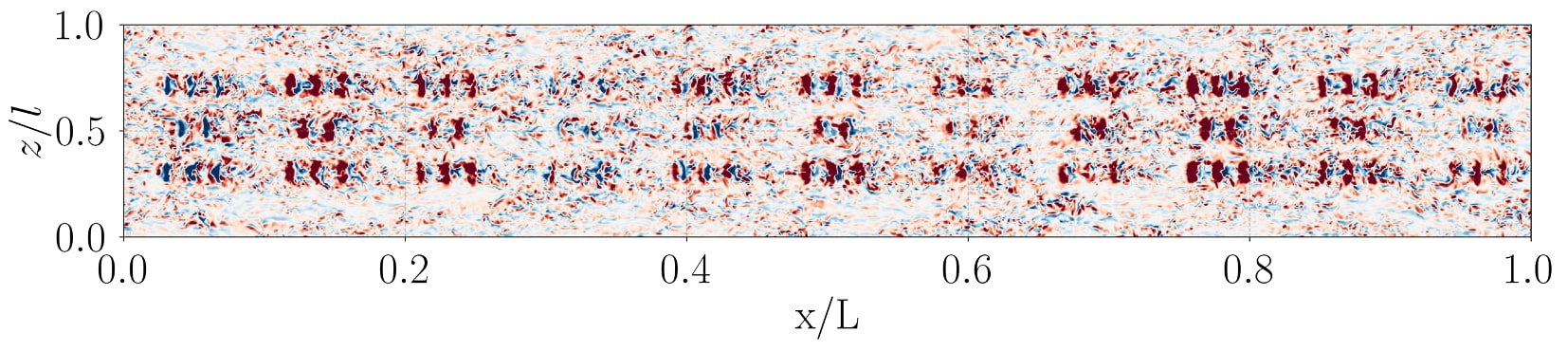}
        \caption{}
        \label{fig:subfig4}
    \end{subfigure}
    \begin{subfigure}[b]{0.2\textwidth}
        \centering
        \raisebox{0.8cm}{ 
            \includegraphics[width=0.6\linewidth]{Figure_new/colorbar_figure13_bc.jpg}
        }
    \end{subfigure}
    \vspace{0.5cm}
    \begin{subfigure}[b]{0.75\textwidth}
        \centering
        \includegraphics[width=\textwidth]{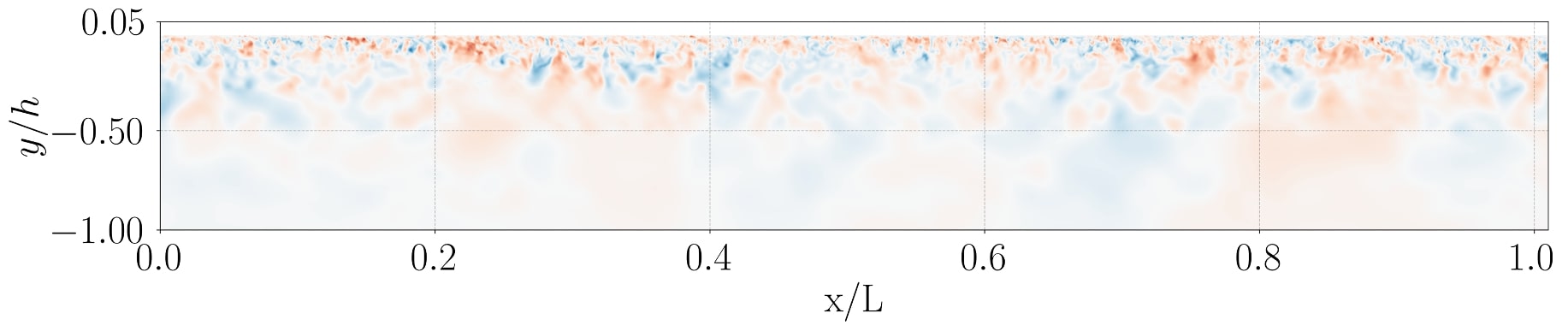}
        \caption{}
        \label{fig:subfig5}
    \end{subfigure}
        \begin{subfigure}[b]{0.2\textwidth}
        \centering
        \raisebox{0.8cm}{ 
            \includegraphics[width=0.6\linewidth]{Figure_new/colorbar_figure13_bc.jpg}
        }
    \end{subfigure}
    \begin{subfigure}[b]{0.75\textwidth}
        \centering
        \includegraphics[width=\textwidth]{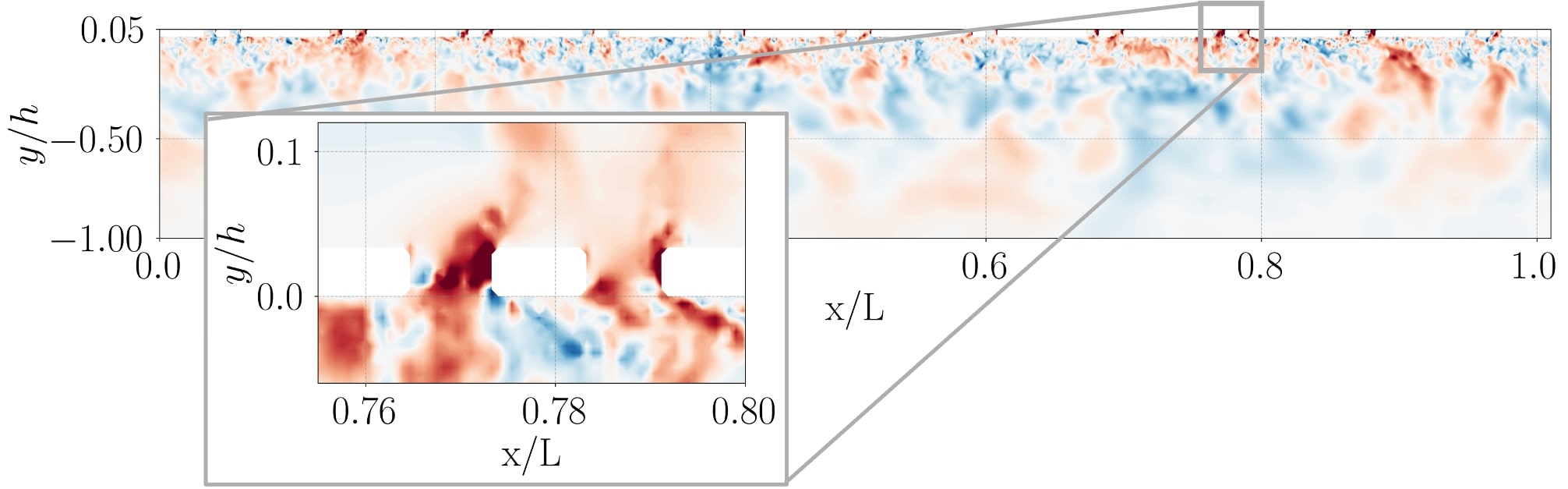}
        \caption{}
        \label{fig:subfig6}
    \end{subfigure}
    \begin{subfigure}[b]{0.2\textwidth}
        \centering
        \raisebox{0.8cm}{ 
            \includegraphics[width=0.6\linewidth]{Figure_new/colorbar_figure13_bc.jpg}
        }
    \end{subfigure}
    \caption{(a, b) Instantaneous streamwise $u^{\prime}$ (c, d) and wall-normal $v^{\prime}$ velocity components at $y/h = 0.003$. (e, f) Instantaneous vertical velocity $v^{\prime}$  component at $z/l = 0.5$. (a, c, e) Represent the smooth wall case while (b, d, f) represent the lined wall case. \blue{All figures show the lined wall case without incident sound.}}
    \label{fig:fig1}
\end{figure}

Before digging into the mean flow quantities, a qualitative description of the instantaneous flow features is reported in Figure \ref{fig:fig1}. More in detail, Figures \ref{fig:fig1} (a, b) present the contour plots of the instantaneous streamwise velocity component, and Figures \ref{fig:fig1} (c, d) the wall-normal velocity component in a wall-parallel plane at $y/h=0.003$, corresponding to $y^+ \approx 15$ in the smooth wall case. Comparisons are made between the smooth wall and the lined surface. The streamwise velocity component exhibits low and high-speed streaks, as for the smooth wall case, but with higher intensity fluctuations and larger spanwise size. \blue{Higher wall-normal velocity fluctuations are observed where the liner orifices are located}, caused by the flow recirculation within the orifices \citep{Shahzad2023TurbulenceLiners}. This is more evident in the wall-normal plane at $z/l = 0.5$ shown in Figures \ref{fig:fig1} (e, f). The higher velocity fluctuations, caused by the flow within the orifices, are not only localised close to the surface but also extend towards the center of the duct. This suggests that the acoustic liner modifies the flow near the wall and in the outer portion of the flow.

\subsection{Mean velocity profiles}
\label{sec:meanvelocityprofile}
\begin{figure}
    \centering
    \includegraphics[width=0.5\linewidth]{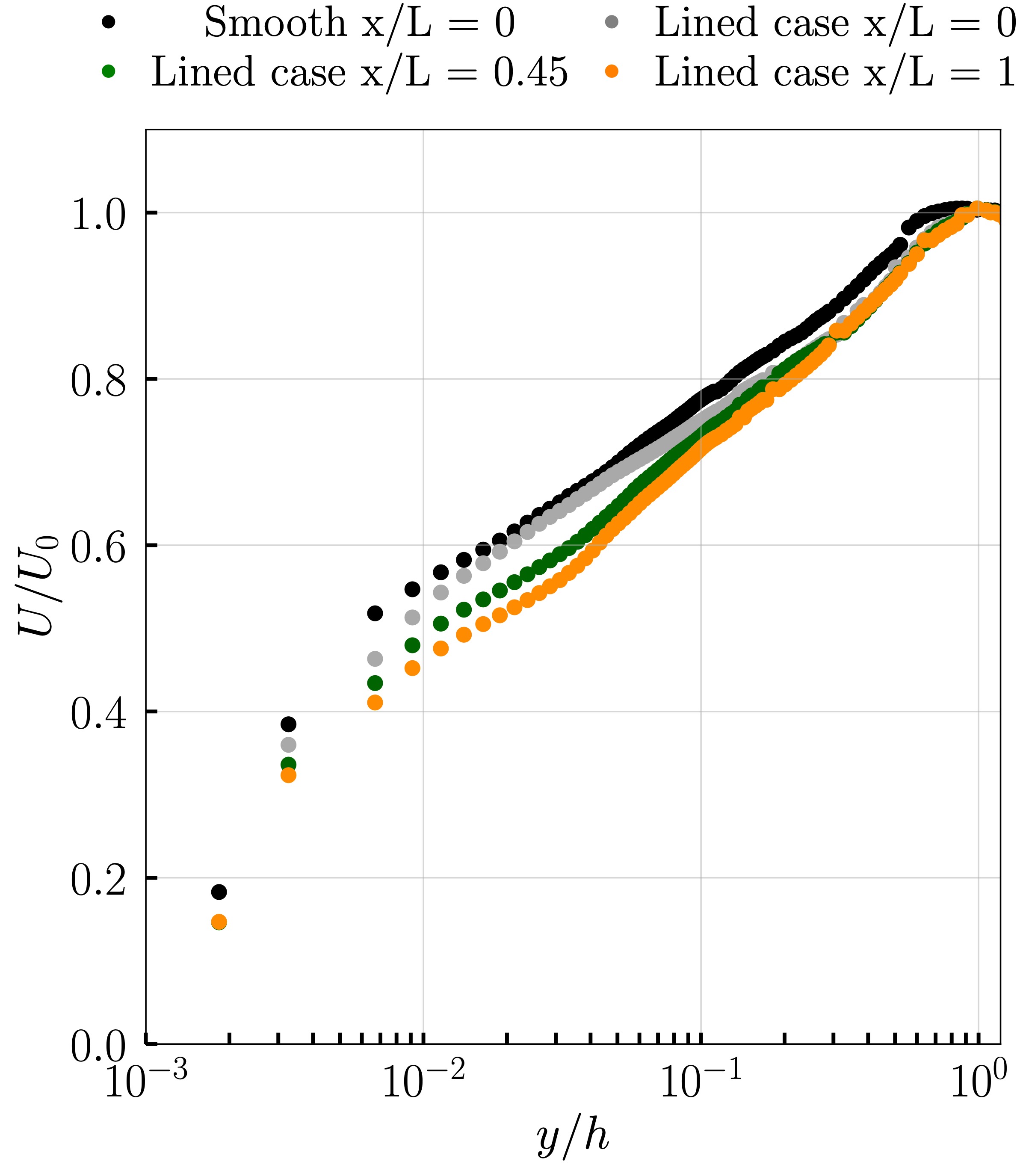}
    \caption{Time-average of streamwise velocity profile in outer coordinates sampled at three locations over the liner compared with the smooth case. \blue{No incident sound wave.}}
    \label{fig:mean_velocity_noac}
\end{figure}

Figure \ref{fig:mean_velocity_noac} compares the smooth wall mean streamwise velocity profiles at $x/L=0$ with the ones at three streamwise positions over the liner $x/L=0,0.45$ and $1$. At the beginning of the liner ($x/L=0$), the introduction of the lined section \blue{changes} the pressure gradient compared to the smooth-wall configuration, primarily affecting the outer-layer velocity profiles. The velocity profile at $x/L=0.45$ reveals a pronounced reduction of the velocity in the wall-normal region ($10^{-2} <y/h < 5 \cdot 10^{-2}$). The average streamwise velocity reduction in this region, relative to the inlet profile, is $\Delta U =- (U_{x/L = 0.45} - U_{x/L=0})/U_{x/L=0} = 14\%$. This result is consistent with previous findings for turbulent flows over rough surfaces \citep{Jimenez2004} and acoustic liners \citep{Shahzad2023TurbulenceLiners}. The streamwise velocity near the wall continues to reduce moving at $x/L = 1$, but the rate of reduction is less strong than for the first half of the lined section ($\Delta U =- (U_{x/L = 1} - U_{x/L=0.45})/U_{x/L=0.45} = 3\%$).

\begin{figure}
    \centering
    \includegraphics[width=\linewidth]{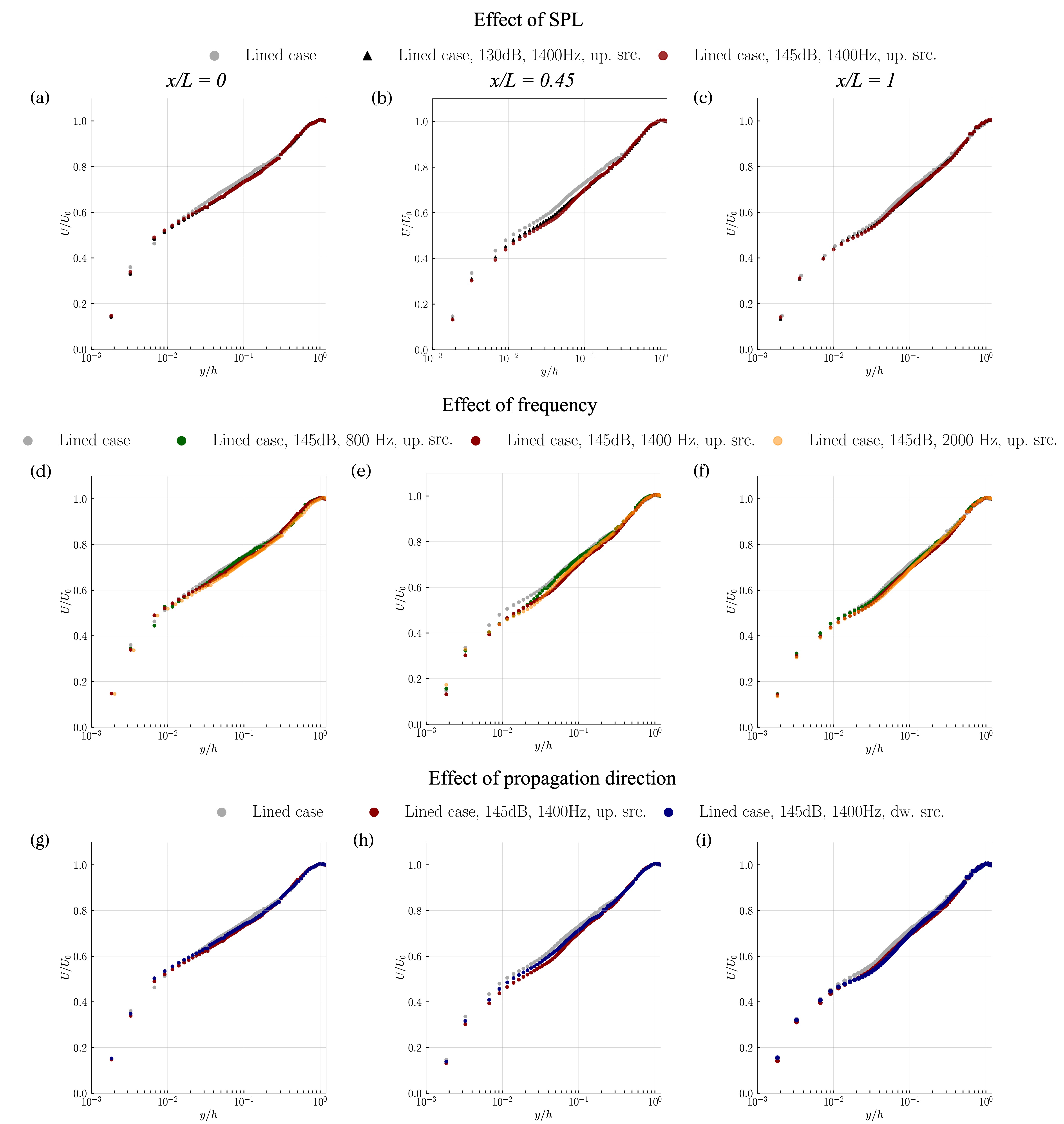}
    \caption{Time-averaged velocity profiles sampled at (a, d, g) $x/L = 0$, (b, e, h) $x/L = 0.45$, and (c, f, i) $x/L = 1$. The impact of the plane wave properties: (a, b, c) SPL, (d, e, f) frequency,  and (g, h, i) propagation direction of the acoustic wave is described.}
    \label{fig:fig2}
\end{figure}

The impact of the grazing acoustic wave on the streamwise development of the channel flow is shown in Figure \ref{fig:fig2}. \blue{Here, the mean flow velocity profiles over the liner with acoustics are compared to those without acoustics.} Each column reports a streamwise location, moving downstream from left to right. Each row describes the impact of a different parameter of the acoustic wave. At the beginning of the lined section, $x/L=0$, the grazing acoustic waves weakly affect the mean velocity profiles, independently of the SPL (Figure \ref{fig:fig2} (a)), the frequency of the plane wave (Figure \ref{fig:fig2} (d)), and the propagation direction of the acoustic wave (Figure \ref{fig:fig2} (g)). The acoustic waves modify the velocity profile in the region between 0.01 $< y/h < $ 0.04; these differences are more evident, even if weak, i.e., $\Delta U < 4\%$, when increasing SPL and near the resonance frequency of the liner, estimated without flow. Furthermore, the upstream acoustic source causes a slightly higher velocity deficit with respect to the downstream case.

The most relevant impact of the acoustic waves on the mean flow is found at $x/L = 0.45$. Here, the mean flow profile varies depending on the characteristics of the acoustic wave. The trends are similar to those found at $x/L = 0$ but enhanced. The observed dependence on the excitation frequency may be explained by the different proximity of the excitation to the liner's resonance frequency in the presence of grazing flow (Figure \ref{fig:impedance_145}).  The dependency on the propagation direction is more evident at this streamwise location, with a larger velocity deficit for the upstream source.

At the end of the lined section, the mean velocity profiles do not depend on the characteristics of the plane acoustic wave. This might be due to the flow recovering immediately from the perturbations caused by the acoustic-induced field.

\subsubsection{Streamwise distribution of the boundary layer integral quantities}

Previous analyses have shown that the presence of the liner affects the mean streamwise velocity profiles in the region $10^{-2} <y/h < 5 \cdot 10^{-2}$. To investigate this effect in more detail, the streamwise evolution of the boundary layer thickness $\delta$ and the boundary layer displacement thickness $\delta^*$ is described.  The two quantities are defined as:
\begin{align}
    U(y=\delta) = 0.99 U_{0}, \\
    \delta^* = \int_0^{h} \left(1 - \frac{U(y)}{U_{0}}\right) \, dy.
\end{align}

\begin{figure}
    \centering
    \begin{subfigure}[b]{\textwidth}
        \centering
        \includegraphics[width=\textwidth]{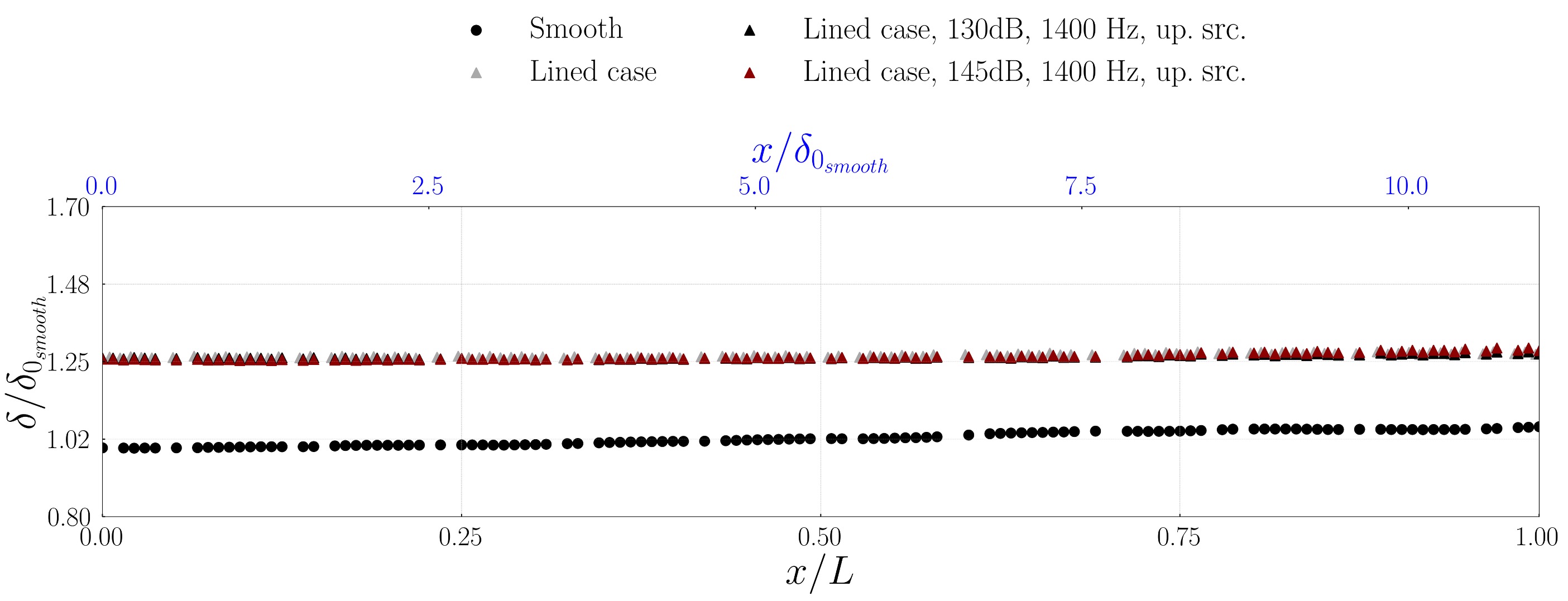}
        \caption{}
        \label{fig:subfig5-1}
    \end{subfigure}
    \hfill
    
    \begin{subfigure}[b]{\textwidth}
        \centering
        \includegraphics[width=\textwidth]{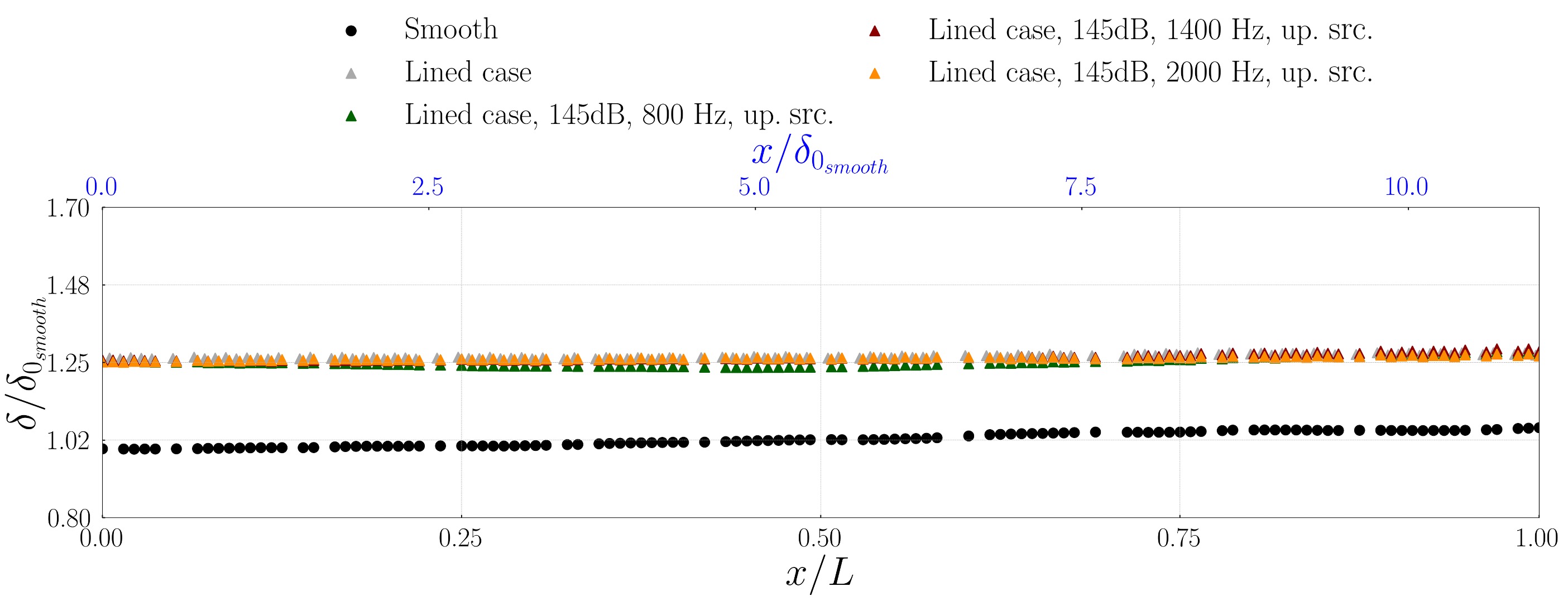}
        \caption{}
        \label{fig:subfig5-3}
    \end{subfigure}
    \hfill
    
    \begin{subfigure}[b]{\textwidth}
        \centering
        \includegraphics[width=\textwidth]{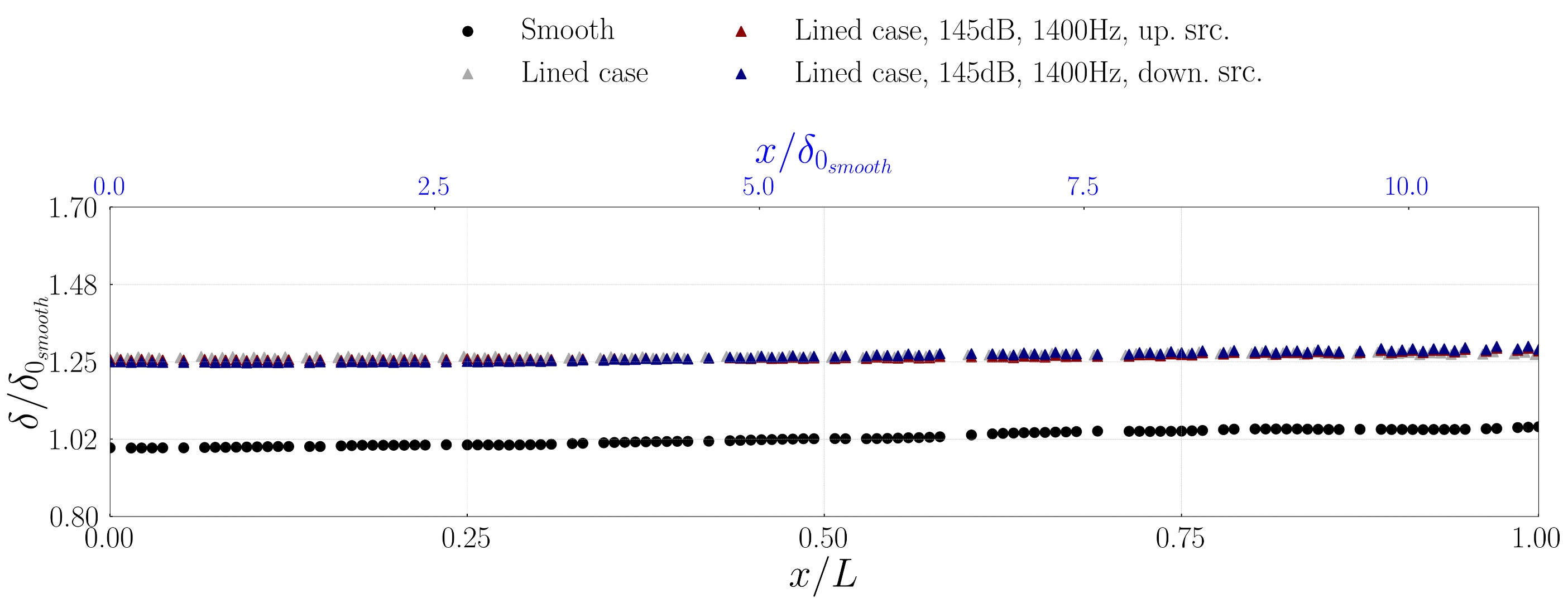}
        \caption{}
        \label{fig:subfig5-5}
    \end{subfigure}
    \caption{Streamwise development of $\delta$ over the liner. (a) Effect of acoustic waves' amplitudes at a fixed frequency of 1400 Hz; (b) effect of the acoustic waves' frequency at a fixed amplitude of \SI{145}{dB}; (c) effect of different propagation directions at a fixed frequency of 1400 Hz and amplitude of 145 dB.}
    \label{fig:fig5}
\end{figure}

\begin{figure}
    \centering
        \begin{subfigure}[b]{\textwidth}
        \centering
        \includegraphics[width=\textwidth]{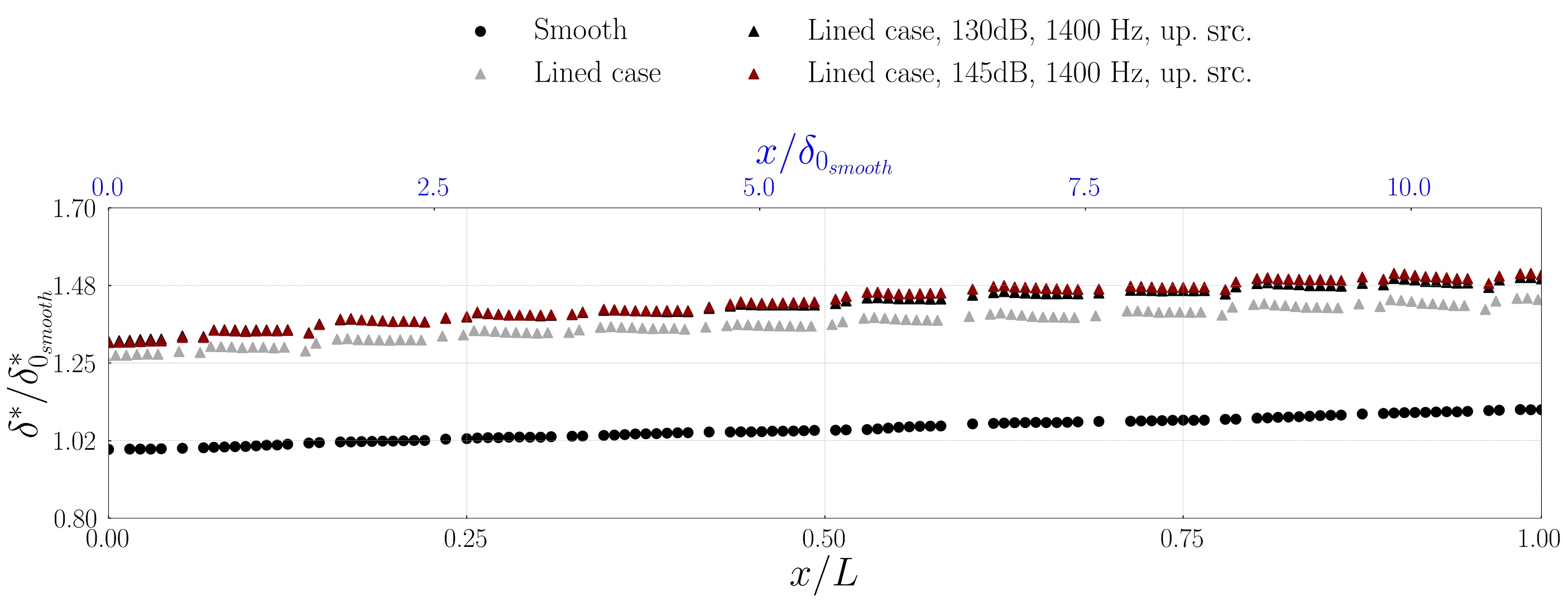}
        \caption{}
        \label{fig:subfig5-2}
    \end{subfigure}
    \hfill
        \begin{subfigure}[b]{\textwidth}
        \centering
        \includegraphics[width=\textwidth]{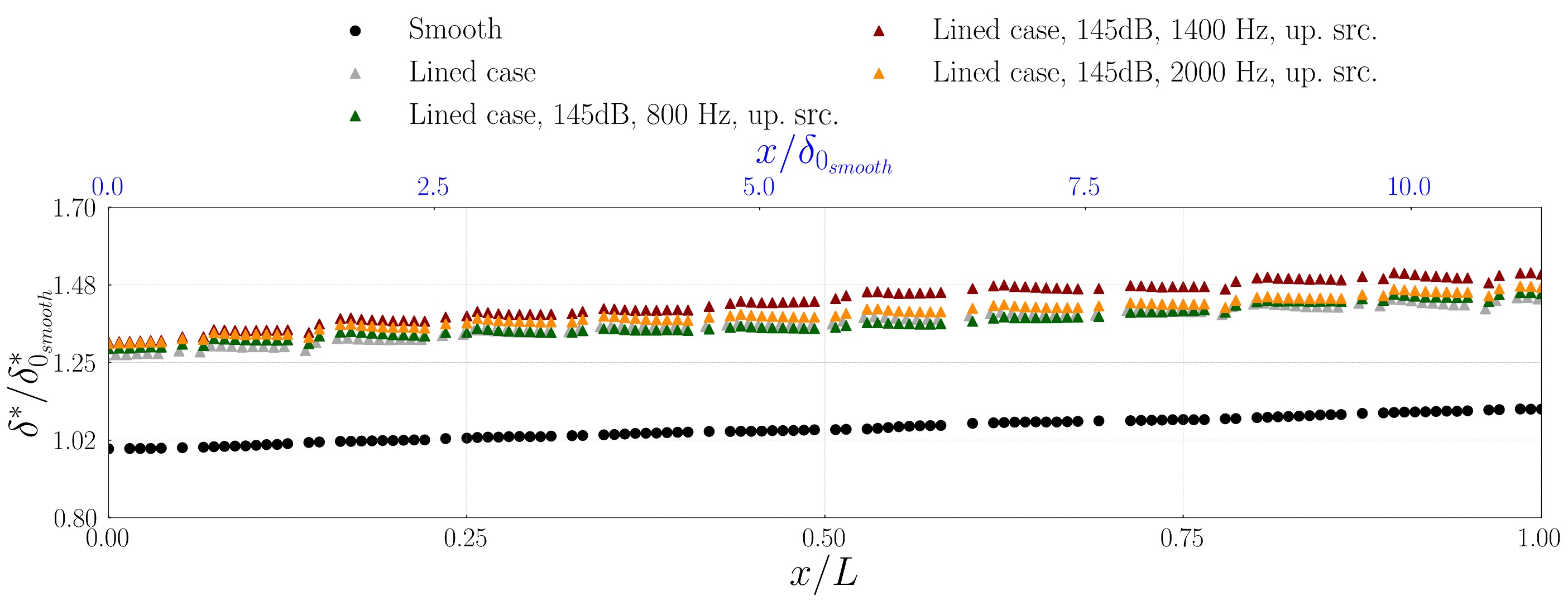}
        \caption{}
        \label{fig:subfig5-4}
    \end{subfigure}
    \hfill
      \begin{subfigure}[b]{\textwidth}
        \centering
        \includegraphics[width=\textwidth]{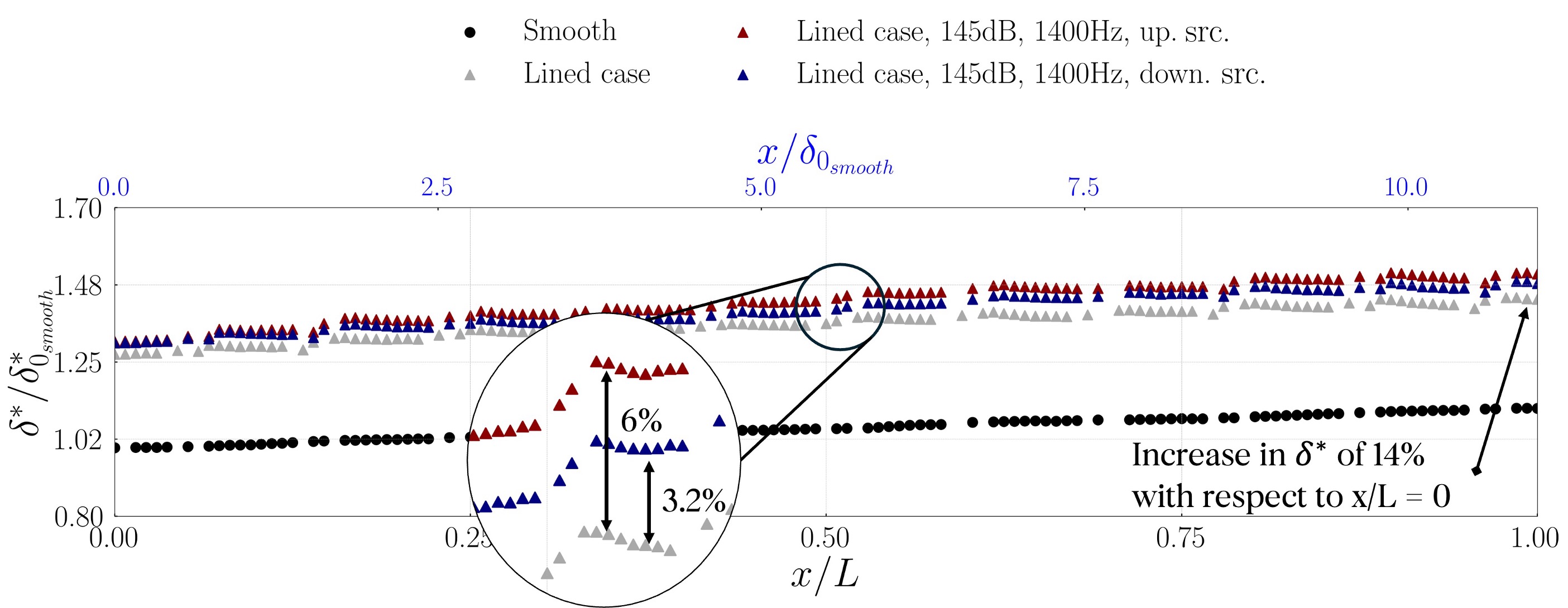}
        \caption{}
        \label{fig:subfig6}
    \end{subfigure}

    \caption{Streamwise development of $\delta^*$ over the liner. (a) Effect of different acoustic waves' amplitude at a fixed frequency of \SI{1400}{Hz} ; (b) effect of acoustic waves' frequency at a fixed amplitude equal to \SI{145}{dB}; (c) effect of different propagation directions at a fixed frequency of 1400 Hz.}
    \label{fig:delta_star_dev}
\end{figure}

The boundary layer thickness, in this context, is used to measure the impact of the liner on the velocity profile. The reasons are: (i) the flow within the channel is not symmetric with respect to the centerline because the acoustic liner is present only on one side; (ii) the flow is not yet fully developed, as also happens in the reference experiment. \blue{However, since $\delta$, is by definition an arbitrary quantity, we quantified also the streamwise variation of the displacement thickness $\delta^*$, which better indicates near-wall flow gradients and is an} important parameter for characterising the flow profile in the semi-empirical models used for computing impedance \citep{NAYFEH1974413, Yu2008ValidationData}. These quantities are reported in Figure \ref{fig:fig5} and Figure \ref{fig:delta_star_dev} and are made non-dimensional with respect to the corresponding value for the smooth wall at $x/L=0$.

The streamwise development of $\delta$ in the smooth wall case (Figure \ref{fig:fig5}) confirms that the flow is not yet fully developed in the lined section. As a matter of fact, $\delta$ shows a plateau from $x/L\approx0.75$. The presence of the liner increases $\delta$ by approximately 25\%. The streamwise development of $\delta$ varies weakly when the acoustic wave is present. It is interesting to notice that the effect of the SPL, frequency, and direction of the acoustic waves are visible up to $x/L\approx 0.8$ or $x/\delta_{smooth} \approx 10$. Beyond this point, the curves tend to converge, suggesting that the influence of the acoustic perturbation may diminish once the turbulent flow starts adapting to the lined surface. The weak dependence of $\delta$ on the features of the acoustic wave aligns with the previous observations (Figure \ref{fig:fig2}) that the presence of the acoustic liner dominates the mean flow variation.

The streamwise development of $\delta^*$ (Figure \ref{fig:delta_star_dev}) shows a more pronounced increase with respect to the smooth wall up to 30\% and depends on the features of the acoustic plane wave. More in detail, it increases in the presence of the acoustic wave, it grows more with higher SPL (Figure \ref{fig:delta_star_dev}(a)), at frequencies close to the resonance frequency of the liner (Figure \ref{fig:delta_star_dev}(b)) and it depends on the propagation direction of the acoustic wave (Figure \ref{fig:delta_star_dev}(c)). In contrast to $\delta$, the $\delta^{*}$ emphasises the differences between upstream and downstream acoustic waves. This behaviour highlights the influence of wave direction on near-wall \blue{velocity deficit}.

The step-like behaviour of $\delta^*$, in correspondence with each cavity of the acoustic liner, is due to the near-wall flow modifications induced by the orifices. This effect is further enhanced by the acoustic-induced flow field. These flow features will be visualised and described in the next section.

\subsection{Impact of flow development on the near-wall and in-orifice flow field}
\begin{figure}    
        \centering
        \includegraphics[width=\linewidth]{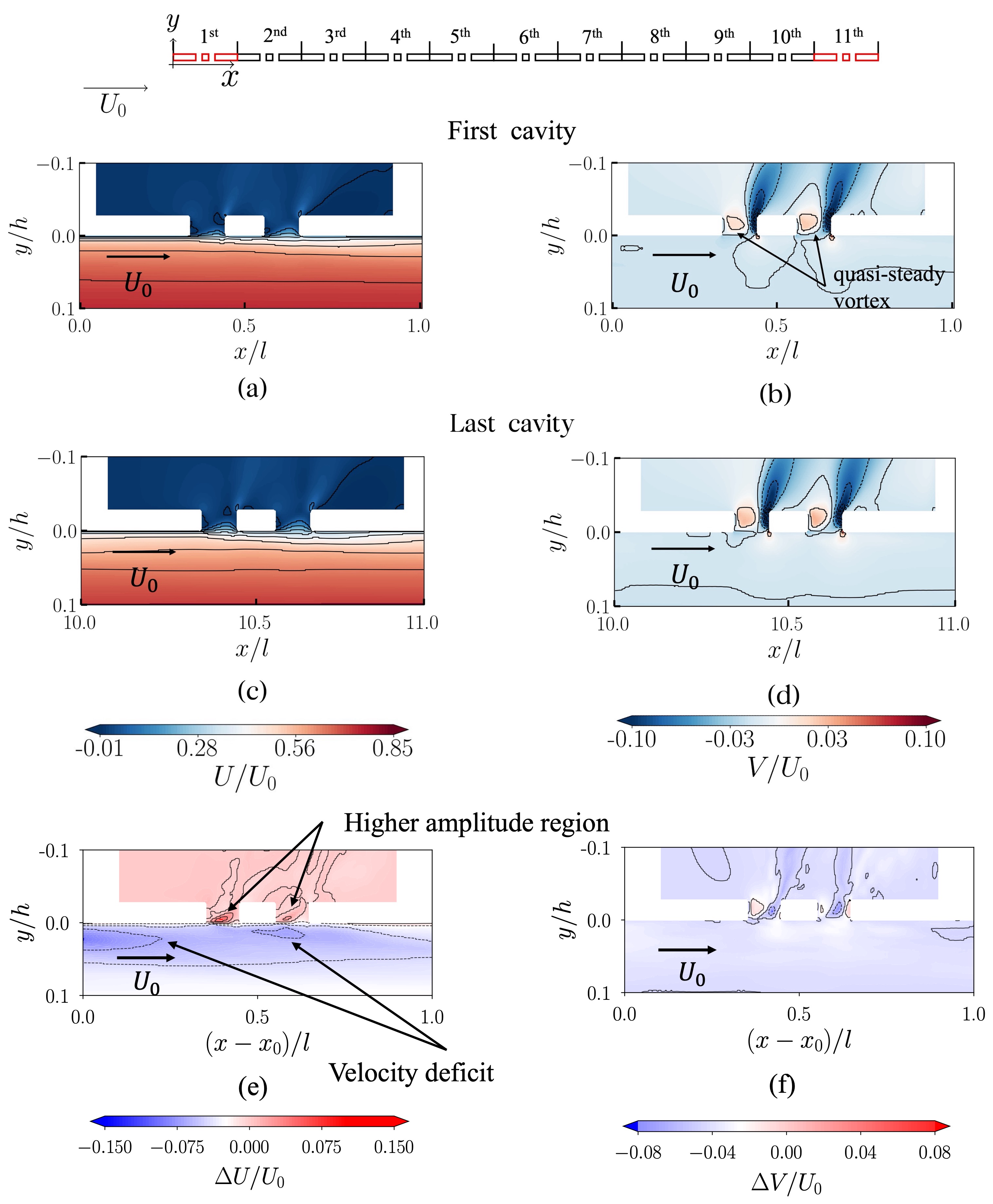}
    \caption{Contour of mean velocity component (a, c) and wall-normal velocity component (b, d) for the flow-only case. (a, b) First cavity, (c, d) last cavity. \blue{(e) Difference in the streamwise velocity defined as $\Delta U = (U_{\mathrm{last\,cav.}} - U_{\mathrm{first\,cav.}})$. (f) Difference in the wall-normal velocity defined as $\Delta V = (V_{\mathrm{last\,cav.}} - V_{\mathrm{first\,cav.}})$. All results refer to the lined wall case without acoustic sources.}}
    \label{fig:contour_effect_of_delta}
\end{figure}

The observed changes in boundary layer integral parameters are driven mainly by the presence of the liner. The introduction of the acoustic liner modifies the near-wall flow, which is reflected in an increase in the boundary layer displacement thickness. To further investigate this aspect, Figure \ref{fig:contour_effect_of_delta} presents contour plots of the streamwise and wall-normal velocity components at the first and last cavity in the absence of an acoustic plane wave. 

In both cavities, the wall‑normal velocity contours reveal a quasi‑steady vortex occupying the upstream portion of each orifice, consistent with earlier observations \citep{tam_microfluid_2000, Avallone2019Lattice-boltzmannFlow}. The vortex strength, however, slightly differs between cavities and between successive orifices. Looking at streamwise velocity components over the two cavities (Figure \ref{fig:contour_effect_of_delta} (a, c)), for both cases, downstream of the orifices, the flow is first displaced away from the wall and then tends to reattach. The displacement in the downstream direction causes a reduction of the near-wall streamwise velocity gradient from the most upstream to the most downstream location. \blue{This is quantified in Figure \ref{fig:contour_effect_of_delta} (e) where the difference between the flow field over the cavities is quantified as $\Delta U = (U_{\mathrm{last\,cav.}} - U_{\mathrm{first\,cav.}})$. Over the last cavity, near the wall, the velocity decreases by up the 12\% $U_0$.} As a consequence, over the orifices, the streamwise velocity component gradient in the wall-normal direction $dU/dy$ is also less strong for the most downstream cavity\blue{, thus resulting in localized high amplitude $\Delta U/U_0$ regions within the orifice of approximately 10\%}. The latter \blue{might affect} the strength of the quasi-steady vortex that is formed within each orifice \blue{as suggested by an increase in the wall‑normal velocity in the upstream part of the last cavity’s orifice by $\Delta V = V_{\mathrm{last\,cav.}} - V_{\mathrm{first\,cav.}} = 0.03\,U_0$ (Figure \ref{fig:contour_effect_of_delta}(f))}.

\begin{figure}
\centering
   \includegraphics[width=\linewidth]{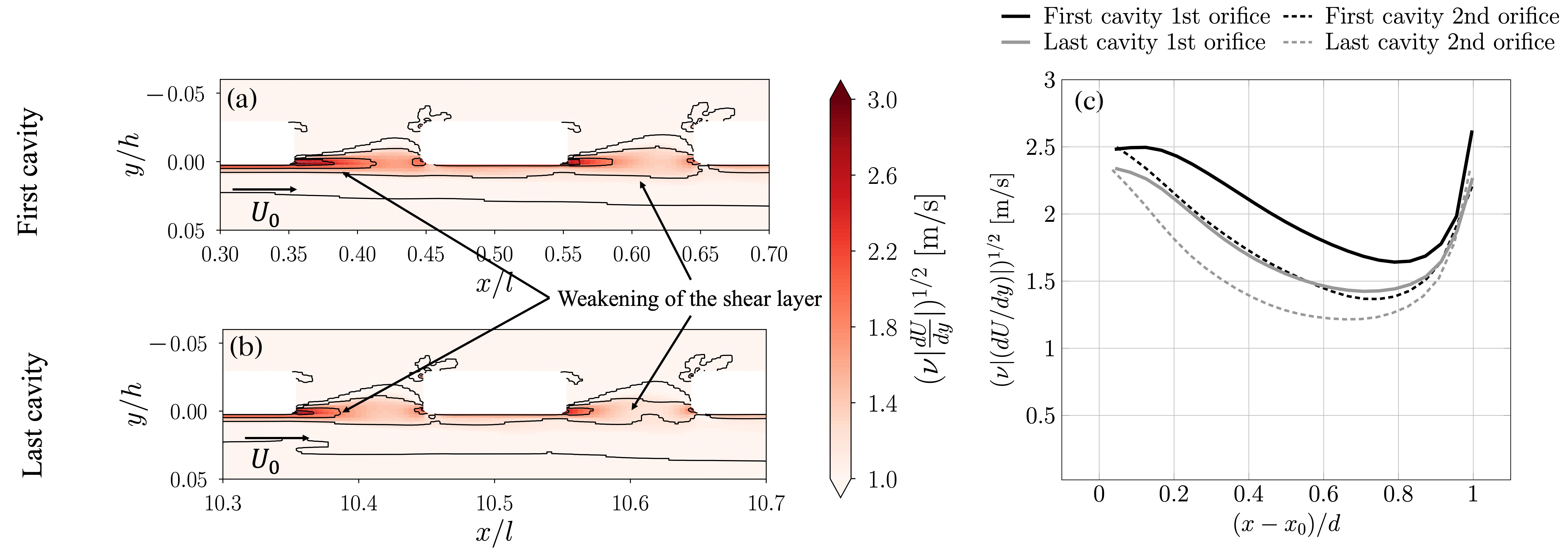}
    \caption{Contour of $\nu (|dU/dy|)^{1/2}$ for (a) the first and (b) the last cavity for the flow-only case. \blue{(c) Line plots of $\nu (|dU/dy|)^{1/2}$ sampled at $y/h=0$ above the orifices, comparing the first and last cavities and highlighting the downstream weakening of the shear layer. Plots show the case without incident sound waves.}}
    \label{fig:shear_contours}
\end{figure}

To quantify this effect, Figure \ref{fig:shear_contours} examines the local shear-related scale $(\nu |dU/dy|)^{1/2}$, which has the same units as the friction velocity, for the first and last cavity in the flow-only configuration. The contours reveal a streamwise attenuation of shear intensity over the orifice in the downstream direction \blue{as clearly shown in Figure \ref{fig:shear_contours} (c), which presents this quantity at $y/h=0$.} In the first cavity, the upstream orifice exhibits the strongest shear layer; at the downstream orifice, the magnitude is roughly halved, and the high-intensity shear region extends less far downstream. This trend is even more pronounced in the last cavity: the shear over its first orifice is already weaker than that in the first cavity, and it diminishes sharply over the second orifice.
\blue{Moreover, the $\nu (|dU/dy|)^{1/2}$ over the last cavities experiences a reduction in magnitude of up to approximately 30\% compared to the first one.}
The stronger shear layer at the first cavity obstructs the penetration of the acoustic wave into the orifice with respect to what happens at the last cavity. This variation can be interpreted as a spatial asymmetry in the flow field, caused by the growth of the boundary layer displacement thickness. \blue{The acoustic wave propagating in the direction opposite to the mean flow will see different near-wall flow gradients and different shear over the orifices with respect to the wave which propagates in the same direction as the flow.} Furthermore, the development of the flow highlights the relevance of near-wall flow features and the need to correctly account for the boundary layer displacement thickness.

\subsection{Impact of SPL on the near-wall and in-cavity flow field}
\label{sec:impactofspl}
The preceding analysis demonstrated that the streamwise development of the grazing flow alters the near-wall and in-orifice velocity fields, primarily through a progressive reduction in the streamwise velocity gradient and weakening of the shear layer over the downstream orifices. These modifications influence the vortex dynamics and flow penetration within the cavities, even in the absence of acoustic excitation. To further understand the coupling between fluid dynamics and acoustics, this section describes how the amplitude of the acoustic wave affects the mean and fluctuating flow fields in the vicinity of the orifices. Analyses were conducted for the sixth cavity (i.e. $x/L$ = 0.45). 

\begin{sidewaysfigure}
        \vspace*{13cm}
        \centering\label{fig:schematic_where_analyzing_effects_of_spl}
    \centering
      \includegraphics[width=0.95\linewidth]{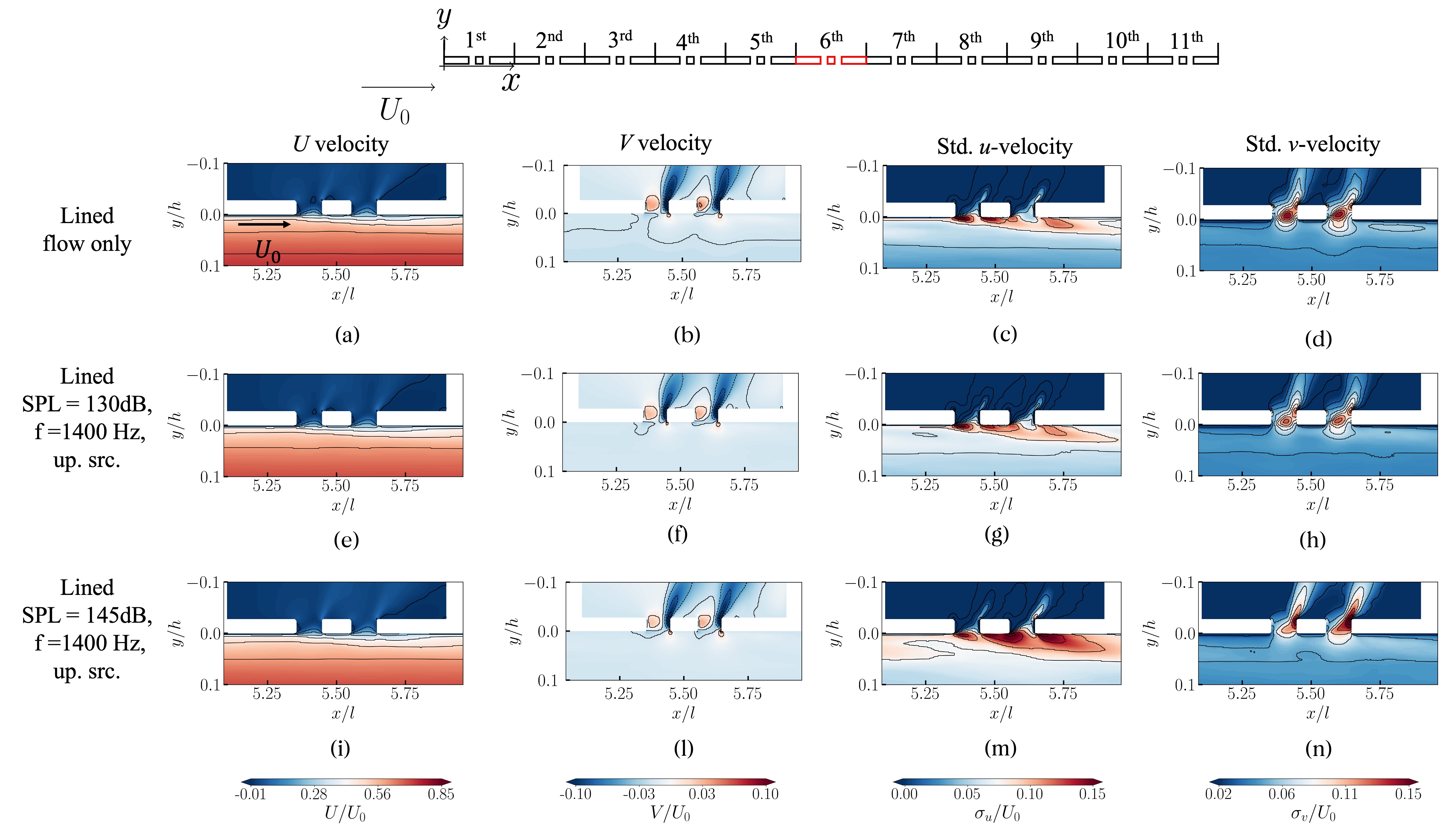}
    \caption{Contour of the mean and standard deviation of the streamwise and wall-normal velocity components for the cases: flow only Lined (first row), flow and upstream acoustic source with SPL = \SI{130}{dB} and $f$ = 1400 Hz (middle row) and flow and upstream acoustic source with SPL = \SI{145}{dB} and $f$ = 1400 Hz (bottom row).}
    \label{fig:theeffectofspl}
\end{sidewaysfigure}

Figure \ref{fig:theeffectofspl} presents the contours of the mean streamwise and wall-normal velocity components, along with their standard deviations. The case with acoustic excitation is compared to the case without acoustic excitation. In the absence of acoustic waves (Figure \ref{fig:theeffectofspl} (a-d)), the contour of the standard deviation of the streamwise velocity component $\sigma_u/U_0$ (Figure \ref{fig:theeffectofspl} (c)) highlights the presence of high-intensity fluctuations over the first orifice and after the second one. These fluctuations are caused by vortices shed near the wall, which are also responsible for the displacement of the flow away from the wall. On the other hand, the turbulent flow causes high-intensity fluctuations of the vertical velocity component $\sigma_v/U_0$ (Figure \ref{fig:theeffectofspl} (d)) at the center of the orifice and the downstream edge of the face sheet. The location of the peak displaces slightly upstream in the second orifice with respect to the first one.

The presence of acoustic waves, with only one frequency reported for the sake of conciseness, weakly alters the mean flow, but largely the velocity fluctuations, \blue{as can be seen by comparing Figure \ref{fig:theeffectofspl} (c) and (m)}. As the amplitude of the acoustic wave increases, the flow is progressively displaced away from the wall, a behaviour analogous to what occurs in turbulent boundary layers subjected to wall-normal blowing or flow over permeable surfaces \citep{Jimenez2004, Zhang2016NumericalLayers, Avallone2019Lattice-boltzmannFlow, Spillere2020ExperimentallyDownstream}. This results in a stronger velocity deficit near the wall and elevated streamwise fluctuations in the outer layer.
Increasing the SPL to 145 dB, the amplitude of the streamwise velocity fluctuations near the wall increases; $\sigma_u/U_0$ shows a large amplitude further away from the wall, which increases at the downstream orifice. Also, $\sigma_v/U_0$ varies substantially with respect to the other two cases; the peak of $\sigma_v/U_0$ is localized at the downstream edge of both orifices, and it increases in amplitude at the more downstream edge.

\subsection{Impact of different directions of propagation of the acoustic wave}

This section investigates how the propagation direction of the acoustic wave affects the flow field.  To isolate this effect, the mean velocity field is analysed at  $x/L=0.54$, corresponding to the seventh cavity. Here, the \blue{incident} SPL is the same for both directions (SPL $\approx 137$ dB), yet the displacement thickness still differs by $\approx 3\%$.

\begin{sidewaysfigure}
 \vspace*{13cm} 
\centering  
\includegraphics[width=0.95\textwidth]{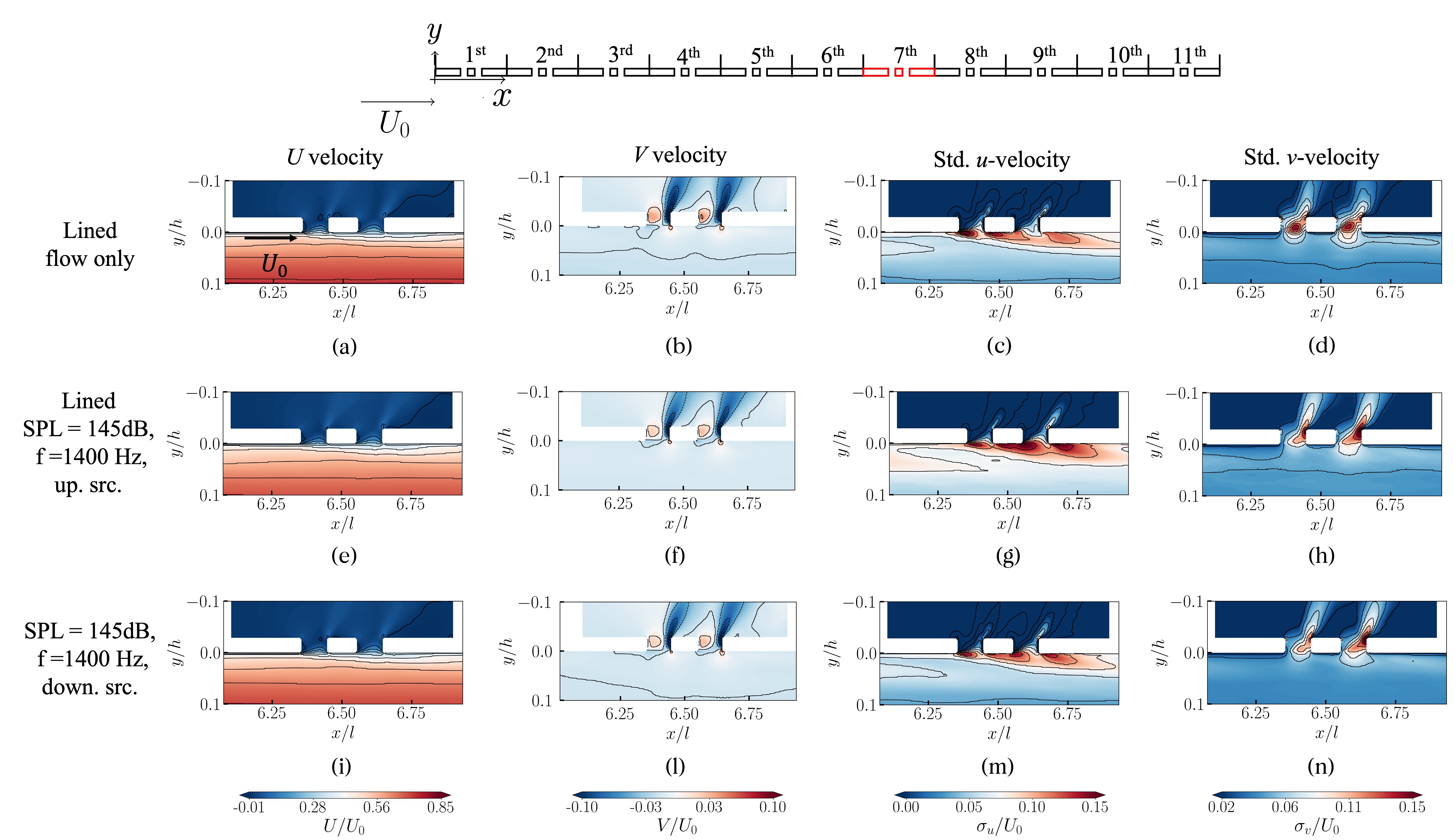}
    \caption{Contour of the mean and standard deviation of both the streamwise and the wall-normal velocities for the cases: flow only Lined (first row), flow and upstream acoustic source with SPL = \SI{145}{dB} and $f$ = 1400 Hz (middle row) and flow and downstream acoustic source with SPL = \SI{145}{dB} and $f$ = 1400 Hz (bottom row).}
    \label{fig:theeffectofpropagationdirection}
\end{sidewaysfigure}

Contours of the time-averaged and standard deviation of both the streamwise and wall-normal velocity components are shown in Figure \ref{fig:theeffectofpropagationdirection} for the cases: flow-only, flow and upstream acoustic source, and flow and downstream acoustic source. Only the case with acoustic plane wave amplitude equal to 145 dB and $f=1400$ Hz is reported for the sake of conciseness, because similar observations can be made for the other cases.

The propagation direction of the acoustic wave with respect to the mean flow has a minor impact on the time-averaged flow field when the local amplitude of the acoustic wave is similar. Contours of $U/U_0$ and $V/U_0$ do not show relevant differences between the two cases. The only visible difference is that, downstream of the second orifice, the flow is slightly less displaced away from the wall when the acoustic wave propagates in the opposite direction to the mean flow, which is consistent with the $\delta^*$ evolution (Figure \ref{fig:delta_star_dev} (c)). The strong similarities between the two cases explain the previous observation that the mean flow is affected more by the presence of orifices than by the acoustic field. However, the propagation direction of the acoustic wave has more impact on the amplitude of the velocity fluctuations than on their spatial distribution. For the upstream acoustic source,  $\sigma_u/U_0$ is higher, and the region with high amplitude extends spatially further downstream and away from the wall than in the case with a downstream source. This might be caused by the convective nature of the flow that tends to drag downstream the velocity fluctuations induced by the acoustic-induced velocity within the orifice. 
\blue{For the downstream acoustic source, by contrast, it appears that the regions with high amplitude velocity fluctuations are confined near the wall.}

\begin{figure}
    \centering
\includegraphics[width=0.85\textwidth]{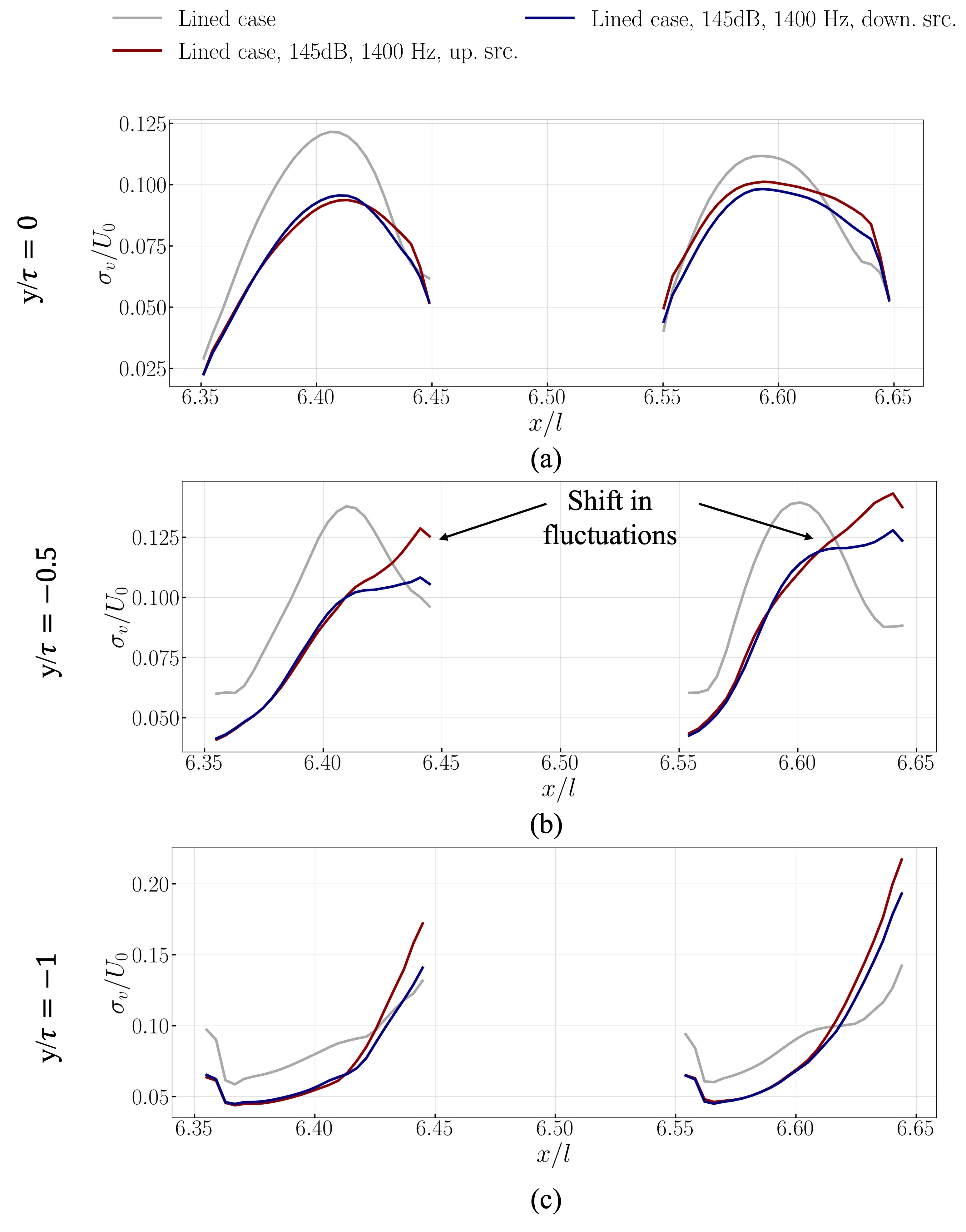}
    \caption{Spatial distribution of $\sigma_{v}/U_0$ into two adjacent orifices of the seventh cavity at (a) $y/\tau = 0$, (b) $y/\tau = -0.5$, (c) $y/\tau =-1$. Lined flow only case, flow and acoustics with SPL = \SI{145}{dB} $f = 1400$ Hz and upstream acoustic source, SPL = \SI{145}{dB} $f$ = 1400 Hz and downstream acoustic source.}
    \label{fig:lineplot_wallnstd_flow_direction}
\end{figure}

The direction of propagation of the acoustic wave also impacts the amplitude of the vertical velocity fluctuations within the orifice. As illustrated in Figure \ref{fig:lineplot_wallnstd_flow_direction}, the peak of $\sigma_v/U_0$ shifts towards the downstream edge of each orifice and penetrates deeper into the cavity when acoustic forcing is present. For the cases with acoustics, the peak value nearly doubles at $y/\tau=-1$ compared with the flow-only case. The distribution remains topologically similar for the two cases. Therefore, it can be concluded that, at similar SPL, the wave direction weakly alters the spatial distribution of the in‑orifice flow fluctuations but modulates their intensity.  The larger amplitude of the velocity fluctuations observed in the upstream-propagating wave case may be linked to the different boundary layer displacement thicknesses.

\subsection{Streamwise evolution of the standard deviation of the velocity fluctuations}
\begin{sidewaysfigure}
 \vspace*{13cm} 
\includegraphics[width=\textwidth]{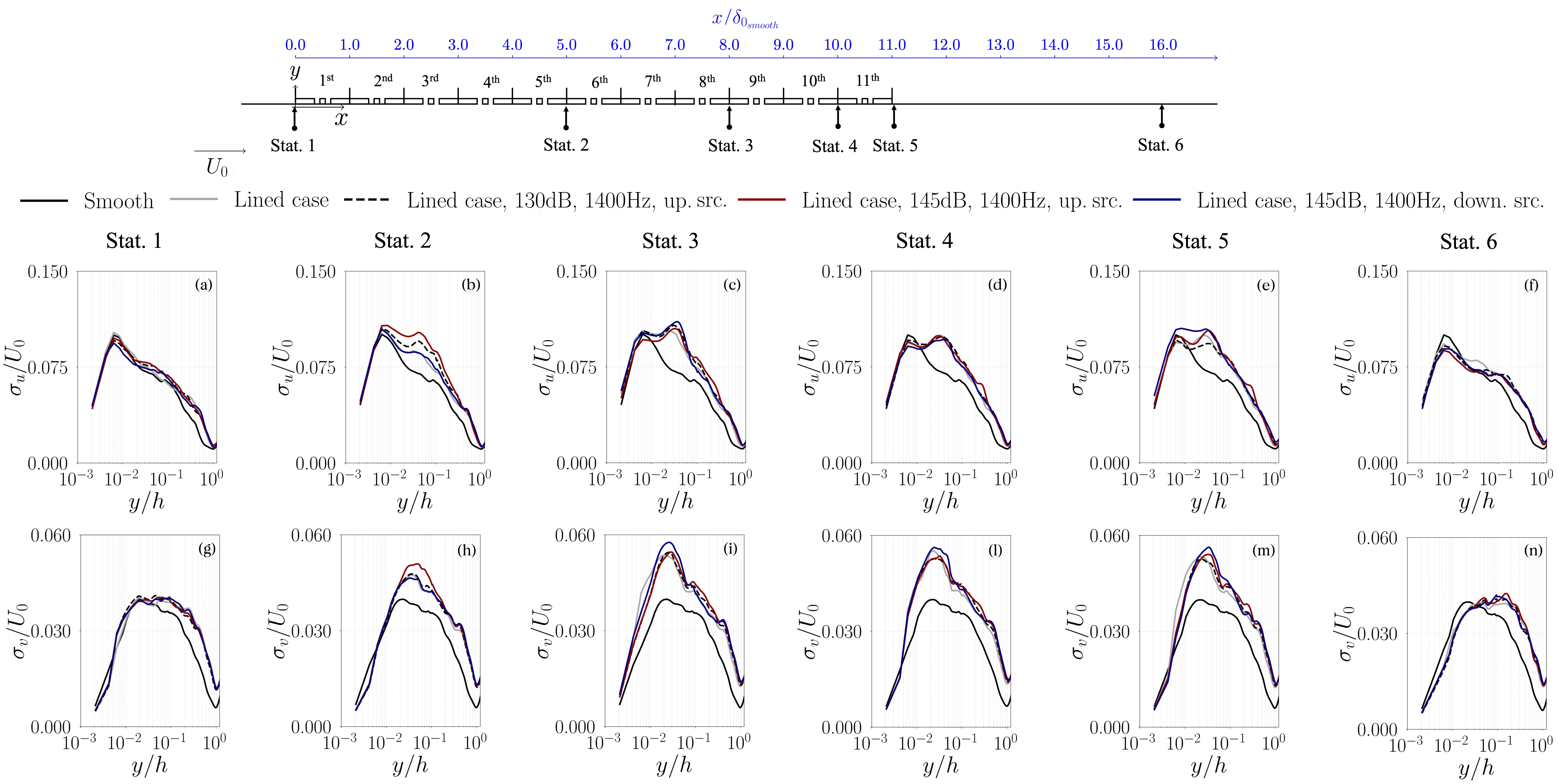}
\caption{Streamwise velocity standard deviation (top row) and wall-normal velocity standard deviation (bottom row) for the different test cases reported in the legend.}
    \label{fig:fig7}
\end{sidewaysfigure}

To assess how the liner surface geometry and acoustic forcing influence the outer layer of the turbulent flow, the standard deviations of the streamwise and wall-normal velocity components are examined at six streamwise stations along the liner, as shown in Figure \ref{fig:fig7}. The analysis focuses on cases with a fixed excitation frequency of 1400 Hz, varying SPLs, and propagation directions. Results are compared against both the flow-only and smooth-wall baselines. Similar trends were observed at other frequencies and are omitted here for brevity.

At $x/L = 0$ (station 1), no appreciable differences are observed in the streamwise velocity standard deviation profiles between the smooth-wall and lined configurations. Similarly, acoustic forcing has no discernible effect on the flow when applied over a smooth wall. In contrast, the wall-normal velocity fluctuations exhibit a slight increase in the region $10^{-2} < y/h < 1$ compared to the smooth-wall case.  However,  differences become more visible at the second streamwise station ($x/L=0.45$) where the lined flow-only case exhibits a distinct hump at $y/h \approx 4 \cdot 10^{-2}$. This outer-layer peak is consistent with previous observations of Zero-Pressure-Gradient (ZPG) turbulent boundary layers over porous or perforated surfaces, where wall-normal transport promotes the formation of secondary motions \citep{SCARANO2024109486}. The hump at $y/h \approx 4 \cdot 10^{-2}$ is amplified by the presence of the acoustic wave. The amplitude of the hump increases as the amplitude of the acoustic wave increases, and when the acoustic wave propagates as the mean flow does. For the case with the acoustic wave propagating in the direction opposite to the mean flow, no difference is visible with respect to the flow-only case. This is because the amplitude of the acoustic wave at this location has decreased to about 130 dB (Figure \ref{fig:spl_decay_145}). Closer to the wall, almost no variation is found with respect to the smooth wall case. Considering the standard deviation of the wall-normal velocity component, a similar conclusion can be drawn. Specifically, an increase in $\sigma_v/U_0$ is observed at the same wall-normal location ($y/h \approx 0.04$), consistent with the upward displacement of fluid from the near-wall region toward the outer layer. This displacement becomes more pronounced in the case with an upstream acoustic source at SPL = 145 dB, highlighting the role of acoustic forcing in enhancing momentum exchange and vertical transport within the near-wall flow.
At  $x/L = 0.73$ (i.e., the third streamwise station), the outer peaks of both the streamwise velocity and wall-normal velocity variance are enhanced with respect to the smooth wall case. At this station, for both the streamwise and wall-normal velocity, it is more difficult to spot differences in the outer hump amplitude between the cases with varying both the amplitude and direction of the acoustic wave, meaning that the effect of the acoustic is less evident. The reason behind this behaviour is believed to be driven by two different causes: (i) the acoustic wave is attenuated when progressing over the liner, (ii) the turbulent flow is adapting to the change in the surface. This is further supported by the standard deviation profiles at station 4. As a matter of fact, at this location, almost no differences are found in the amplitude of the outer hump. The turbulent flow seems to be adapted to the lined surface. Even though the phenomenology of the set-up under investigation slightly differs from a ZPG turbulent boundary layer, the behaviour of the flow over the lined surface resembles the behaviour of turbulent flow over rough surfaces, where outer-layer features stabilise further downstream \citep{Jaiswal_Ganapathisubramani_2024}. A slight reduction in the inner peak amplitude is also observed, consistent with earlier studies of turbulent flow over acoustic liners \citep{Shahzad2023DirectLiners}, and this inner‑layer attenuation is modestly modulated by the acoustic forcing.

At $x/L=1.0$ (station 5), the flow encounters a sudden transition from the lined surface to a smooth wall. Here, the outer-layer hump is modulated once again by the amplitude and direction of the acoustic wave. The largest deviation from the lined flow-only case is found in the downstream case at SPL = 145 dB. This region is characterised by the formation of secondary motions induced by the abrupt surface change, which are further intensified by the acoustic forcing. As a result, both streamwise and wall-normal velocity fluctuations show increased amplitude modulation by the acoustic wave properties.

Finally, at the most downstream location, $x/L = 1.5$, where the smooth condition is recovered downstream of the liner, variance profiles for both wall-normal and streamwise velocity components resemble the ones for the smooth wall. This indicates that after the surface transitions back to a smooth one, the flow tends to recover to a canonical state, regardless of whether acoustic forcing is present or not.
 
\section{Acoustic-induced velocity and mass flow rate}
\label{sec:acousitc_induced_flow}

\subsection{Acoustic-induced velocity field within the orifices}
\begin{figure}
    \centering
    \includegraphics[width=\linewidth]{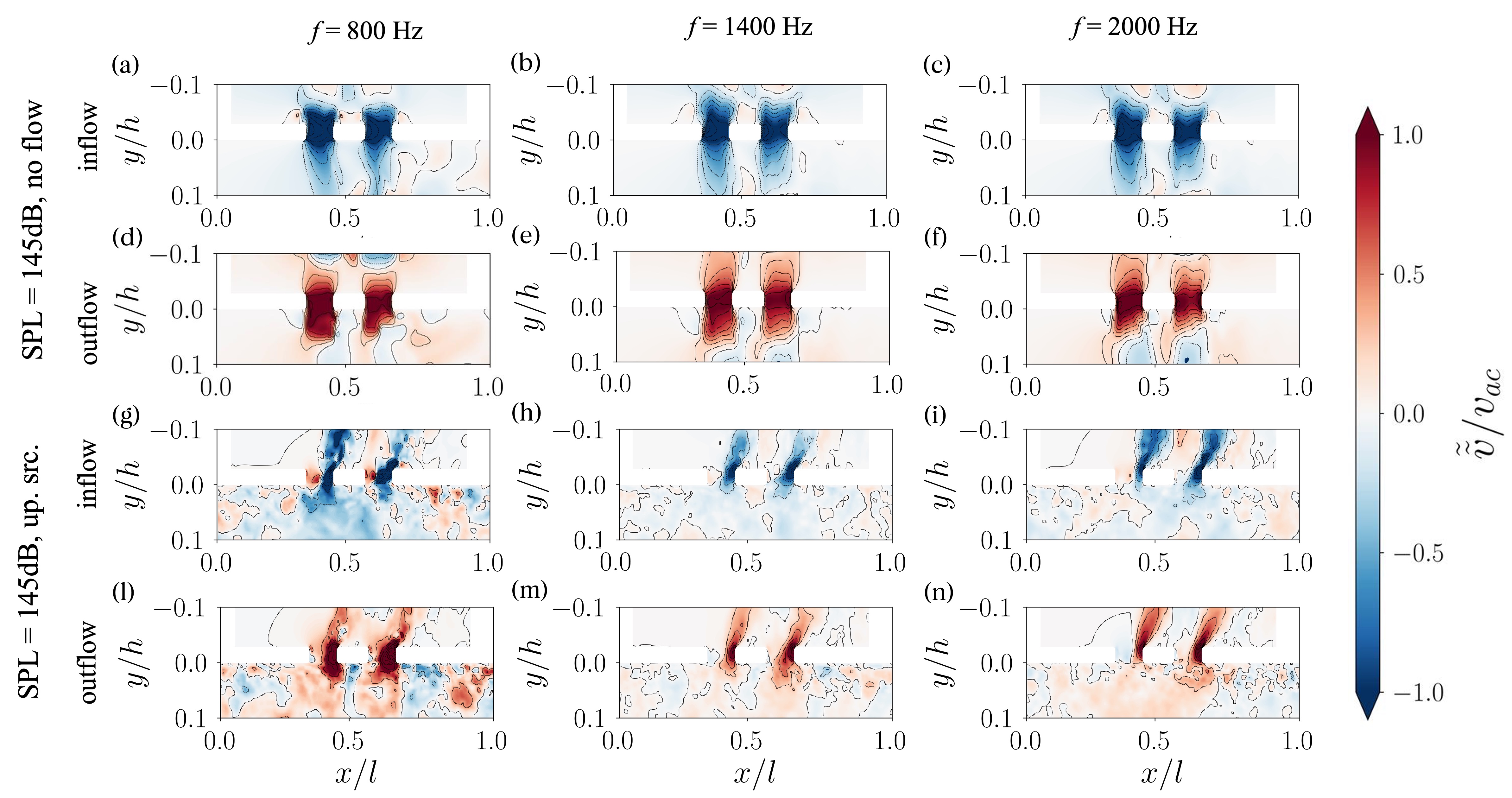}
    \caption{Contour plots of the acoustic-induced velocity at the peak of the inflow phase and peak of the outflow phase. All cases have an incident sound with amplitude equal to \SI{145}{dB} and the three frequencies analysed.}
    \label{fig:sound_induced_flow_grazing_no_flow}
\end{figure}

Building on the previous findings, it is evident that the combined effects of the lined surface and acoustic excitation influence both the time-averaged and fluctuating flow fields near the wall. These modifications, in turn, alter the flow dynamics within the orifices. The behaviour of the flow in and around the orifices \blue{might serve}  as a direct indicator of acoustic energy dissipation. In this section, details of the acoustic-induced velocity within the orifices are described. For this purpose, the acoustic-induced velocity fields have been obtained using the triple value decomposition approach described in Section \ref{sec:Methodology}. They are shown in Figure \ref{fig:sound_induced_flow_grazing_no_flow} for the case with SPL equal to 145 dB and the three frequencies. For each case, the inflow and outflow phases are shown, corresponding to the maximum and minimum velocity within the upstream orifice.

The reconstructed velocity is normalised by the theoretical acoustic velocity proposed by  \cite{morse_theoretical_1968} as done in previous studies \citep{Leon2019Near-wallFlow, Zhang2016NumericalLayers}, even if valid in the absence of grazing flow:

\begin{equation}
v^*_{ac}=\frac{|\blue{p^\prime}|}{\rho \omega (\tau+0.8d)} \frac{1}{\sqrt{   \left [(\frac{\omega_H}{\omega})^2-1 \right ] + \left( \frac{\omega_H}{\omega Q}\right)^2    }}
\end{equation}
where $\blue{p^\prime}$ is the pressure fluctuation, obtained from the SPL, $\omega$ and $\omega_H$ are the forcing and resonant frequency, $Q$ is the quality factor, $\rho$ is the density. A constant value of $Q = 7$ was used for both the no-flow and flow conditions. This value was determined as the best fit to match the acoustically induced velocity in the no-flow case, resulting in velocity values of 6.76 m/s, 23.66 m/s, and 12.37 m/s for acoustic wave amplitudes of 145 dB at frequencies of 800 Hz, 1400 Hz, and 2000 Hz, respectively.

In the absence of grazing flow, the velocity contours exhibit a slightly asymmetric spatial distribution for all frequencies and for both the inflow and outflow phases. The asymmetry concerning the center of the orifice $x/d=0$ is due to the grazing acoustic wave causing a flow reversal region near the upstream edge. This is true for both the inflow and outflow cases.  Additionally, the orifices exhibit mirrored velocity distributions, suggesting a mutual interaction between them. As expected, the amplitude of the acoustic-induced velocity is higher near the resonance frequency. 

When the grazing flow is introduced, the spatial distribution of the acoustic-induced velocity in the orifice changes. The quasi-steady vortex formed on the upstream edge of the orifice opposes the penetration of the acoustic velocity. Therefore, only the downstream half of each orifice is largely subjected to the periodic acoustic-induced motion. The spatial distribution is slightly affected by the frequency of the acoustic wave. Interestingly, for all the frequencies, the amplitude of the maximum velocity is very similar between the no-flow and the grazing flow cases. These observations agree with previous observations that one of the effects of the grazing flow is to reduce the effective porosity of the liner \citep{Zhang2016NumericalLayers, Avallone2021Acoustic-inducedLayer}. Finally, even though the wavelength exceeds the size of the cavity, there is still a delay in the time at which maximum velocity is reached at the two successive orifices. 
 
By comparing the spatial distributions of the acoustic-induced velocity, it can be noticed that while for the inflow phase, the acoustic-induced velocity penetrates within the cavity, for the outflow phase, the acoustic-induced velocity weakly penetrates within the turbulent grazing flow, thus suggesting that the flow opposes the ejection of acoustic-induced velocity from the cavity.

\begin{figure}
    \centering
     \centering
    \includegraphics[width=\linewidth]{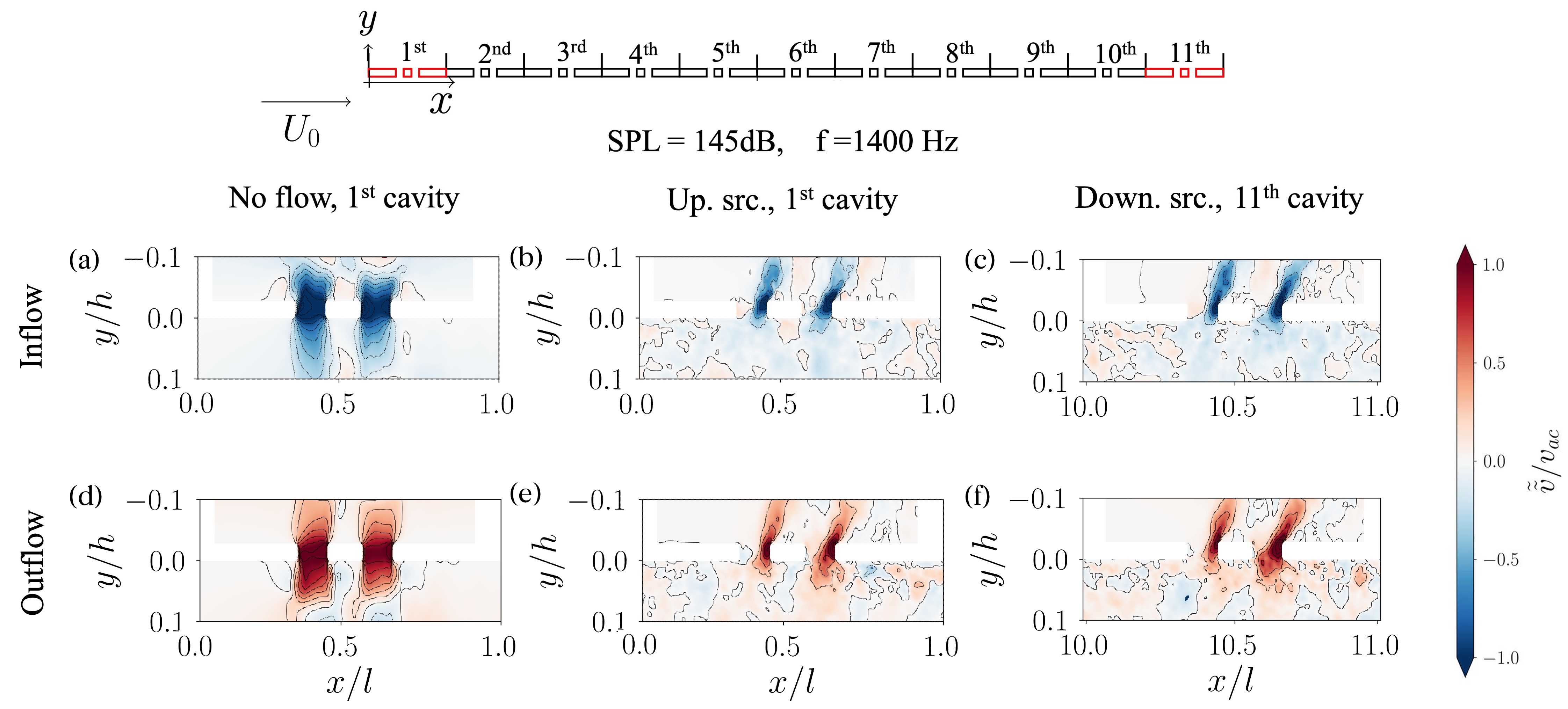}
    \caption{Contour plots of the acoustic-induced velocity inside the orifices for different acoustic source locations relative to the grazing flow. The top row represents the peak of the inflow phase, while the bottom row corresponds to the peak of the outflow phase. Acoustic source: SPL = \SI{145}{dB}, \textit{f} = 1400 Hz. The phase locking has been performed on the first orifice of each cavity with respect to the acoustic source propagation direction.}
    \label{fig:contours_effect_wave_direction}
\end{figure}

As described earlier, the streamwise evolution of the flow modifies the shear layer over the orifices and, therefore, \blue{depending on the shear strength and the local SPL, the way the acoustic wave penetrates within the orifice might differ.} For this reason, the acoustic-induced velocity fields are shown for the first and last of the cavities in Figure \ref{fig:contours_effect_wave_direction}. The contours of the acoustic-induced flow field are compared for the upstream and downstream-propagating acoustic waves. Only the case with the acoustic wave amplitude equal to 145 dB and frequency equal to 1400 Hz is shown for the sake of brevity.

The figure highlights the role of the shear layer over the orifices. As a matter of fact, for a given SPL, over the last cavity, the shear layer strength is lower, as it was shown in Figure \ref{fig:shear_contours}. Consequently, both during the inflow and outflow phases, the high-intensity region of the acoustic-induced velocity is more spread within the orifice. This behaviour is particularly evident when comparing the first and last orifices — i.e., those initially exposed to the acoustic wave. This suggests the streamwise development of the flow over the liner has an impact on how each cavity contributes to the acoustic dissipation. 

\subsection{Mass flow rate within the orifices}

To quantify the impact of the flow development, and in particular of the displacement of the near-wall flow over the liner, the phase averaged mass flow rate \blue{$\dot{m} (\phi)$} through each cavity has been calculated as:

\blue{
\begin{equation}
\dot{m} (\phi)
=
\int_{A_o}
\rho\!\left(\mathbf{x,\phi}\right)\,
\tilde{\mathbf{v}}\!\left(\mathbf{x, \phi} \right)
\cdot \mathbf{n}\,
\mathrm{d}A_o,
\end{equation}
}

\blue{where $A_o$ is the area of the eight orifices. Therefore, the mass flow rate is computed as the integral of the mass flux through the eight orifices of the cavity.}
\begin{figure}
    \centering
    \includegraphics[width=\textwidth]{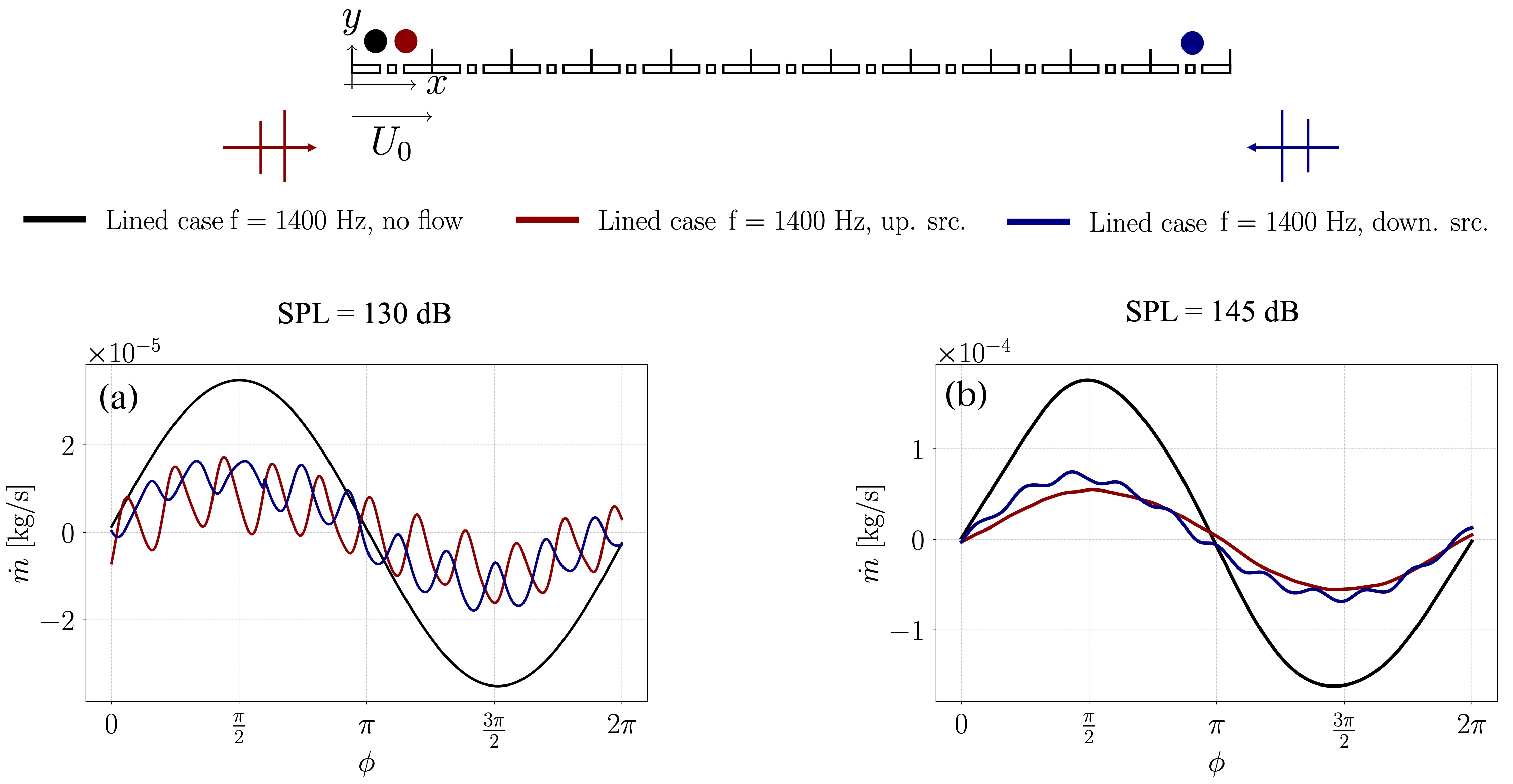}
    \caption{\blue{Comparison of the phase-locked mass flow rate over a cycle for a whole cavity. (a) SPL = 130 dB, (b) SPL = 145 dB.}}
    \label{fig:massflowrate}
\end{figure}

\begin{figure}
    \centering
    \includegraphics[width=\textwidth]{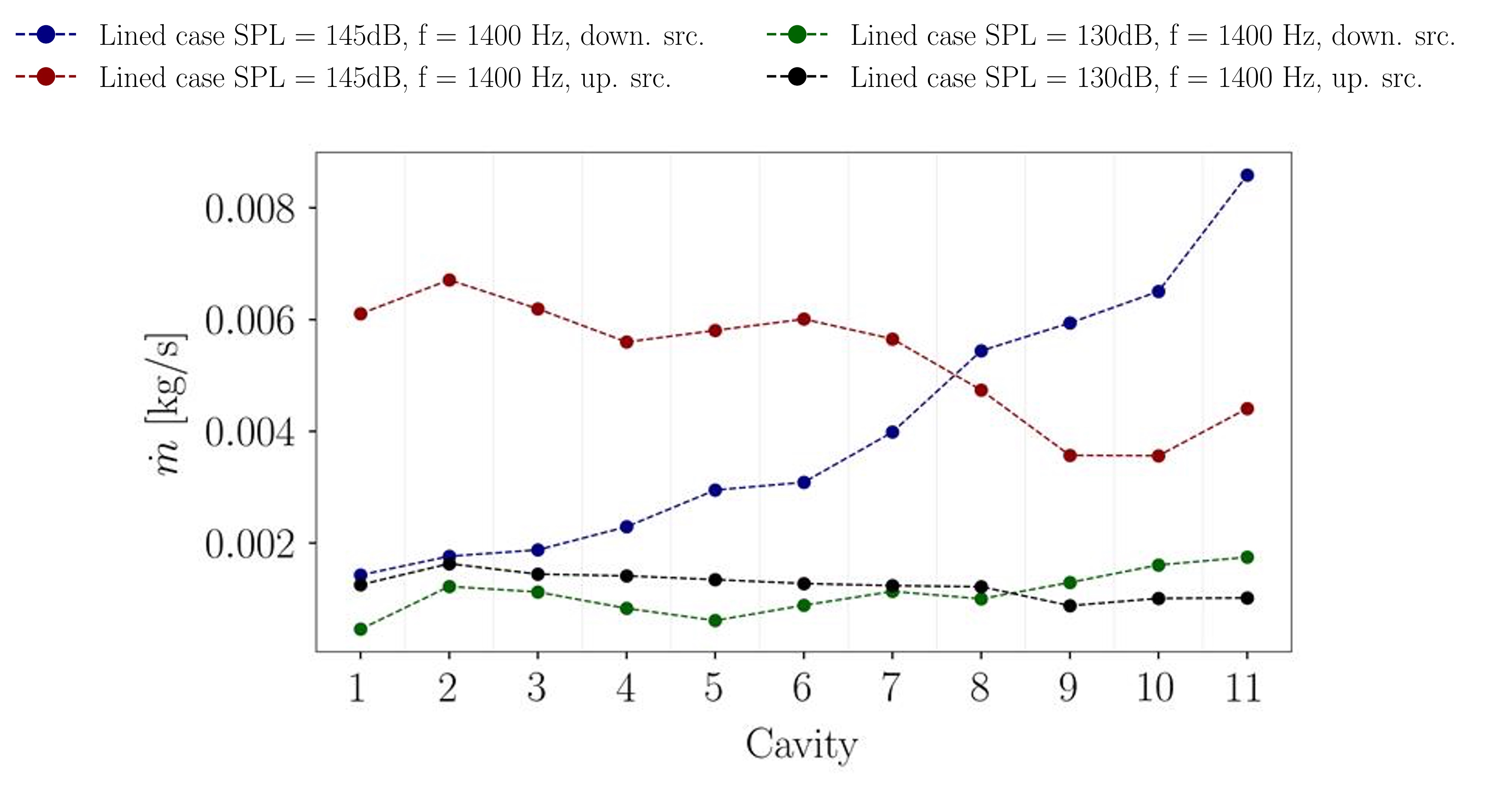}
    \caption{\blue{Streamwise evolution of the mass flow rate computed as the integral over the inflow phase. }}
    \label{fig:massflowrate_streamwise}
\end{figure}
Figure \ref{fig:massflowrate} (a) shows the phase-averaged mass flow rate over one acoustic cycle for plane waves with SPL equal to 130 dB and 145 dB over the first and last cavity. For the lowest SPL, the mass flow is one order of magnitude lower than at 145 dB. However, for both values of SPLs, the mass flow decreases in the presence of grazing flow, because of the reduction of the effective area due to the quasi-steady vortex. At 130 dB, there is almost no difference between the upstream and downstream cases. Based on the previous findings, this is likely related to the fact that the amplitude of the acoustic source is of the same order as turbulent velocity fluctuations. With SPL = 145 dB, \blue{the mass flow rate is higher} over the last cavity.  The weaker shear layer over the last cavity leads to an increased capability of the acoustic wave to penetrate inside the cavity, as \blue{described} earlier. The higher frequency fluctuations visible for both cases in the presence of flow are due to the acoustic fluctuation associated with the cavity depth mode \blue{that triggers a shear-layer instability over the orifice, thereby generating a Rossiter-like mode. This feature is identified at both SPLs, and it also persists in the absence of acoustic forcing. This confirms that it is not driven by the external excitation but rather corresponds to an intrinsic aerodynamic mode of the system, influenced by the cavity geometry and orifice shape \citep{Baars2024Smallflow}.

Theoretical background on such cavity modes is provided in \citet{PantonResonantresonators1975}, while recent experimental evidence of similar high-frequency tones has been reported by \citet{Baars2024Smallflow}. Moreover, a modal decomposition analysis, performed on a subset of the present dataset \citep{Scarano_decompositions}, revealed a mode corresponding to the self-excited tone at \SI{14}{kHz}.}

\blue{Figure \ref{fig:massflowrate_streamwise} reports} the streamwise distribution of \blue{mass flow computed as the integral over the inflow phase (0 to $\pi$ in Figure \ref{fig:massflowrate} (a-b)).} \blue{It is evident that, for a given SPL, i.e. 145 dB, the mass flow is higher of about 22\% at the last cavity for the downstream acoustic source with respect to the first cavity for the upstream acoustic source}. This is \blue{more} evident for the 145 dB case because the amplitude of the acoustic fluctuations is higher; it is less evident for the 130 dB case, but the trend is comparable. 

Referring to the SPL decay presented in Figure \ref{fig:spl_decay_145}, \blue{the different slopes between the different acoustic source locations may be explained by the different shear strength over the orifices along the liner.} Therefore, when the acoustic wave propagates in the direction opposite to the mean flow, it first encounters a thicker boundary layer displacement thickness and orifices with weaker shear strength, and then ones with larger shear strength and smaller boundary layer displacement thickness. The opposite happens for the upstream-located acoustic source. This supports the idea that, in conventional facilities, the boundary layer cannot be assumed thin. Its development over the liner must be accounted for, as it plays a crucial role in the interaction between the acoustic field and the flow. Furthermore, this finding suggests that the flow development must be considered in eduction methods, which is not the case for the methods currently adopted.

\section{Conclusions}
\label{sec:conclusion}
Lattice-Boltzmann Very Large Eddy simulations of a single-degree-of-freedom locally reacting liner in the presence of grazing flow and grazing acoustic waves were performed. The geometry and turbulent flow conditions replicated experiments carried out in the flow impedance tube at the Federal University of Santa Catarina. Simulations of a single row of 11 cavities were carried out. The impact of the acoustic wave amplitude, frequency, and propagation direction was analysed. The resulting database provides a comprehensive benchmark for future modelling, physical interpretation, and data-driven analysis of flow–acoustic coupling in these systems. 
Numerical simulations, validated against experimental measurements of impedance, confirm that the presence of the grazing flow increases the resistance component of impedance and that, when using eduction techniques, differences are present when varying the propagation direction of the acoustic wave with respect to the mean flow. 

Numerical simulations were used to compute the local impedance over the entire liner surface using the in-situ Deans' method. It was found that using this method, the value of impedance changes largely depending on the sampling location over the face sheet. Furthermore, differences are found when comparing impedance computed using the in-situ and the mode matching method, in agreement with experimental findings.  Interestingly, although the in-situ method is sensitive to the sampling location, when impedance is averaged over the surface, it does not depend on the propagation direction of the acoustic wave. On the other hand, this discrepancy is more evident when looking at the results obtained with the mode matching method. 
The streamwise distribution of impedance obtained with the in-situ method suggested that local flow features near the wall play a role by enhancing or reducing the local SPL. This was confirmed by looking at the flow field near the face sheet surface. 

The presence of orifices, coupled with the periodic injection of momentum due to the reaction of the liner to the acoustic excitation, causes the displacement of flow further away from the face sheet surface. Therefore, the boundary layer displacement thickness \blue{($\delta^*$)}, already recognised as a key parameter in semi-empirical models, has been shown to vary significantly along the liner surface.
The increase of the boundary layer displacement thickness in the streamwise direction causes the shear layer over downstream orifices to be weaker, thus allowing a larger portion of the orifice to be subjected to the periodic acoustic-induced flow field. This was confirmed by looking at the spatial distribution of acoustic-induced velocity and mass flow rate. The latter increases \blue{, for the downstream cavity when the acoustic wave propagates against the turbulent flow}. As a consequence of an increase in the effective area and a weaker shear layer, the near-wall field seen by the acoustic wave over the cavities changes if the acoustic wave propagates in the same direction as or opposite to the mean flow. 
\blue{Moreover, the differences in the turbulent flow and the local $\delta^*$ lead to distinct shear-flow profiles and velocity fluctuations seen by the acoustic wave depending on their propagation direction, which in turn might modify the portion of the acoustic field that is scattered and refracted at the wall impedance discontinuity.}

The findings of this paper suggest that acoustic energy is dissipated differently along the liner because the development of the flow near the wall affects the shear layer strength over the orifices. This aspect is not accounted for in the boundary conditions adopted for educing impedance and \blue{may} therefore impact the measured impedance. It is essential to account for the finite boundary layer; in particular, the value of the boundary layer displacement thickness and its growth cannot be neglected. This also suggests that eduction methods based on streamwise varying impedance might be a good choice for a proper characterisation of the acoustic liners.

Future studies shall investigate a longer liner and compare two boundary layer profiles at similar bulk or centerline Mach numbers \blue{to evaluate the impact of a higher $\delta^*/L$ and different  $\delta^*/d$ on the acoustic response of the liner.}

\appendix
\section{}\label{app:Friction coefficient for the smooth wall case}
\blue{\subsection{Numerical validation}}
The friction coefficient $C_{f}$ has been computed for the smooth wall case to validate the simulations. In a channel flow, the friction coefficient is defined as $C_{f} = 2\tau_{w}/(\rho u_{b}^{2})$ where $\tau_{w} = -dp/dx \cdot \delta$ and $u_{b} = \int_{0}^{2h} u dy/(2h)$ is the bulk velocity. 

\begin{figure}
    \centering
    \includegraphics[width=0.6\linewidth]{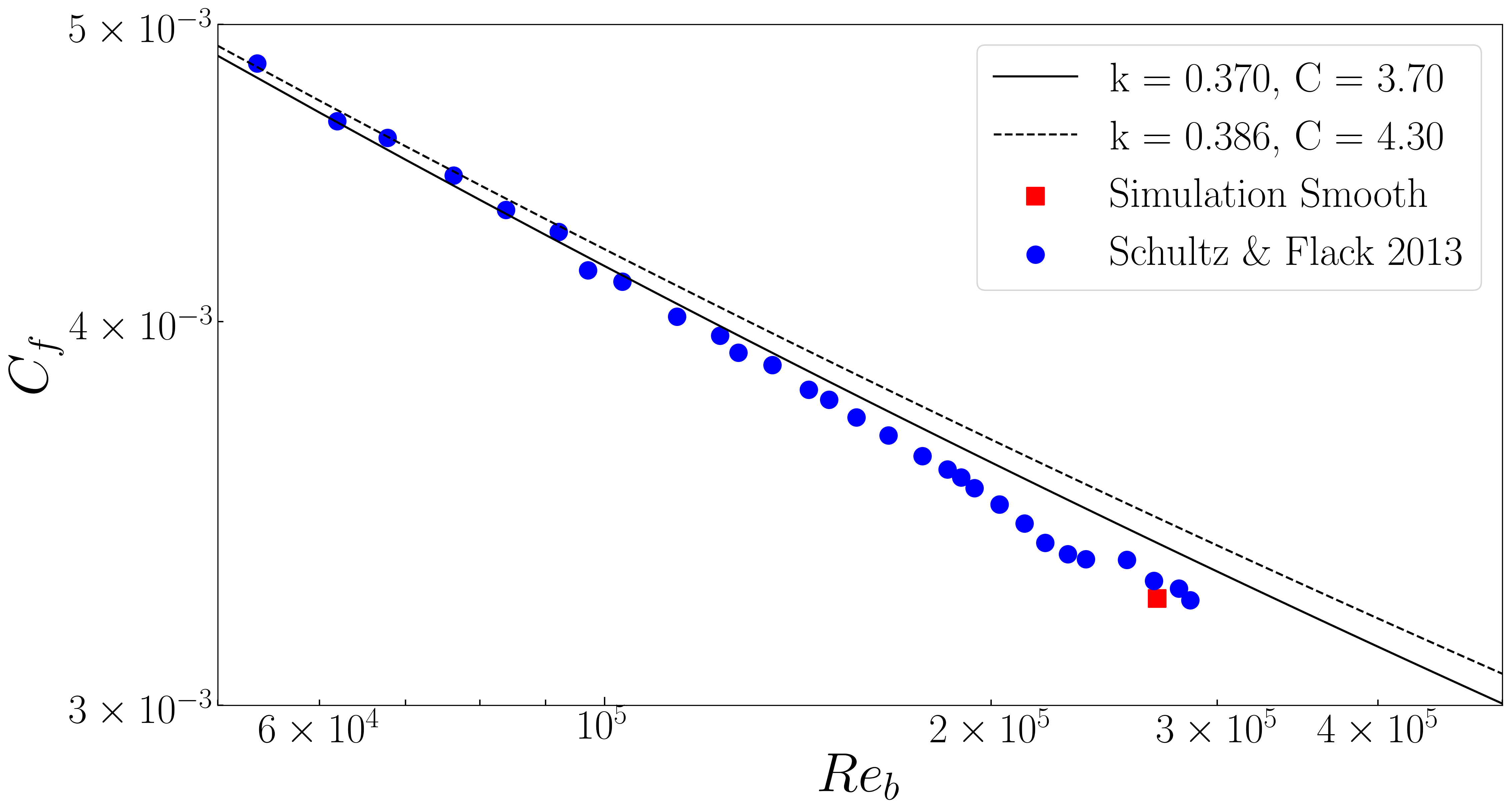}
    \caption{Comparison of skin friction coefficient with friction law Eq. \ref{eq:frictionlaw} and experiments.
    The solid line indicates the friction law \ref{eq:frictionlaw}, while experimental data are ones by \cite{Schultz2013}.}
    \label{fig:friction_coefficient}
\end{figure}

In figure \ref{fig:friction_coefficient}, the simulation results are shown together with the experimental findings from \cite{Schultz2013} and the results from  Prandtl's smooth wall flow formula:

\begin{equation}
    \sqrt{\frac{2}{C_{f}}} = \frac{1}{k}log\Biggl(\frac{Re_{b}}{2} \sqrt{\frac{C_f}{2}}\Biggr) + C -\frac{1}{k}
    \label{eq:frictionlaw}
\end{equation}
where $Re_{b} = 2hu_{b}/\nu$ is the Reynolds bulk. Prandtl's formula has been tested with two different sets of log-law coefficients: those suggested by \citet{Nagib2008}, $k$=0.370, $C$ = 3.70; and those suggested by \citet{Bernardini2014}, $k$=0.386, $C$ = 4.30.  While the discrepancies are minor for $Re_{b} < 2 \cdot 10^{5}$, those tend to increase with increasing $Re_{b}$. A closer agreement is found between simulation data and Nagib's set of coefficients. Overall, results from the numerical simulations are in line with expectations.

\blue{\subsection{Computational cost}
\label{app:computational cost}
To provide the reader with an indication of the computational effort required to build the present numerical database, an overview of the associated cost is given below. The preliminary flow-convergence simulations, run on 10 computer nodes with 48 processors each, required a total of 341.86 wall-clock hours. To optimise computational resources, coarser simulations were first carried out to develop the flow field within the channel; their solutions were then used to initialise the fine-grid acoustic simulations.

The acoustic simulations were performed on 8 compute nodes (383 solver processes in total) connected via InfiniBand, using single precision and AVX2 vectorisation. Each case contained approximately $1.1\times10^9$ voxels with 8 refinement levels and was advanced for $3.84\times10^5$ time steps, requiring 86 hours of wall-clock time and $3.3\times10^4$ CPU-hours, corresponding to an average performance of 1.25 time steps per second. }
\section{}\label{appA}
\begin{figure}
    \centering
        \centering
        \includegraphics[width=0.45\textwidth]{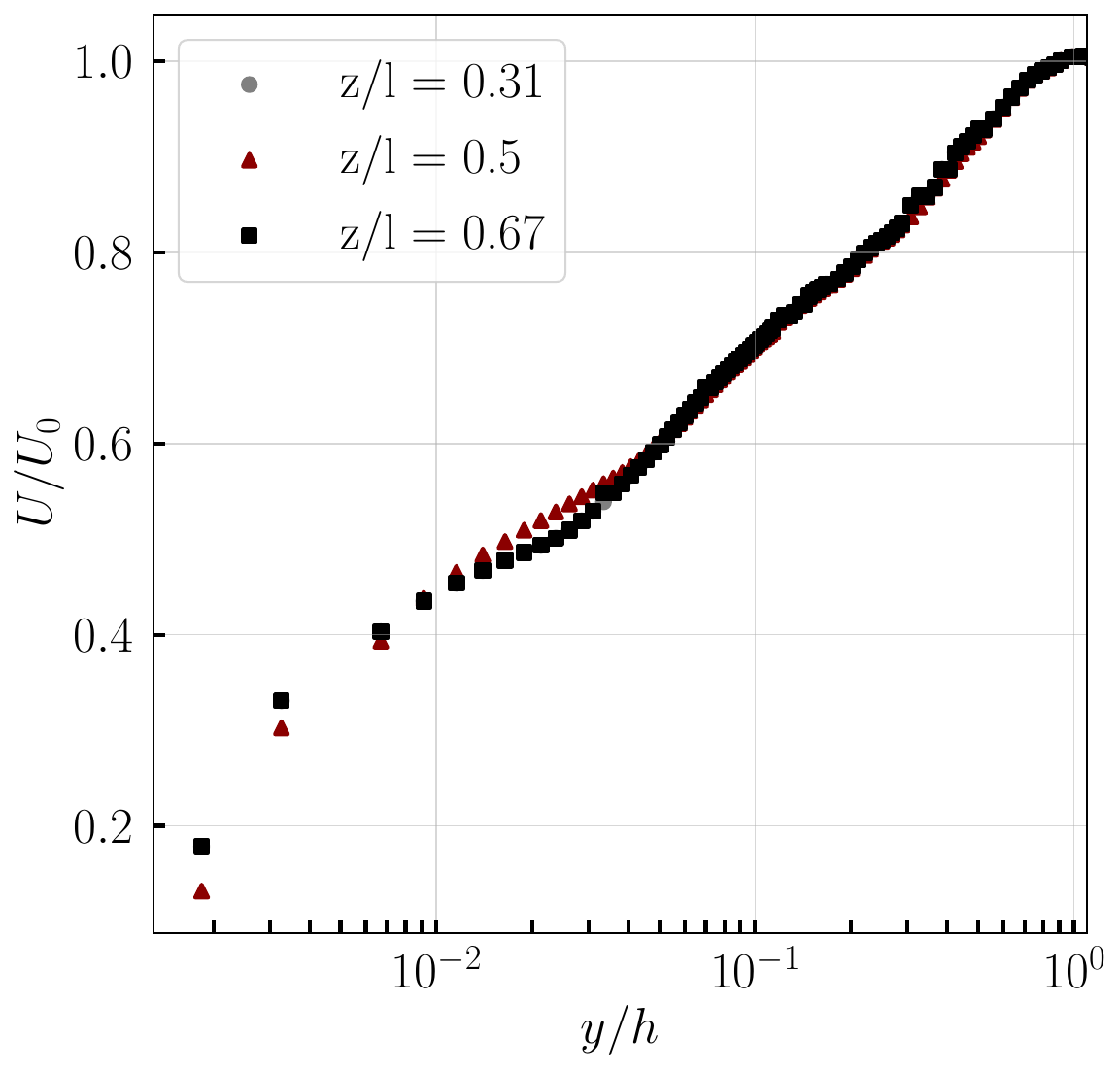}
        \caption{Boundary layer velocity profiles upstream of the liner and at three spanwise locations in the absence of acoustic waves.}
        \label{fig:Spanwise_profile}
\end{figure}

The analysis \blue{concerning the streamwise evolution of the flow profile and second-order statistics has been conducted on a streamwise plane located at $z/l=0.5$, i.e., at the mid-span of the channel flow with periodic boundary conditions.} Since the orifices do not have the same distribution in the span, Figure \ref{fig:Spanwise_profile} shows that the velocity profile upstream of the liner is weakly affected by the span. Differences are present in terms of streamwise evolution because of the large number of orifices, but these do not affect the overall conclusions of the paper.

\section*{Acknowledgment}
The work of A.P., F.S. and F.A. is co-funded by the European Union (ERC, LINING, 101075903). Views and opinions expressed are, however, those of the author(s) only and do not necessarily reflect those of the European Union or the European Research Council. Neither the European Union nor the granting authority can be held responsible for them. The work is partially supported by the AeroAcoustics Research Consortium (AARC), a government-industry partnership supporting pre-competitive research for aircraft noise reduction. JAC acknowledges funding from the Brazilian funding agency CNPq (National Council for Scientific and Technological Development) through the process 315000/2021-0.

\section*{Declaration of Interest}
The authors report no conflict of interest.

\section*{Data availability statement} The simulations data are available at https://zenodo.org/records/19565103.
\newpage
\bibliographystyle{jfm}
\bibliography{jfm}

\end{document}